\documentclass[preprint,showpacs,preprintnumbers,amsmath,amssymb]{revtex4}

\usepackage{graphicx}% Include figure files
\usepackage{bm}% bold math

\begin{document}
%\preprint{APS/123-QED}

\title{Kondo effect of cobalt adatom on zigzag graphene nanoribbon}

\author{Damian Krychowski}
%\email{krychowski@ifmpan.poznan.pl}
\author{Jakub Kaczkowski}
%\email{kaczkowski@ifmpan.poznan.pl}
\author{Stanislaw Lipinski}
%\email{lipinski@ifmpan.poznan.pl}
\affiliation{%
Institute of Molecular Physics, Polish Academy of Sciences\\M. Smoluchowskiego 17,
60-179 Pozna\'{n}, Poland
}%
\date{\today}% It is always \today, today,

\begin{abstract}
Based on ab-initio calculations we discuss Kondo effect due to Co adatom on graphene
zigzag nanoribbon.  Co atom located at hollow site behaves as spin ${\cal{S}} = 1/2$ impurity with
$d_{xz}$ and $d_{yz}$ orbitals contributing to magnetic moment. Dynamical correlations are analyzed with the use of complementary approximations: mean field slave boson  approach, noncrossing approximation and equation of motion method. The impact of interplay between spin and orbital degrees of freedom together with the effect of peculiarities of electronic and magnetic structure of nanoribbon on many-body resonances is examined.
\end{abstract}

\maketitle

\section{\label{sec:level1}INTRODUCTION}
Graphene possesses spectacular electronic, optical, magnetic, thermal and mechanical properties, which make it an exciting material for  technological applications \cite{Novoselov, Geim, Castro Neto, Das Sarma, Katsnelson}. Graphene is a semimetal. For the use  in logic devices a controllable band gap is very much desired. Presence of a gap would increase tremendously the on-off ratio for current flow that is needed for many electronic applications. For example, lack of the  gap prevents the use of graphene in making transistors. Band gap opening is caused by symmetry breaking \cite{Avouris, Zhou}. The most effective way within the realm of single-layer graphene physics is electron confinement e.g. in nanoribbons (partial breaking of translational symmetry) \cite{Wakabayashi, Palacios, Pisani, Yang}. The graphene nanoribbons (GNR) with varying widths can be realized either by cutting \cite{Hiura}, mechanically exfoliated graphenes \cite{Novoselov1} or by pattering epitaxially grown graphenes \cite{Zhang}. The edge geometry is the key factor which determines the electronic properties of nanoribbon. There are two types of nanoribbons, based on their edges shapes, called zigzag (ZGNR) and armchair (AGNR) \cite{Palacios, Nakada}. Recently, electronic devices, such as field effect transistors, have been formed from graphene nanoribbons \cite{Ozyilmaz, Han}. ZGNRs are of particular interest, because due to topological reasons  they are forming edge states \cite{Fujita} i.e. states decaying exponentially into the centre of the ribbon  \cite{Castro Neto, Brey}. The decay lengths are in the range of a few nanometer \cite{Niimi}. The edge states has been observed in scanning tunneling microscopy \cite{Kobayashi}. The localized nature of these states gives rise to a flat band extending over one-third of the one-dimensional Brillouin zone and correspondingly also to a sharp peak in the density of states right at the Fermi level. As a consequence magnetic ground state emerges from a Fermi instability \cite{Son, Wu}. Recently the spin splitting of the edge density of states of ZGNRs has been confirmed experimentally \cite{Tao}. Theoretical studies have shown that the spins on each edge are ferromagnetically ordered, and those between the edges are antiferromagnetically coupled, the later resulting from the interaction of the tails of the edge states \cite{Son2, Wu, Kunstmann, Kan1, Pisani}.  Modification of the electronic structure can be also introduced  by chemical functionalization, what allows the band gap engineering and designing different types of magnetic order. Based on density functional calculations Son et al. \cite{Son}  have shown  that one can modify the band gap of ZGNR by applying transverse electric field and that the electric field  closes the gap for one of the directions selectively (half metallicity). This conclusion has been confirmed by calculations of Kan et al. \cite{Kan1} with the use of hybrid functional potential (B3LYP), which is viewed as one of the most accurate methods for estimation of the gap. The predicted critical fields of transition into half metallicity are much higer in this method than those from normal DFT calculations. It is well known, that graphene nanostructures are promising for spintronics due to their long spin relaxation and decoherence times owing to the low intrinsic spin-orbit interaction \cite{Tombros}.  The mentioned possibility of band gap tuning and controlling magnetism and spin transport  of the ribbons  by electric field is the principal advantage of these systems. The pure nanoribbon has no net magnetic moment. The functionalities of the ribbons can be enriched by doping the magnetic adatoms. Due to the open surface controlled adatoms manipulation is within reach of atomic force microscopy in these systems \cite{Seo, Gross}. In the last few years several studies focused on understanding structural, electronic and magnetic properties of $3d$ impurities in  graphene nanoribbons \cite{Kan3, Longo, Power, Sevincli, Rigo, Cocchi}. Also vacancies and defects have been predicted to give rise to magnetic moments \cite{Lehtinen, Palacios2}. The relative stability of local moments depends on the balance between the Coulomb repulsion, exchange interaction, position of  $3d$ levels and hybridization with the neighboring carbon atoms. Especially two latter factors are strongly affected by impurity location, one expects  different energetics, structural, and electronic properties nearly the edge sites of GNRs and different when adatom is located inside the ribbon. The electronic structure of nanometer-wide ribbon is dominated by confinement effects and Van Hove singularities and this strongly affects the hybridization path. As opposed to normal metals, the damping of the local levels is energy dependent and the hybridization self energy  acquires also significant real contribution near singularities causing effective shift of local energy levels.   Since the chemical potential of GNRs can be tuned, a formation of local moment can be controlled by gate voltage and particularly strong gate dependence is expected near singularities. At low temperature, the localized spin is screened by conduction electrons and a narrow Kondo peak appears near the Fermi level. Most of the early studies in Kondo effect were carried on for the metallic systems with constant density of states at the Fermi surface, in the case of graphene structures the details of the band structure play the decisive role in screening. Recently Kondo effect has been observed in graphene both in resistivity measurements \cite{Chen} and by scanning tunneling microscopy (STM) \cite{Mattos}.  As opposed to transport measurements STM probes local electronic properties of Kondo impurities.  The Kondo resonce observed in tunneling spectroscopy usually does not show up as a peak but rather as a dip. This is a consequence of interference of direct channel into the localized orbitals of impurity and an indirect one to the bands of the host \cite{Madhavan}. The Kondo temperature in graphene is tunable with carrier density from $15-90$ K \cite{Chen, Mattos}. A number of interesting theoretical studies have been published on this topic discussing specificity of Kondo screening for the gapless system, where a critical hybridization is necessary for the occurrence of this effect \cite{Chao, Uchoa, Vojta}. Due to valley degeneracy of the Dirac electrons in perfect graphene the possibility of  multichannel Kondo effect has been also discussed \cite{Zhu}. Recently appeared two fundamental, realistic studies of Kondo effect of single Co adatom in graphene based on first principles calculations \cite{Wehling, Jacob}. These papers expose the role of orbital symmetry on dynamical correlations.  Along this line is also analysis presented in the present paper.

The topic of our study is Kondo effect in zigzag graphene nanoribbon. The crucial requirement of the occurrence of Kondo effect is that the adatom should retain its magnetic  moment in the presence of electrons of the host.   We open our analysis with  presentation of the  first-principles electronic structure calculations of Co impurity in   narrow zigzag GNRs  discussing  energetics, geometry of adsorption, magnetic moments and magnetization densities for different positions of impurities. We  discuss which adsorption site is most favorable  and show the result of optimization of the adsorption height and indicate which orbitals most strongly hybridize with nanoribbon states and which contribute to impurity magnetic moment. Both the binding energies of the impurity  and the magnitude of the moment strongly depend on the location of the adatom across a ribbon. Due to the strong variation of  ZGNR  density of states with chemical potential an interesting question arises of possibility of driving the magnetic impurity in and out of the Kondo regime. Another important problem is how the  Kondo  screening is affected by ZGNR edge states and what is the role  of polarization of these states in spin-orbital Kondo effect. Performing the calculations for different locations of chemical potential with respect to the band gap, also for the case when it crosses the low energy singularities of density of states, allows us to analyze different coupling regimes and  track  an impact of symmetry breaking in  both orbital and spin  sectors. In general more than one orbital effectively contribute to magnetic moment and in the  Kondo screening apart form spin also orbital degrees of freedom are involved. The role of orbital of a given symmetry  changes  both with geometrical location of impurity and  with position of the Fermi level.  Static mean-field methods like density functional calculations (DFT) cannot describe dynamical electron correlations. Therefore for simple and intuitive analysis of many-body correlations  we use the multiorbital Anderson-like  model in which impurity is described by parameters, but nanoribbon electronic structure  and hybridization function are calculated within  DFT.  This Hamiltonian is then solved in the next step by commonly used many-body approximate methods with the well known applicability regimes and limitations.  The principal method used in the present work, the slave boson mean field approach (SBMFA) best describes systems close to the Kondo fixed point i.e. for the case of fully degenerate deep  atomic levels at low temperatures \cite{Hewson}, but often is also used for a qualitative insight away from this limit.  We adopt the Kotliar-Ruckenstein formulation  \cite{Ruckenstein, Dong}, which is convenient tool for  discussing   finite Coulomb interaction case and for analysis of effects introduced by polarization. Two other complementary methods used by us:   equation of motion method (EOM) \cite{Lacroix, Entin, Kashcheheyevs, Wingreen} and noncrossing approximation (NCA) \cite{Bickers, Pruschke, Grewe, Haule, Kuramoto, Gerace} allow to get a deeper insight into the role of charge fluctuations in many-body physics and are better adopted for higher temperatures. EOM works in the whole parameter space except the close vicinity of Kondo fixed point but it breaks at low temperatures \cite{Kashcheheyevs} and NCA  gives reliable  results  in the wide temperature range, including the region close to $T_{\cal{K}}$ and in the  range of the lowest temperatures down to fraction of $T_{\cal{K}}$.  It is claimed that this method is not suitable for spin polarized systems due to the well known artifacts resulting from the neglect of vertex corrections \cite{Wingreen}.

The paper is organized as follows: Section II presents density functional theory calculations of electronic and magnetic properties of zigzag graphene nanoribbons in transverse electric fields and analyzes adsorption of Co adatom in these structures. In Sec. III the generalized Anderson model  with DFT hybridization function is described. Next we present numerical results and analyze the impact of  confinement and band gap  singularities of electronic structure  as well as the role of orbital physics and magnetic polarization   on the Kondo effect. Finally, we give  conclusions  and some final remarks in Sec. IV.

\section{DENSITY FUNCTIONAL STUDY OF Co ADATOM ON ZGNR}

\subsection{Computational details}

Zigzag nanoribbons  are quasi-one dimensional structures with infinite length and nanometric widths, the latter being defined by the parameter N indicating the number of zigzag lines along the ribbon widths.  Most of our considerations are addressed to 4ZGNR ($N=4$, Fig.1), but we also present some comparative calculations for wider ribbons.  To saturate the edge C dangling bonds the ribbons are passivated by hydrogen atoms. The following first principles analysis of Co adatom on graphene nanoribbon provides the necessary input information for  analysis of correlation effects, which we undertake in the next chapter. Here we discuss which are  the most favorable adsorption sites for Co atom, the corresponding electron configurations and magnetic moments, as well as  impurity induced  magnetic polarization of the  ribbon.  For simulation of Co impurity we have used a supercell consisting of four graphene unit cells, which  contains one adatom. To check whether this supercell is  sufficiently large to obtain reliable results, especially  concerning  magnetic moment, we have  also performed testing calculations for larger supercells obtaining similar results. We consider  three classes of high symmetry  adsorption sites presented in Fig. 1: hollow - in the centre of the carbon hexagon (h), top - at carbon atom (t) and bridge (b) - between two carbon atoms. Unlike graphene, where infinite plane ensure the equivalence of lattice sites, in nanoribbons the number of inequivalent position of impurities within each class increases with the width of the ribbon.  For convenience of the discussion the different carbon atoms spaced across the ribbon  are also marked in Fig. 1. To get an insight into the interaction of Co adatom on nanoribbon we performed spin-polarized density functional calculations. The main idea of DFT is to describe the interacting system of fermions via its density and not via its many-body wavefunction \cite{Hohenberg}. The key problem of DFT formalism  is a choice of exchange-correlation potential. Most of our calculations have been performed using semilocal  generalized gradient approximation (GGA) with  Perdew, Burke and Ernzerhof (PBE) formula for the exchange-correlation \cite{PBE}. The inclusion of gradient corrections  is of special importance for the considered systems,  because  large gradients in the charge density  occur at the nanoribbon edges. Since it is known, that  local approaches often underestimate magnetic moments and  band gaps, we have also done some test calculations using  hybrid non-local exchange potential HSE \cite{HSE, Paier, Gillen, Xiao, Park, Barone}.   The mixing of nonlocal and semilocal exchange  overcomes the major flaws of LDA or GGA \cite{Pisani, Kan2}.
\begin{figure}
\includegraphics[width=7.2 cm,bb=50 0 390 220,clip]{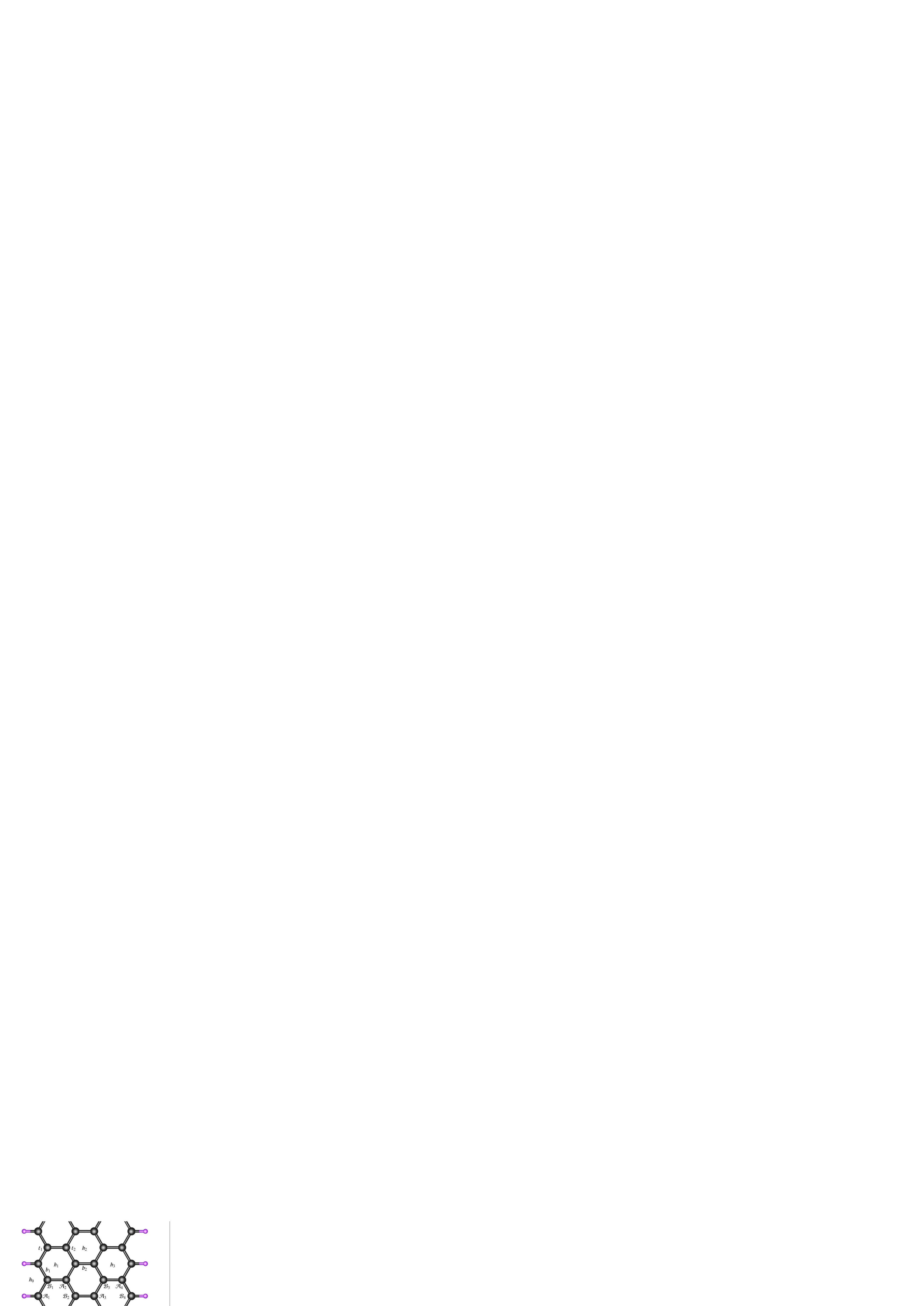}
\caption{\label{fig:epsart1}(Color online) Structure of H passivated 4ZGNR, where the dark grey and purple/light grey spheres represent the C and H atoms respectively. Labeling of carbon atoms across the ribbon with division into sublattices is introduced. The possible adatom locations: $h_{1}$, $h_{2}$, $h_{3}$ (hollow positions - most favorable), $t_{1}$, $t_{2}$ - top sites, $b_{1}$, $b_{2}$ - bridge positions and perturbed hollow like site of the edge $h_{0}$ are depicted.}
\end{figure}
Concerning the choice of the wave function basis set   two codes have been employed: Vienna simulation package (VASP) \cite{VASP} with the projector augmented wave basis sets (PAW) \cite{PAW}  and OPENMX, which uses basis set of localized pseudoatomic orbitals (LCPAOs) \cite{OPENMX}. In the latter case for the geometrical optimization and the electronic band structure calculations the LCPAO basis functions were specified by the choice of two primitive orbitals for $s$ component  and one primitive orbital for $p$ component for hydrogen ($H5.0-s2p1$) and three $p$ orbitals for carbon ($C5.0-s2p3$). The cutoff radius of $5.0$ Bohr has been assumed.  VASP code is widely used, but  due to  the plane wave picture it is difficult to describe the effects of edges and to discuss  field induced charge accumulation or dipole moments.  In both codes the GGA-PBE exchange-correlation potential has been  adopted \cite{PBE}, which is  specified not only by spin densities, but also by their gradients.  In comparison with LSD  GGA's  tend to improve total energies and structural difference \cite{PBE}.  In VASP, where smooth pseudopotentials are  used  a kinetic energy cutoff of $400$ eV was found to be sufficient to achieve a total energy convergence of the energies of the systems to within $1$ meV. In OPENMX real-space grid technique was adopted in numerical integration with energy cutoff up to $150$ Ry.
In both methods the structures were relaxed until the Hellman-Feynman force became smaller than $10^{-4}$ Ha/bohr. Brillouin integration was carried out at $8\times1\times1$ Monkhorst-Pack grid and Gaussian smearing of $0.03$ eV was chosen to accelerate electronic convergence in both codes. For band structure calculations $50$ and $200$ uniform k points along the one-dimensional BZ were used in VASP and OPENMX respectively.
To avoid interaction between images made by periodic boundary conditions the vacuum region was set up to $18$ {\AA} in y- and up to $20$ {\AA} in z-directions, in x-direction ribbon was treated as infinite. The adatom - ribbon system lacks inversion symmetry and therefore has a net electric magnetic moment perpendicular to the surface. To remove spurious dipole interaction between periodic images, we selfconsistently applied corrections to the local electrostatic potential and total energy \cite{Chan}.
To test an impact  of correlations on the adsorption energy and magnetic moments we have performed also some   GGA+U type calculations using rotationally invariant LDA+U functional proposed by Lichtenstein et al. \cite{Liechtenstein}. The stability of  adatom  on the relaxed GNR was examined analyzing adsorption energy defined as:
\begin{eqnarray}
{\cal E}_{ads}={\cal E}_{{\cal{ZGNR}}+{\cal{C}}o}-{\cal E}_{{\cal{ZGNR}}}-{\cal E}_{{\cal{C}}o},
\end{eqnarray}
where the first term is total energy of ZGNR with Co adatom, and second and third are total energies of clean ZGNR and isolated Co atom.

\subsection{Electronic and magnetic properties of ZGNR}

It is now  well established, that zigzag edge GNR is a semiconductor with two electronic edge states, which are ferromagnetically (F)  ordered, but antiferromagnetically (AF) coupled to each other \cite{Kan1, Son, Son2}. This configuration is consistent with the Lieb theorem \cite{Lieb}. It is also well understood, that magnetism of the edges arises from a Fermi instability of the edges \cite{Gillen}.
\begin{table}
\caption{\label{tab:table1}Differences between energies  of NZGNR states  with  antiparallel (AF) and parallel (F) alignment of magnetic moments on the left and right edges compared with corresponding differences of energies of AF  and nonmagnetic states ($N = 3 - 10$) (VASP).}
\begin{ruledtabular}
\begin{tabular}{c|c|c}
${\cal{N}}$&${\cal{E}}_{{\cal{AF}}-{\cal{F}}} [meV]$&${\cal{E}}_{{\cal{AF}}-{\cal{NM}}} [meV]$\\
\hline
$3$ & $-18.61$ &  $-39.01$\\
$4$ & $-13.23 (-14\footnote[1]{Quantum-Espresso, PBE, Ref. \cite{Marzari}}, -15\footnote[2]{VASP, PBE, Ref. \cite{Jiang}}, -11\footnote[3]{SIESTA, PBE, Ref. \cite{Martins}})$ &  $-55.74 (-59\footnotemark[1])$\\
$5$ & $-10.67$ &  $-65.10$\\
$6$ & $-11.39 (-11.9\footnote[4]{SIESTA, PBE, Ref. \cite{Sun}}, -15\footnotemark[1])$ &  $-69.41 (-85.7\footnotemark[4], -81\footnotemark[1])$\\
$7$ & $-8.83 (-11.2\footnotemark[4])$ &  $-71.64 (-89.4\footnotemark[4])$\\
$8$ & $-5.88 (-8.4\footnotemark[4], -7\footnotemark[1])$ &  $-73.73 (-91.8\footnotemark[4], -83\footnotemark[1])$\\
$9$ & $-4.23$ &  $-75.75$\\
$10$ & $-3.34 (-5.5\footnote[5]{SIESTA, PBE, \cite{Zheng}})$ & $-77.47$\\
\end{tabular}
\end{ruledtabular}
\end{table}
Our VASP  calculations show  that for $N = 4$ the unpolarized solution has energy by   $\Delta{\cal{E}}=55.74$ meV per edge carbon atom   higher compared to  AF state and $13.23$ meV higher than F state. The energy difference between parallel and antiparallel orientations of magnetizations at the edges   decreases with the width (Tab. I)  indicating that the increase of the overlap of edge states is responsible for relative ordering of polarizations.  The obtained values are in good agreement with results reported by other groups.
\begin{figure}
\includegraphics[width=7.2 cm,bb=70 0 580 330,clip]{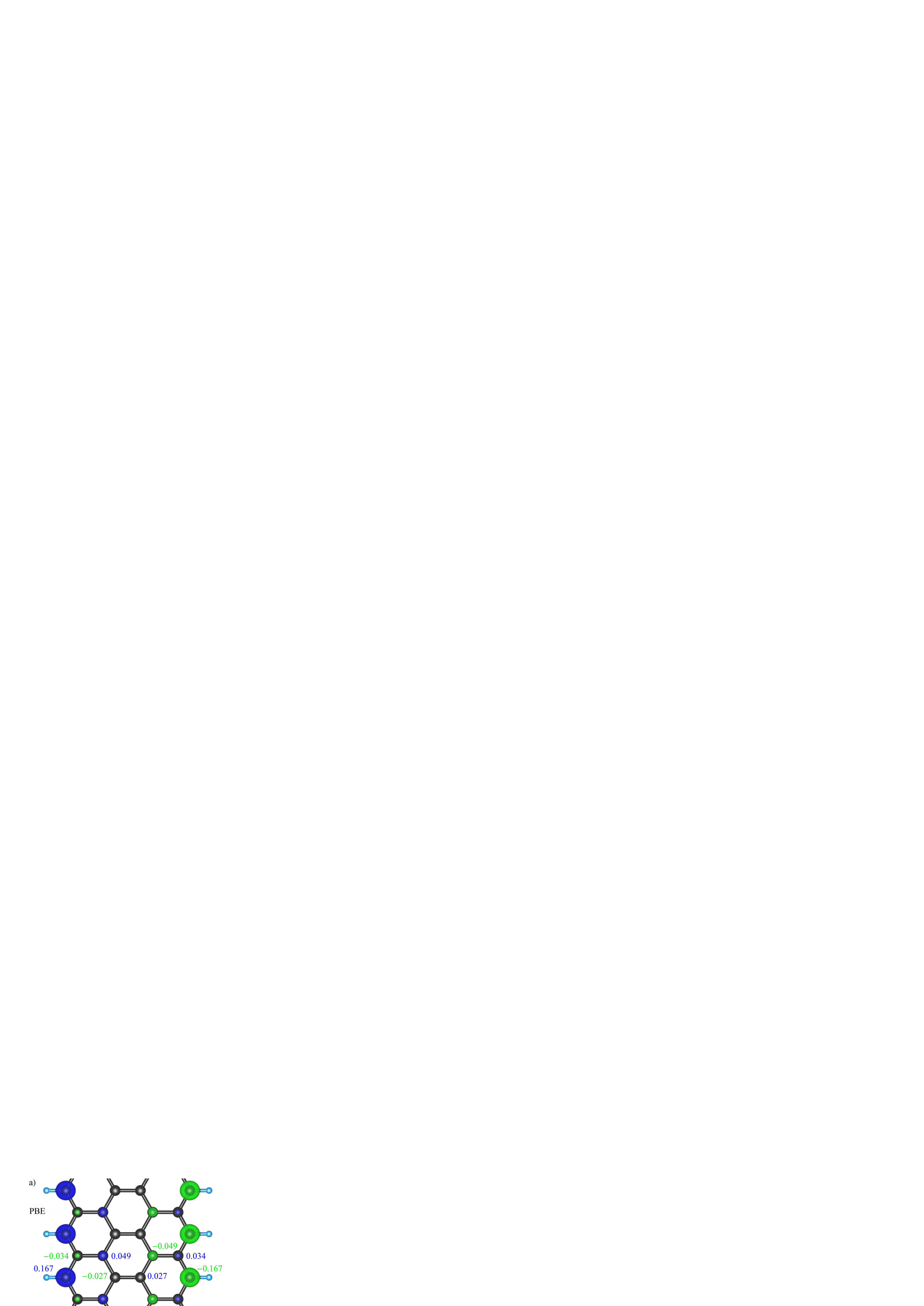}
\includegraphics[width=7.2 cm,bb=70 0 580 330,clip]{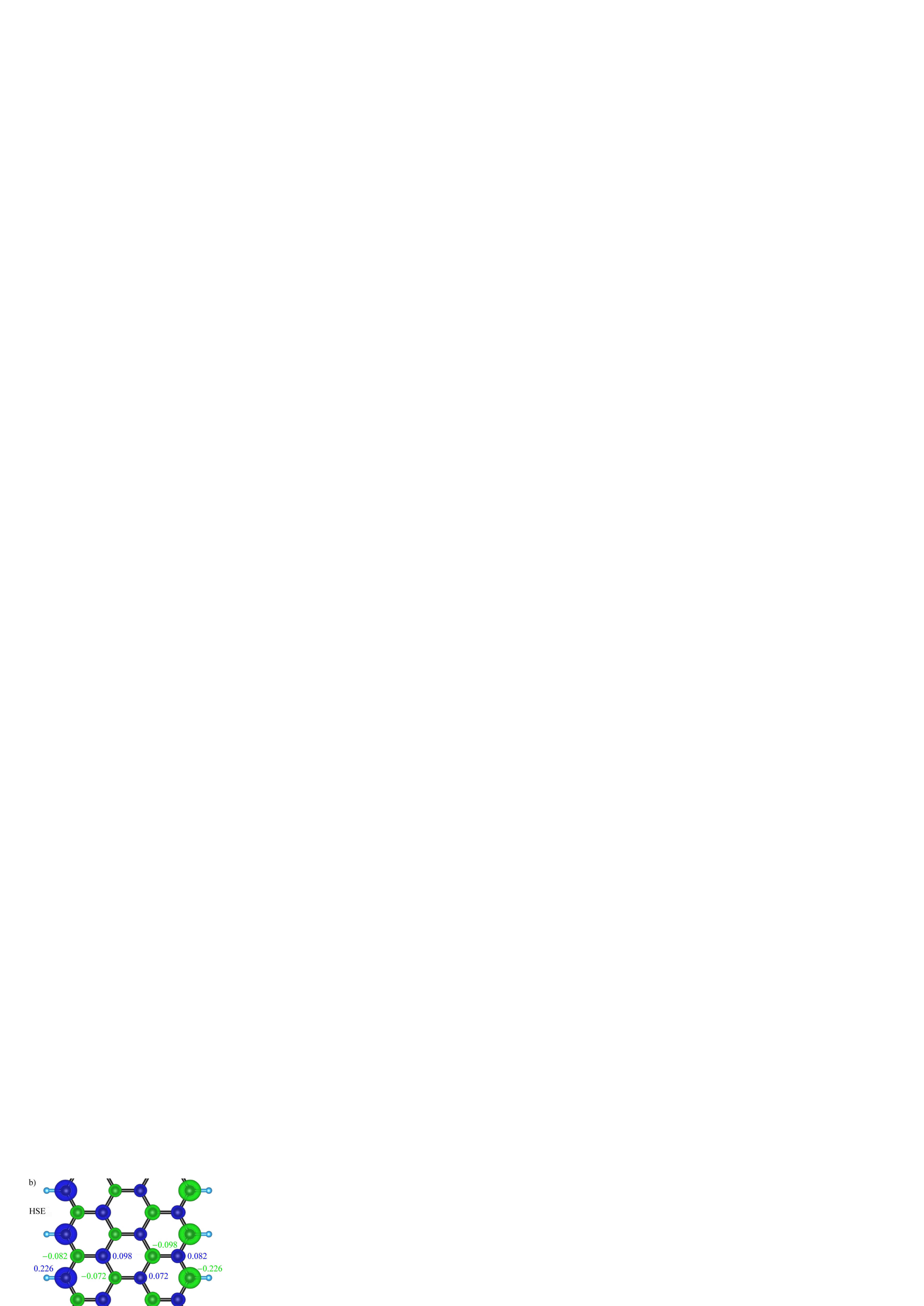}
\caption{\label{fig:epsart2}(Color online) Spin density plots of 4ZGNR showing up (blue/dark grey) and down (green/light grey) spin densities together with the corresponding values of local magnetic moments calculated with VASP code using a) GGA-PBE exchange correlation potential b) non-local GGA-HSE potential.}
\end{figure}

Fig. 2 compares  spin densities plot of 4ZGNR  calculated with local exchange potential (GGA-PBE) with corresponding picture obtained within non-local approach (HSE). Estimation of magnetic moment is sensitive to the choice of exchange potential, for   non-local functional HSE (Fig. 2b) much higher values are obtained and slower decay towards the centre of the ribbon. These  trends can be understood as a consequence of the well known property of non-local potentials, which  localize electronic states  more strongly compared to local potentials \cite{HSE, Paier}.
It is  clearly seen, that spin moments are mainly distributed at the edge carbon atoms.  The magnetic moment fluctuation across the ribbon arises from quantum interference effects caused by edges.  Due to topology of the lattice, the atoms of the two edges belong to different sublattices of the bipartite graphene lattice. The spin moments on the C atoms on one edge are antialigned to the spin moments on the opposite edge and also  the polarizations of neighboring sites belonging to different sublattices are opposite.  Figs. 3a, b present 4ZGNR bands calculated with VASP code  decorated with local spin dependent edge contribution  (overlap of the band eigenstates with $p_{z}$  state localized at ${\cal{A}}_1$). Two observations are  striking, first that the top of the valence band and the bottom of the conduction band are composed  mainly of edge states, especially close to the zone boundary and   second that  in momentum range $2\pi/3a<k<\pi/a$ ($a$ is ZGNR lattice constant) lowest unoccupied conduction band (LUCB) and the highest occupied valence band (HOVB) are characterized by opposite spin polarizations. Of course for the right edge (${\cal{B}}_4$) the spin contributions change the roles. We have also marked in Fig. 3 the direct band gap ($\Delta_{0}$)  and the energy gap at the zone boundary ($\Delta_{1}$).
\begin{figure}
\includegraphics[width=7.2 cm,bb=0 0 660 652,clip]{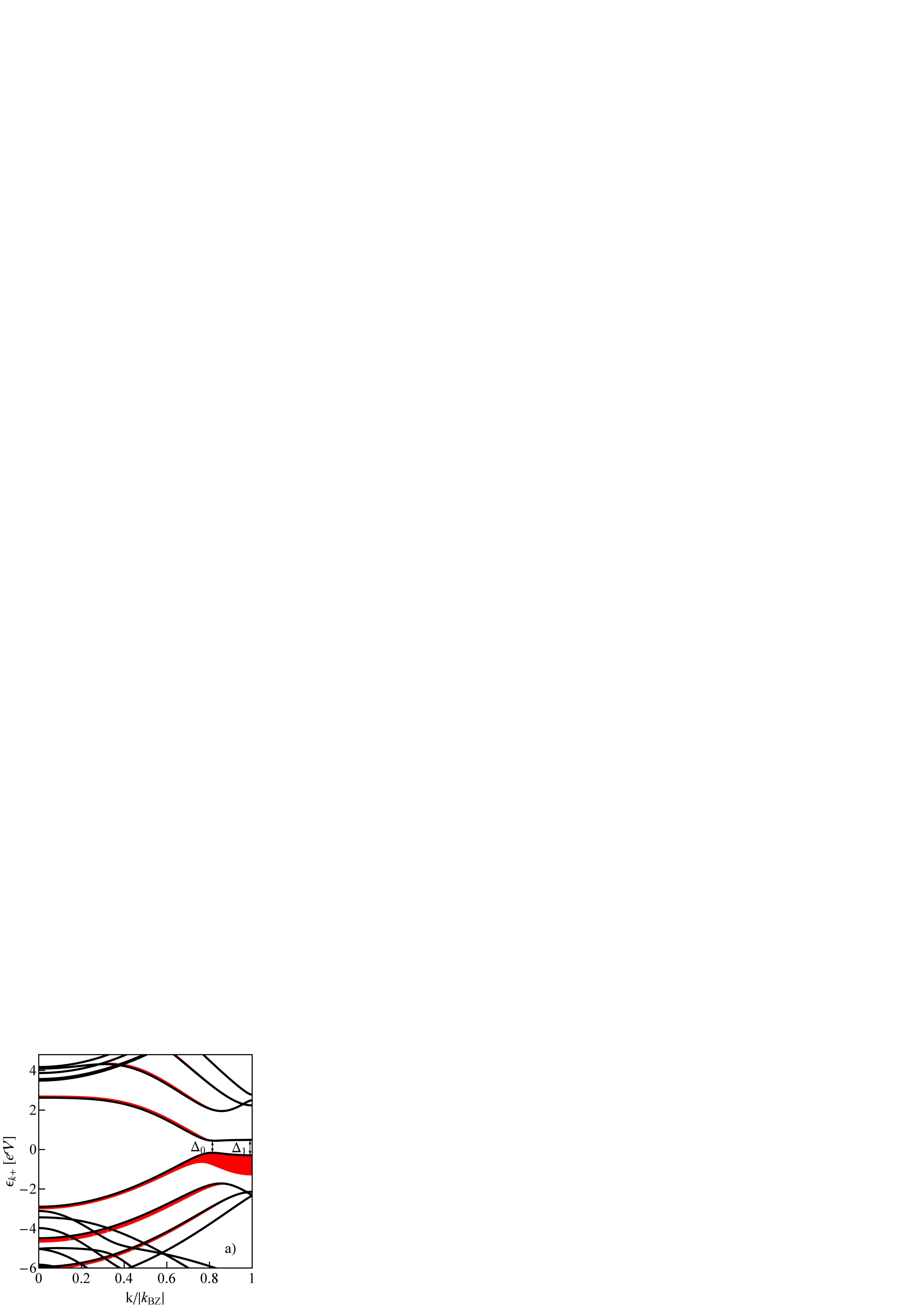}
\includegraphics[width=7.2 cm,bb=0 0 660 652,clip]{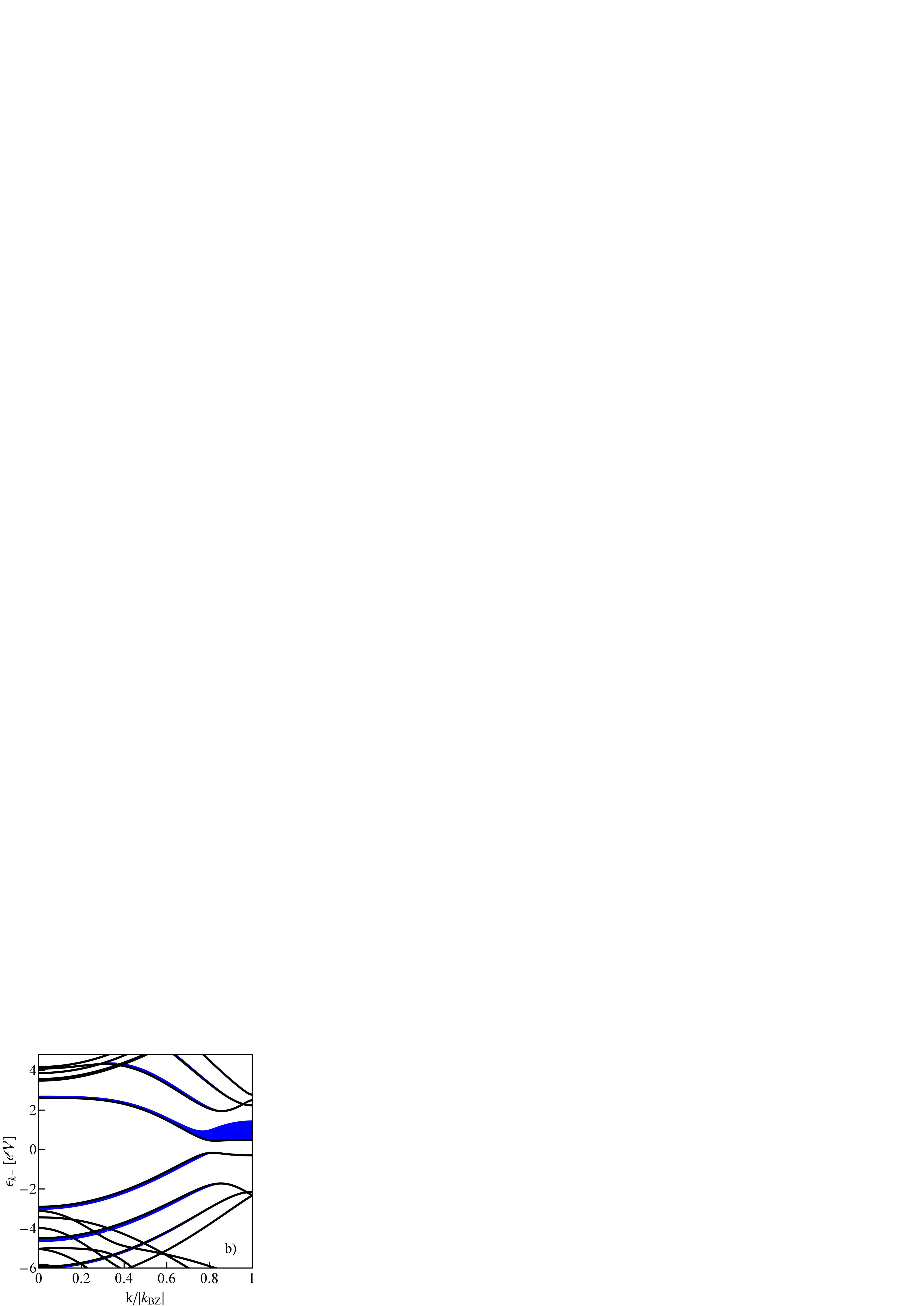}\\
\includegraphics[width=7.2 cm,bb=0 0 288 286,clip]{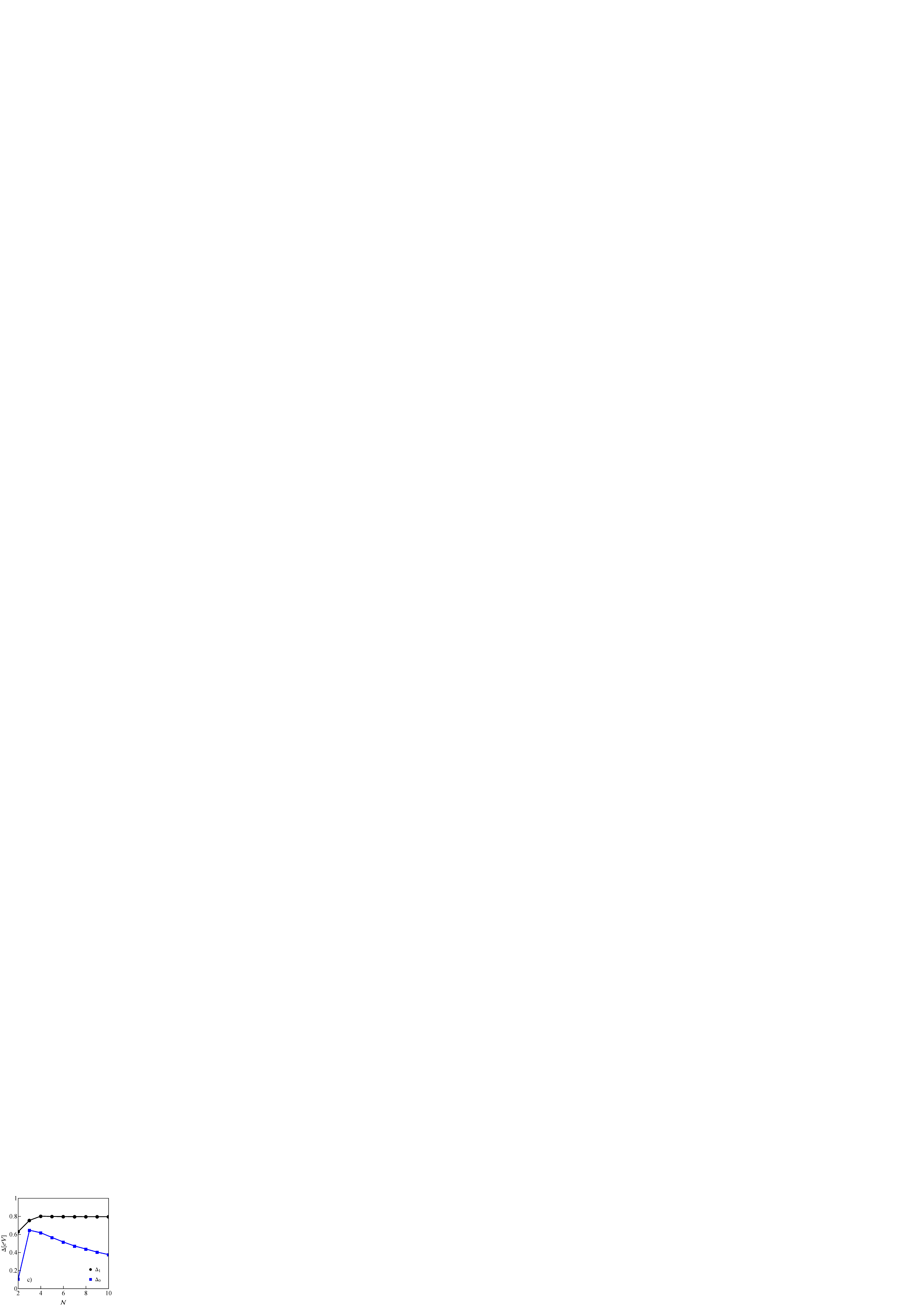}
\caption{\label{fig:epsart3}(Color online) a, b) VASP spin degenerate bands of 4ZGNR with \textit{fat bands} showing the amplitudes of the projection of each band on $p_{z}$ orbital of edge atom ${\cal{A}}_{1}$. $\Delta_{0}$ and $\Delta_{1}$ are direct band gap and gap of the zone boundary c) dependence of the gap on the nanoribbon width.}
\end{figure}
The magnetization induced  staggered  potential opens  a band gap.   The direct band gap  decreases with the increase of the width of the ribbon  due  to confinement and decrease of edge spin polarization  (Fig. 3c).
\begin{figure}
\includegraphics[width=7.2 cm,bb=0 0 660 650,clip]{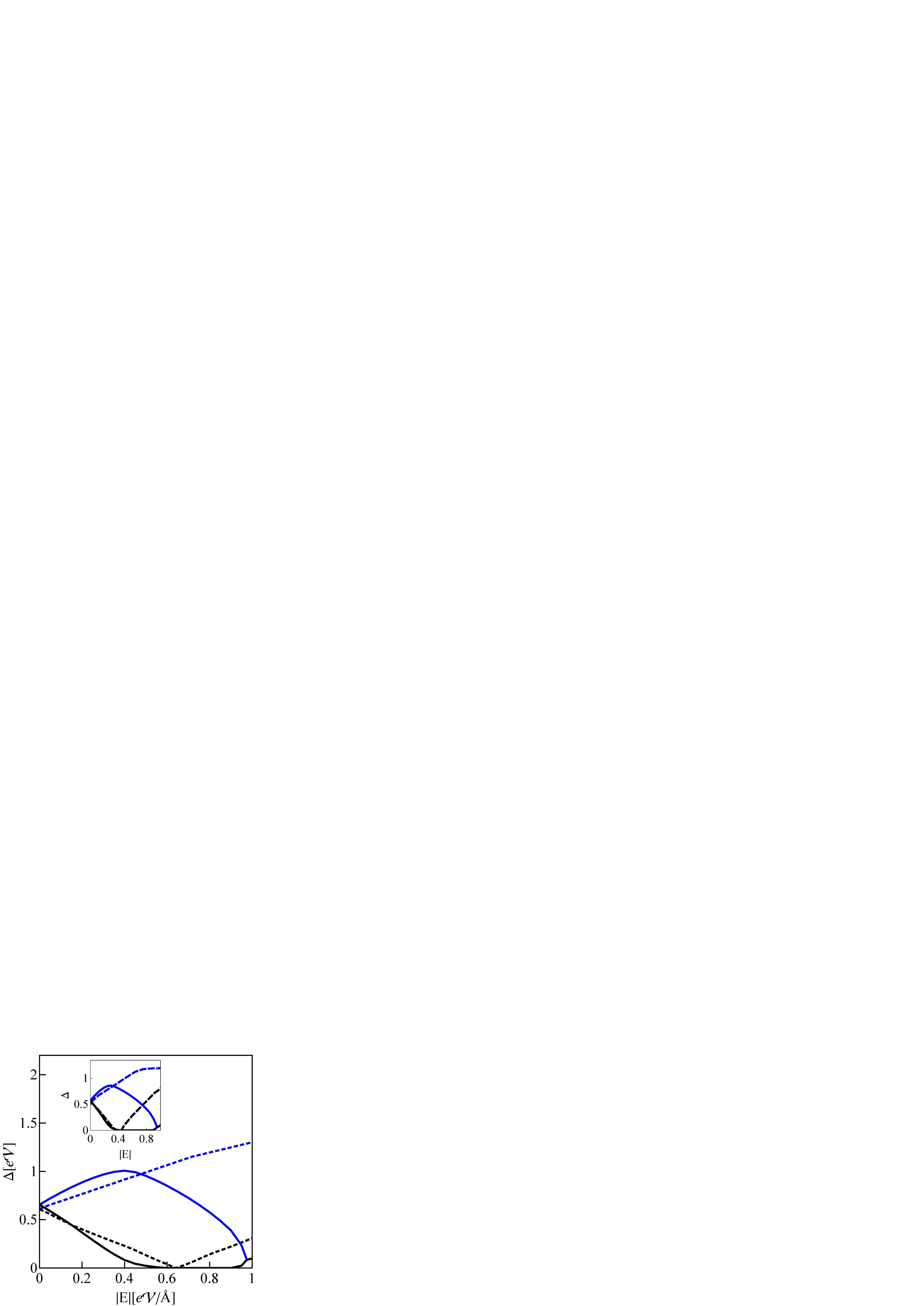}
\caption{\label{fig:epsart4}(Color online) Electric field dependencies of energy gap of 4ZGNR for spin up (lower black curves) and spin down (upper blue curves) calculated with the use of VASP code (dashed lines) and OPENMX (solid lines). Inset shows the corresponding dependencies for 6ZGNR.}
\end{figure}
The energy gap at the zone boundary on the other hand   is almost  not sensitive to the width, because as stated earlier, the edge states close to the zone boundary are highly confined at the edge of ZGNR. It is known, that local or semilocal approximations such as GGA routinely underestimate semiconductor band gaps, due to self-interaction errors \cite{Paier}. For comparison  we have also calculated the band gap with HSE potential, the obtained value is surprisingly high ($\Delta_{HSE}=1.58$ eV for 4ZGNR) but agrees with other HSE calculation \cite{Xiang}. It is general accepted, that band gaps obtained using hybrid functionals are in much better agreement with experimental data, although overestimated \cite{Kummel, Barone}.

\subsection{Evolution of electronic structure with electric field}

Existence of edge states in ZGNR  gives a possibility to tune the  electronic and magnetic properties of these systems and bellow we discuss one way of such modification, the effect of electric field. The external transverse field is simulated in our calculations  by a periodic saw-tooth type potential \cite{Son}, which is perpendicular to the ribbon edge.
\begin{figure}
\includegraphics[width=7 cm,bb=0 0 440 450,clip]{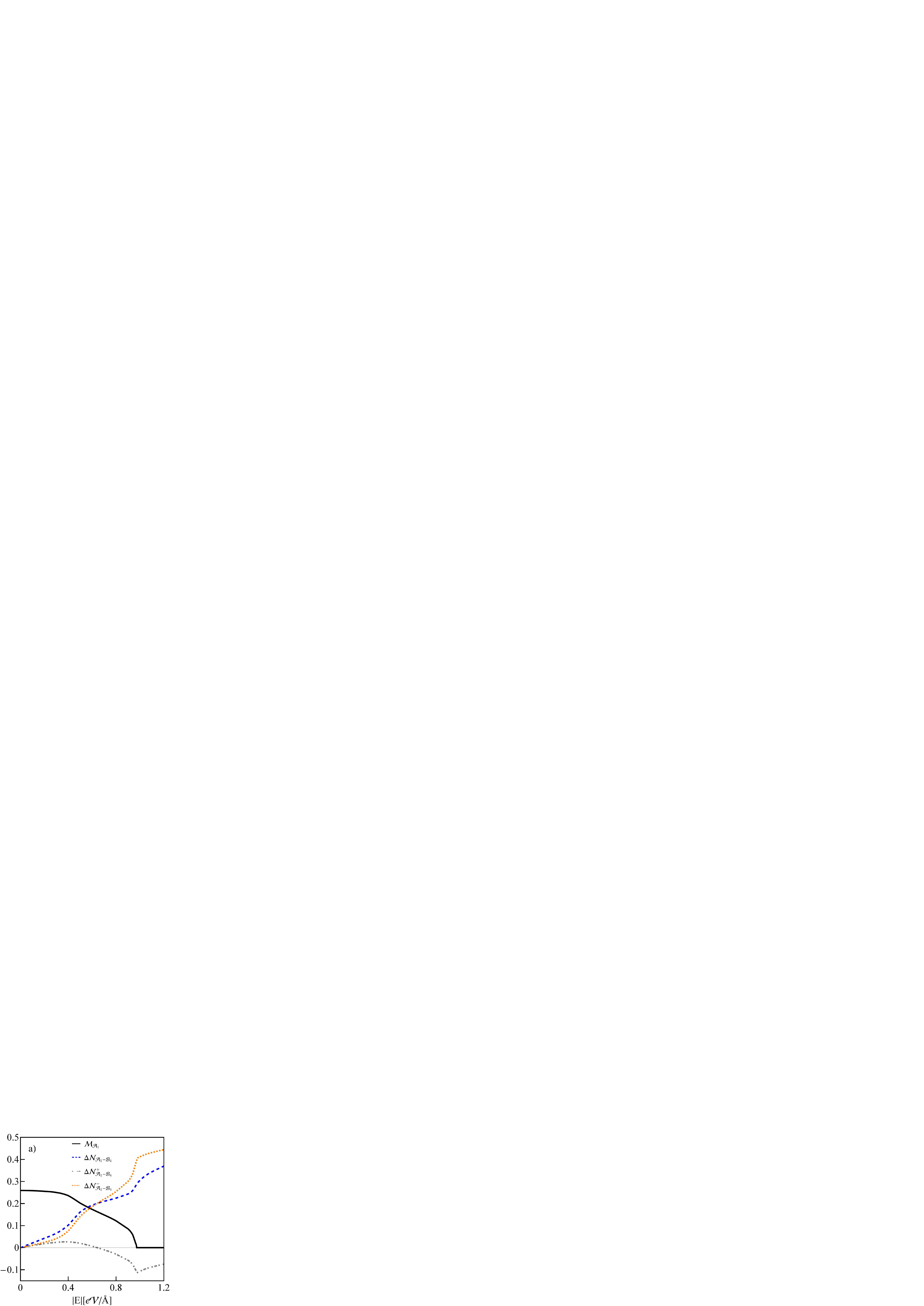}
\includegraphics[width=7.2 cm,bb=0 0 660 635,clip]{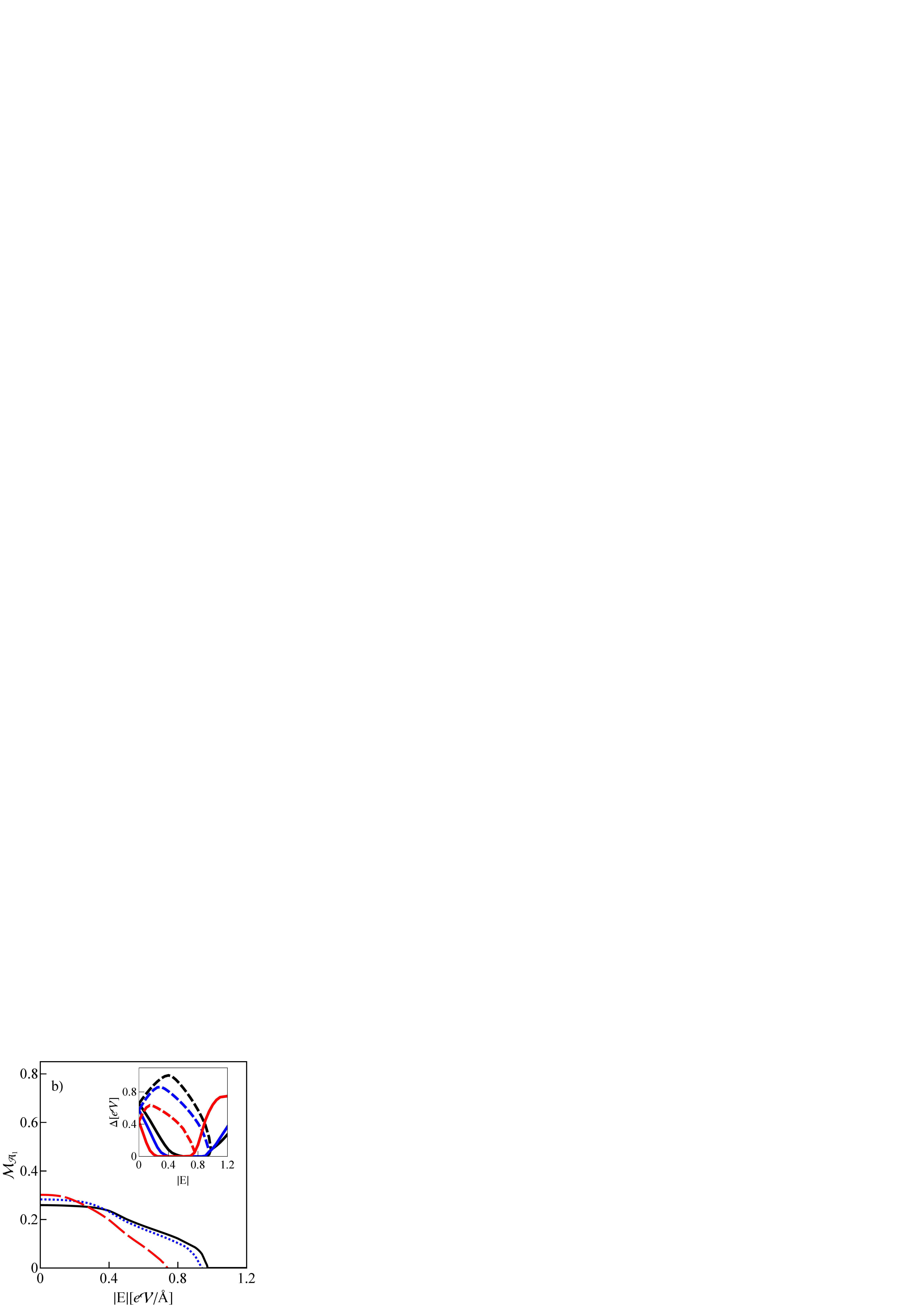}\\
\includegraphics[width=7 cm,bb=0 0 660 636,clip]{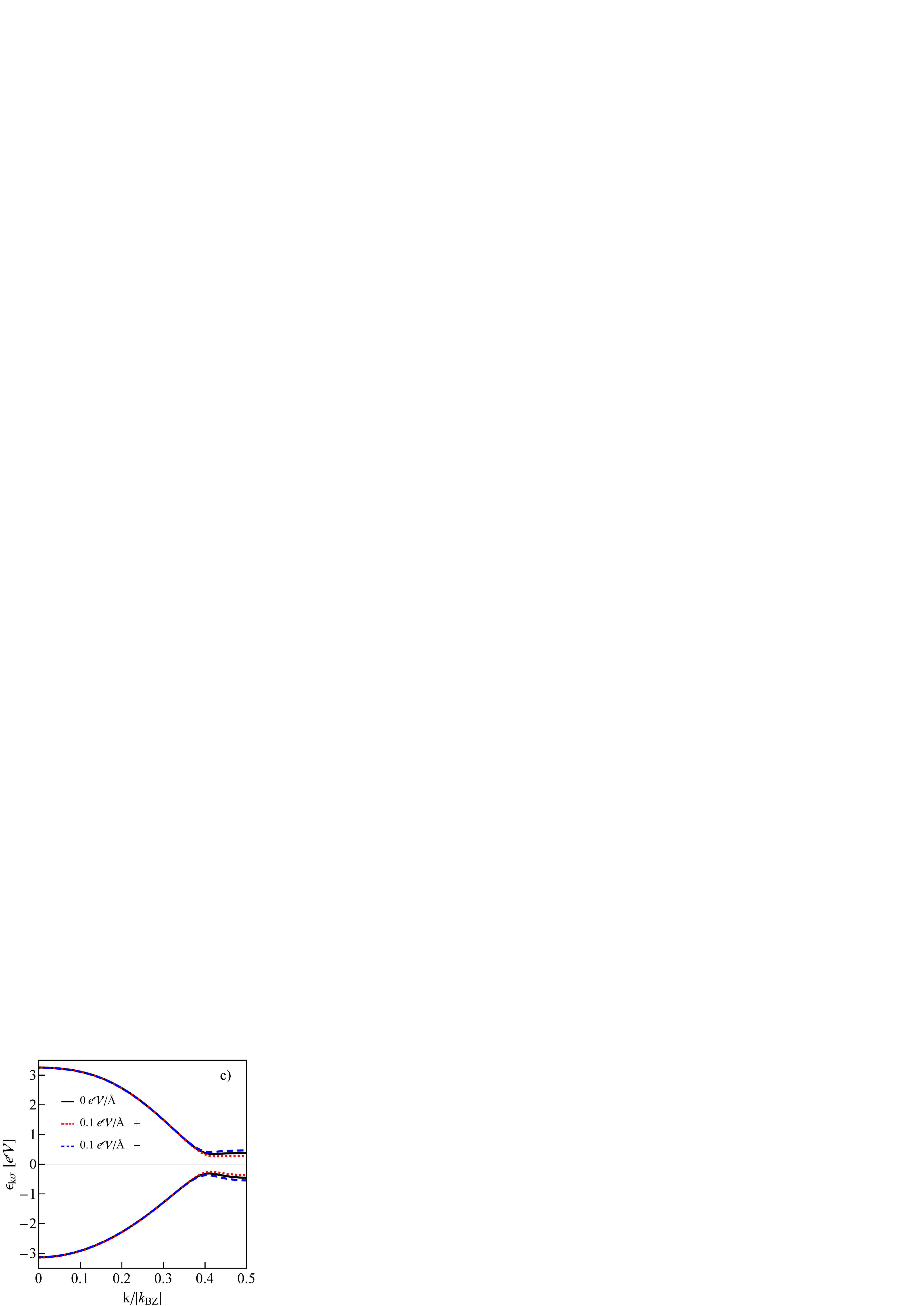}
\includegraphics[width=7.4 cm,bb=0 0 660 610,clip]{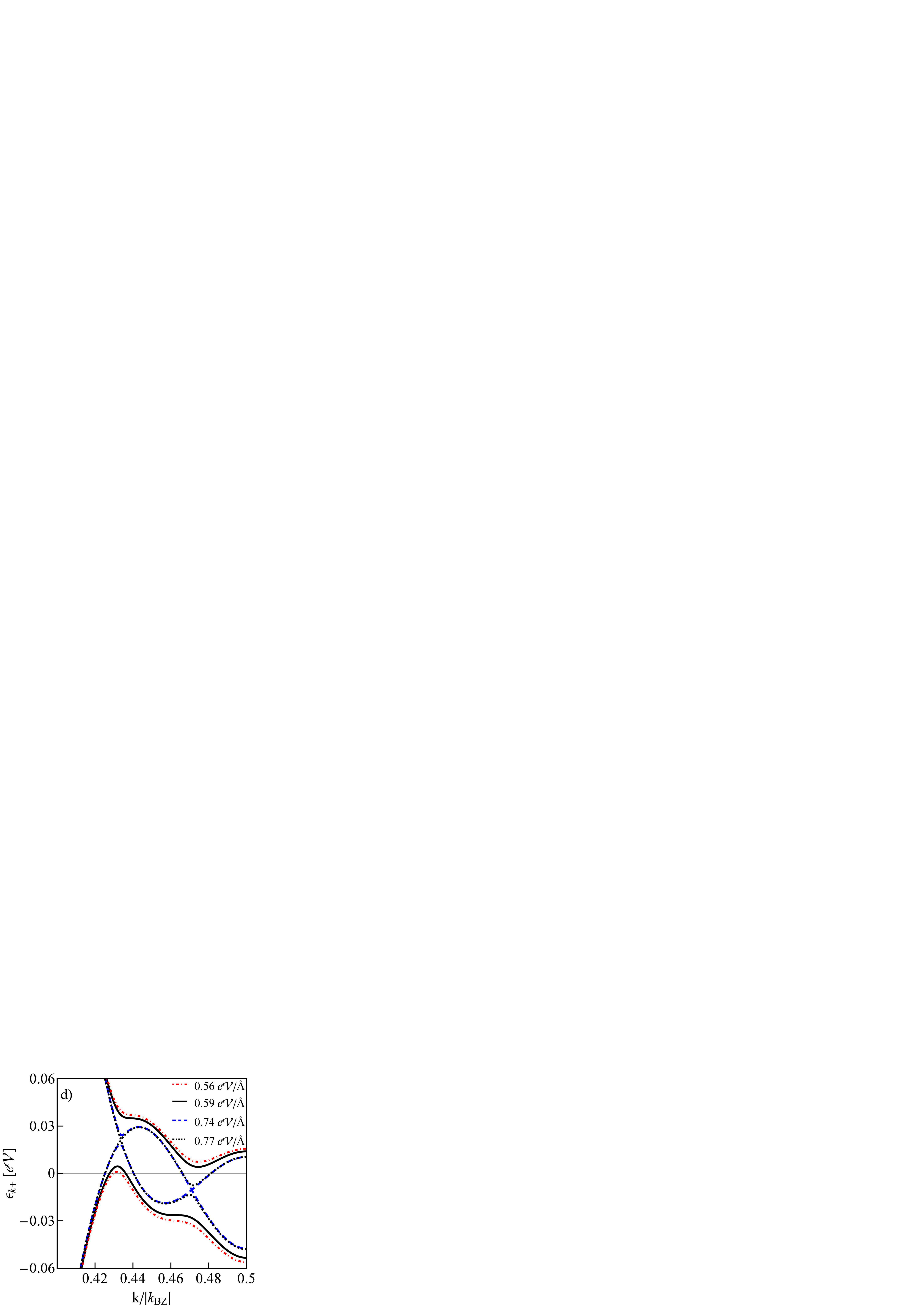}
\caption{\label{fig:epsart5}(Color online) a) Electric field dependence of edge magnetic moment of ${\cal{A}}_{1}$ atom in 4ZGNR (black solid line) together with charge transfer from right to the left edge $\Delta{\cal{N}}_{{\cal{A}}_{1}-{\cal{B}}_{4}}$ (dashed blue line) and corresponding spin up (grey dashed-dot dot line) and spin down (dotted line) contributions to the charge transfer (OPENMX). b) Field evolution of edge magnetic moments of 4ZGNR (solid black line), 6ZGNR (dotted blue line) and 10ZGNR (dashed red line). Inset compares the field evolution of the gaps for the same choice of nanoribbon widths. c) Comparison of LUCB and HOVB bands of 4ZGNR (OPENMX) of zero field case with finite field $|E|=0.1$ eV/{\AA}. d) The zoom-view of spin up LUCB and HOVB in the narrowed momentum range for representative fields in the vicinity of critical field which closes the gap.}
\end{figure}
Field evolution of the gaps obtained by VASP and  OPENMX methods are depicted on  Figure 4. With the increase of electric field, the spin-down band gap decreases  and becomes zero for electric field strength depending on the ribbon width. The critical field  to achieve half-metallicity  decreases with increasing width. The spin-up channel remains semiconducting under all external fields. In agreement with earlier results \cite{Son, Kan2}, our calculations predict that half-metallicity will be destroyed by a too strong electric fields. According to OPENMX  the electric field range at which ZGNR remains half-metallic increases with the ribbon width (Fig. 4 and inset of Fig. 5b), in VASP calculations this range is very narrow. The observed differences are probably due to the different choice of the wave function basis sets and consequently different treatment of screening in both codes.
As suggested by Son \cite{Son}, the half-metallicity comes from the relative movement in energy of edge states under electrostatic potential, oppositely for a given spin direction  on left and right edges. The field evolution of edge magnetic moment and  the  charge difference between the edges of 4ZGNR are presented on Fig. 5a.  The charge transfer from one edge into the opposite edge suppresses the edge moments and for high enough  fields the moments vanish. The charge imbalance between spin up electron from left and right edges is suppressed ($\Delta{\cal{N}}^{+}_{{\cal{A}}_{1}-{\cal{B}}_{4}} = 0$) for the field $|E|=0.59$ eV/{\AA}, when spin up LUCB and HVOB bands start to overlap (Fig. 5d). For higher fields, when the overlap increases the charge transfer $\Delta{\cal{N}}^{+}_{{\cal{A}}_{1}-{\cal{B}}_{4}}$ changes sign (Fig. 5a).
 Fig. 5b compares the electric field dependencies of edge magnetic moments for different nanoribbon widths. Figs 5c, d    show   the representative  VASP dispersions of low energy bands for the selected values  of the field. It is seen that in the interesting momentum range $2\pi/3a < k <\pi/a$ the LUCB and HOVB  bands for one spin direction become closer and their curvatures  undergo reconstruction due to the effect of the screened electric field.  In the range of extremely  narrow gap, where the electric field mixes the occupied states with unoccupied ones   the single  minimum (LUCB) or  maximum (HOVB) evolves into a pair of close  minima or maxima and the bands come close to the Fermi level asymmetrically. The evolution of these bands is affected by coupling of edges, which depends on interference and confinement effects. Our calculations suggest that HOVB crosses the Fermi level first, for fields slightly smaller than the critical fields required for  closing the gap. Since the rest of the paper focuses only on the zero field case, we postpone more  elaborated analysis of the field dependence of the nanoribbon electronic structure for our future publication.

\subsection{Co adatom}

The computational tools we use (VASP, OPENMX) are developed for  periodic structures and therefore  we simulate the single impurity problem by superstructure calculations. As a consequence of periodicity  the extra features in the generated band structure can occur e.g. additional gaps not related to finite geometry, but to the assumed  superstructure.
\begin{figure}
\includegraphics[width=4.8 cm,bb=0 0 660 1180,clip]{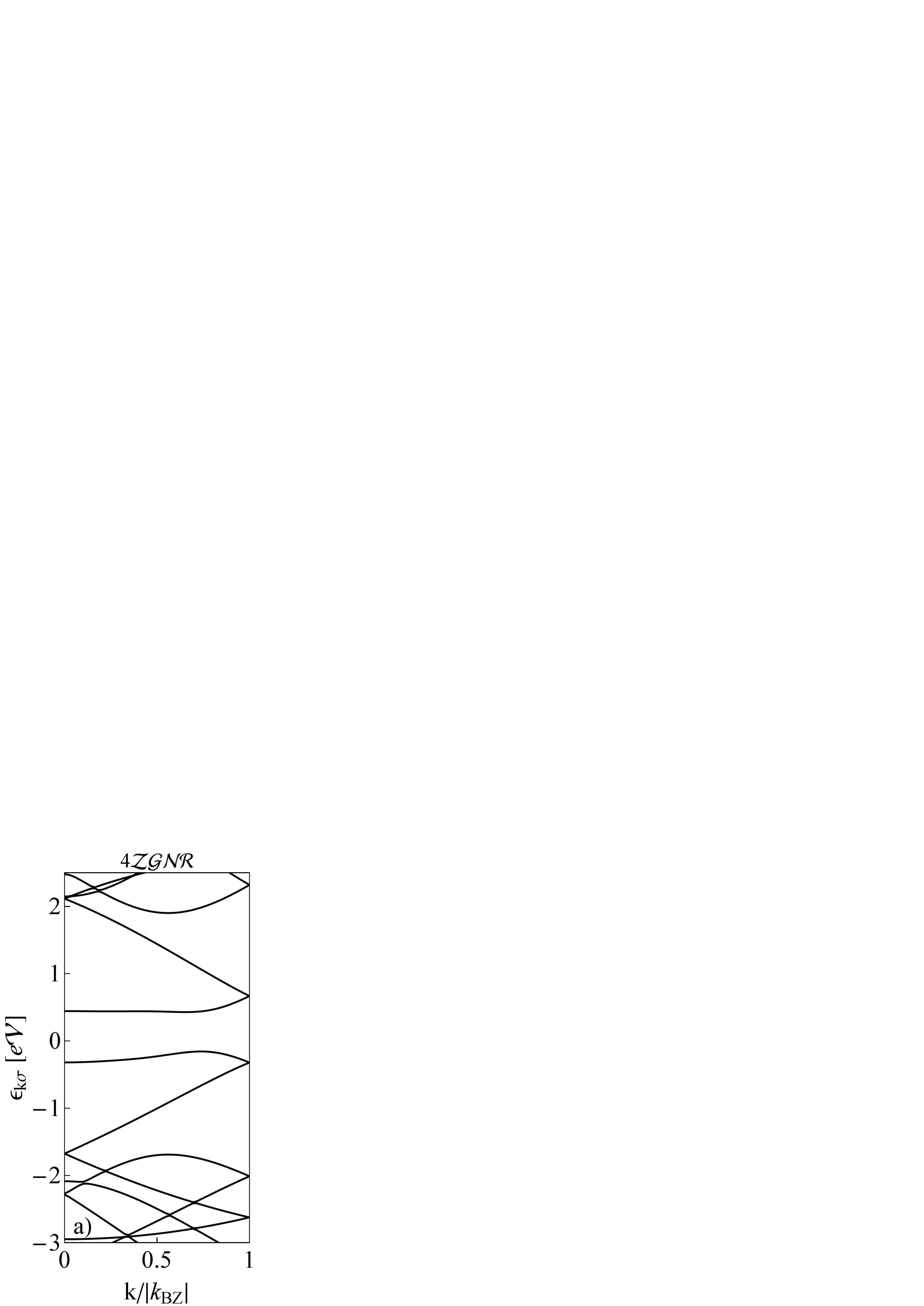}
\includegraphics[width=9.4 cm,bb=0 0 660 600,clip]{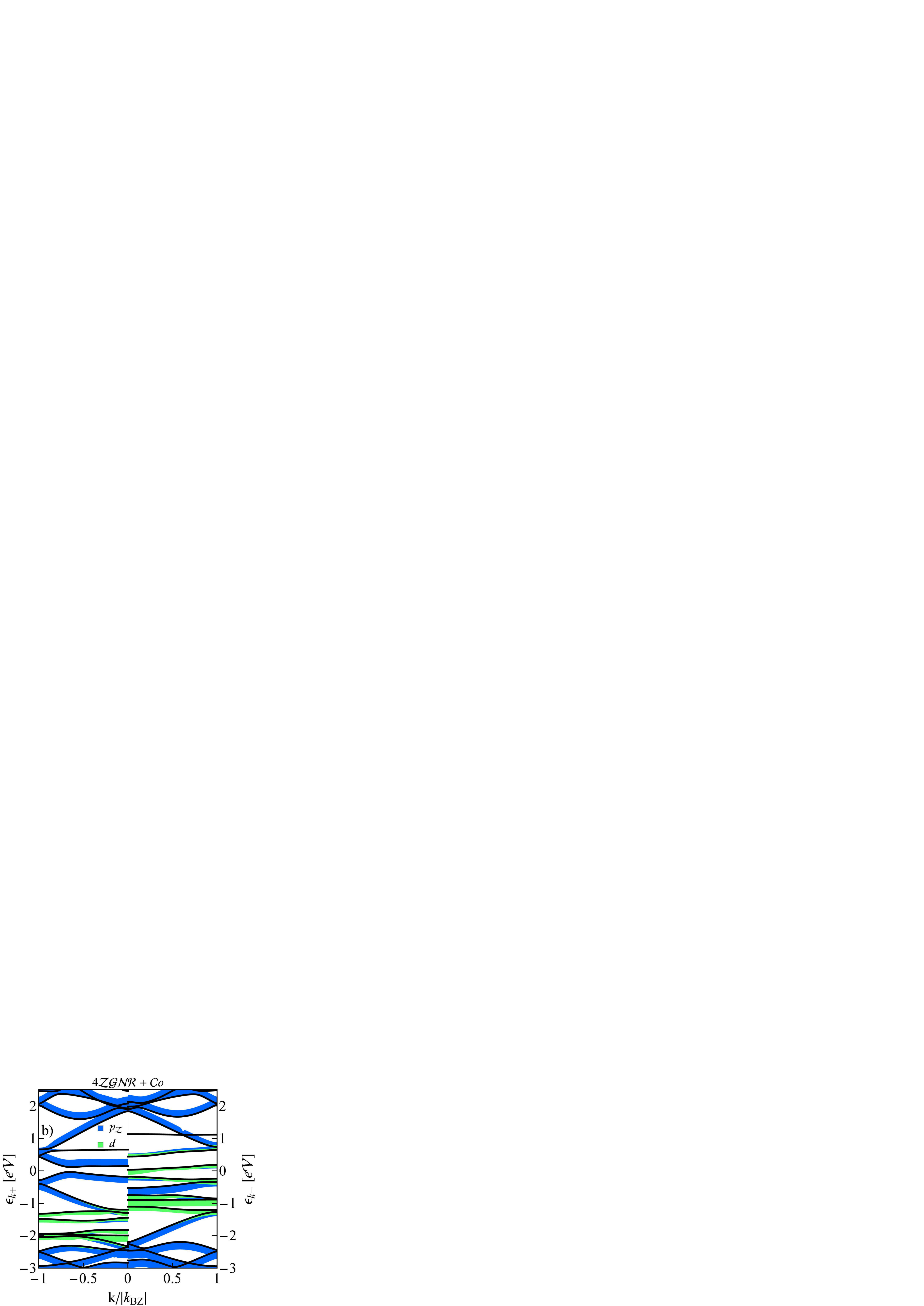}
\caption{\label{fig:epsart6}(Color online) VASP energy dispersion curves of 4ZGNR compared with the bands of 4ZGNR with Co impurity in $h_{1}$ position. The blue/dark grey fat bands highlight carbon $p_{z}$ contribution and green/light grey the Co $d$ contribution. }
\end{figure}
It is believed however, that  using large enough supercells one can still infer about some single impurity properties.  This concerns mainly quantities which depend on the entire density of states and not just on the behavior near the Fermi level, e.g.  occupations or magnetic moments. With some caution one can get also an insight into some parts of  the electronic structure, where superstructure does no interfere considerably. In our study we use a supercell consisting of four replicas of ZGNR unit cell ($4\times1$). This setup corresponds to a coverage of $1$ adatom per $32$ C atoms.  Although the adatom-adatom interaction is not negligible, the distance between adatoms is large enough that the overlap of the electronic states of neighboring atoms is negligible. Several test simulations were also carried out for a $8\times1$ supercell.
Periodic boundary conditions were also used along confined direction assuming   $20$ {\AA}  of vacuum  to prevent unphysical interactions. Different positions, as indicated on Fig. 1  were sampled.  Fig. 6  presents an example of the  band structure  of 4ZGNR$+$Co system with adatom in $h_{1}$ position compared with the band structure of pure nanoribbon.
The bands are decorated by the amplitudes of the projection on atomic $p_{z}$ orbitals of carbon and Co $d$ orbitals. The strong interaction between cobalt and carbon atoms  comes from the mixture of  these states. Carbon $p_{x}$ and $p_{y}$ orbitals are far below ${\cal{E}}_{{\cal{F}}}$ and have weak hybridization with cobalt.
\begin{figure}
\includegraphics[width=7 cm,bb=0 0 438 460,clip]{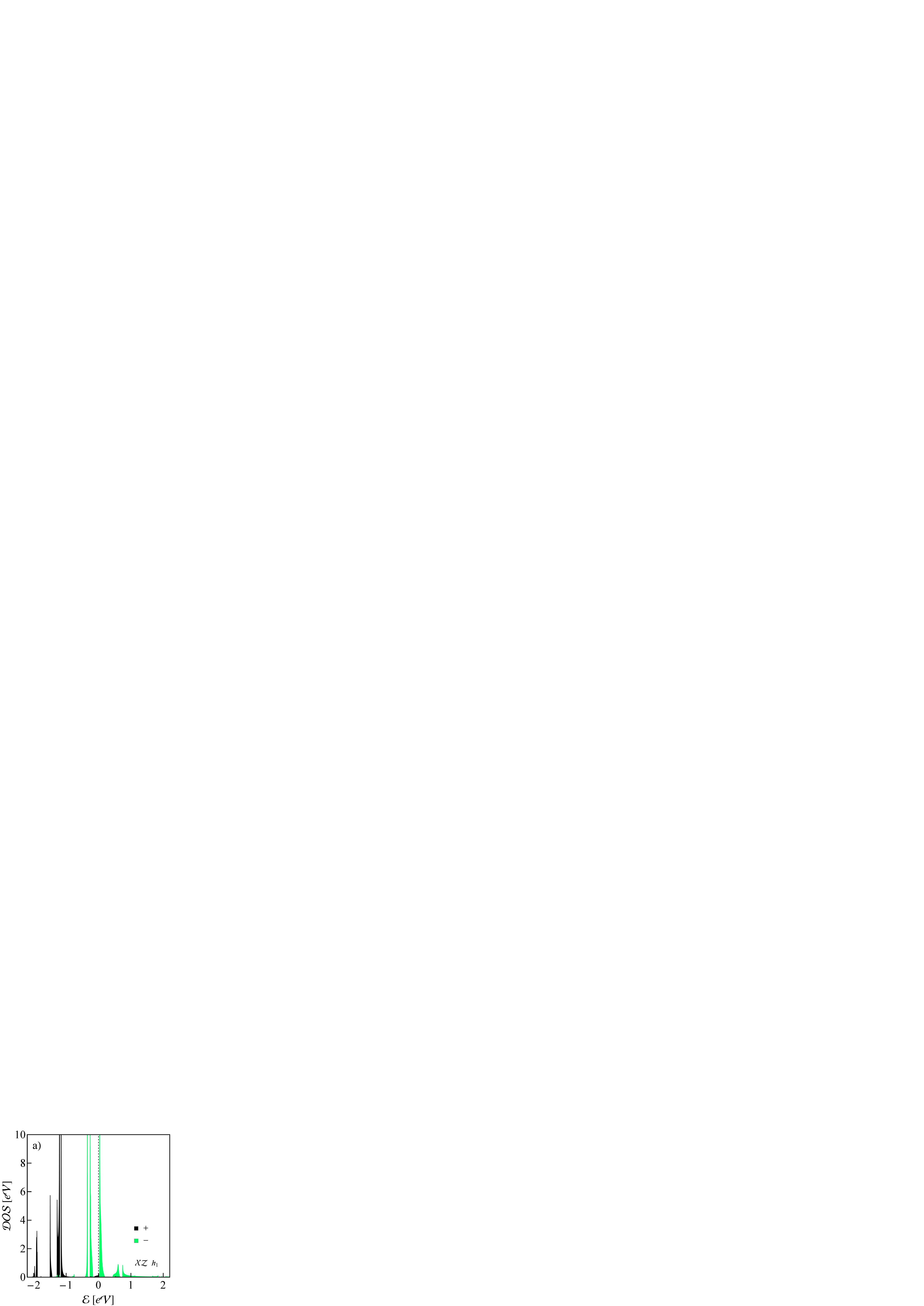}
\includegraphics[width=7 cm,bb=0 0 438 460,clip]{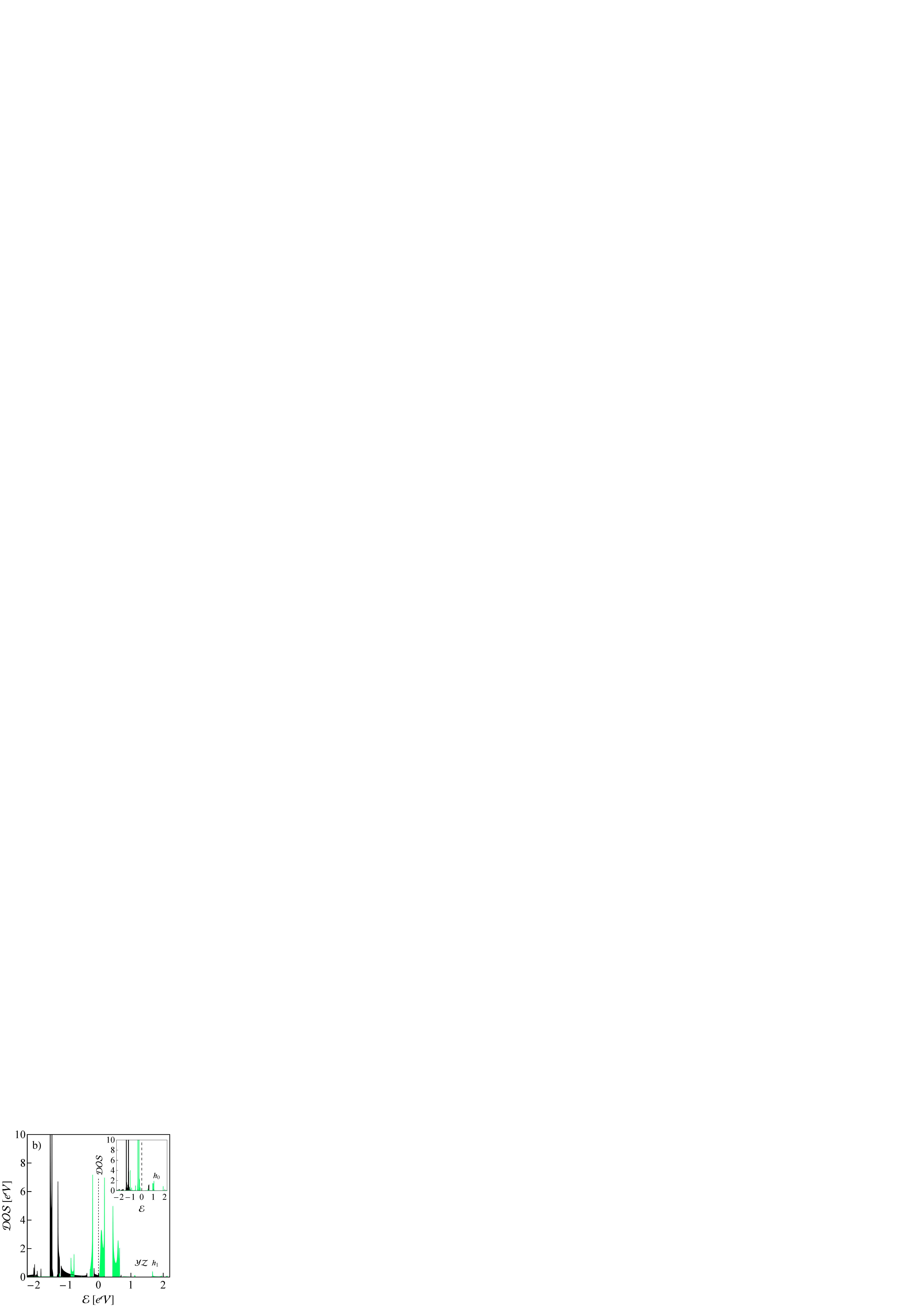}\\
\includegraphics[width=7 cm,bb=0 0 438 460,clip]{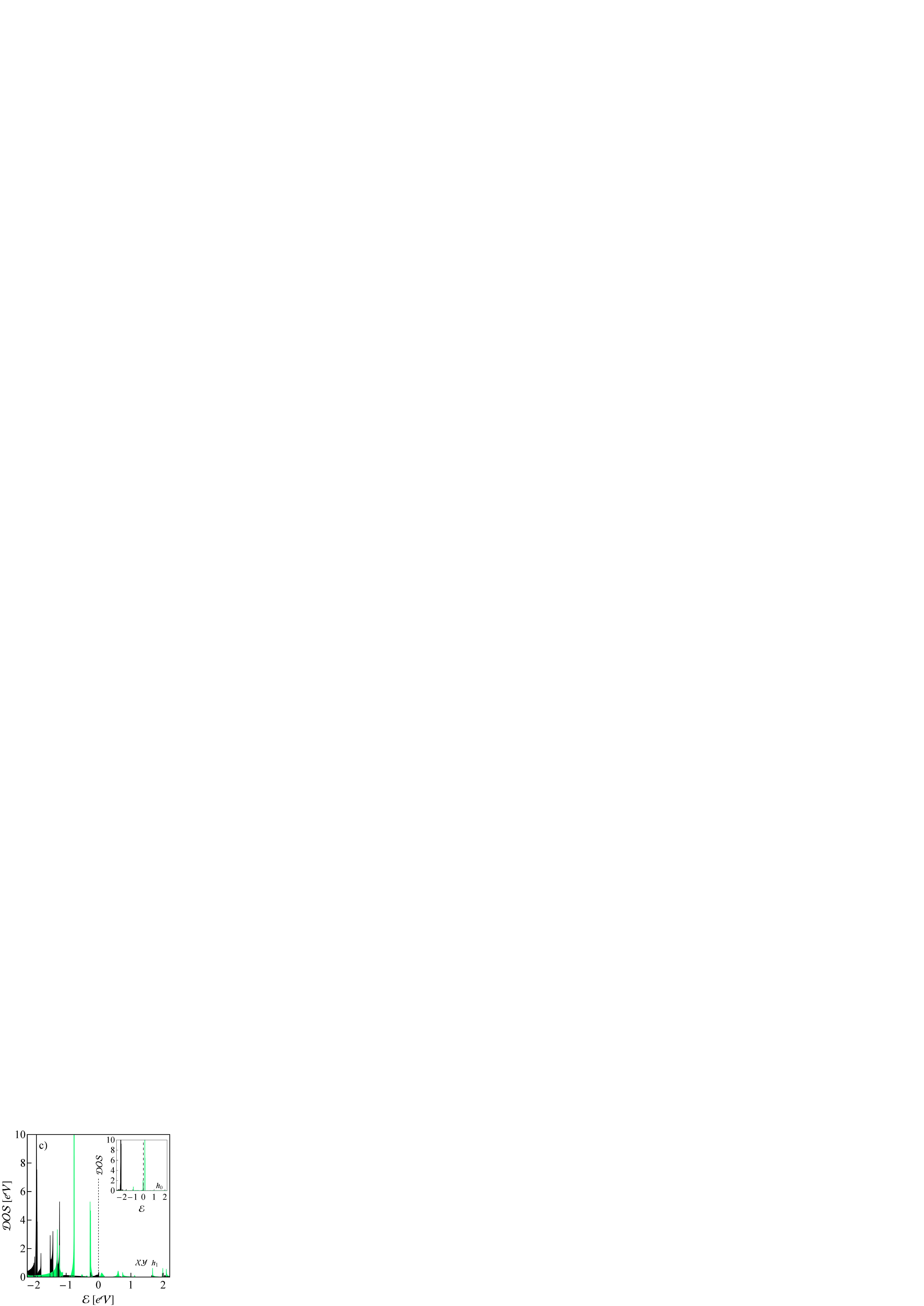}
\includegraphics[width=7 cm,bb=0 0 438 460,clip]{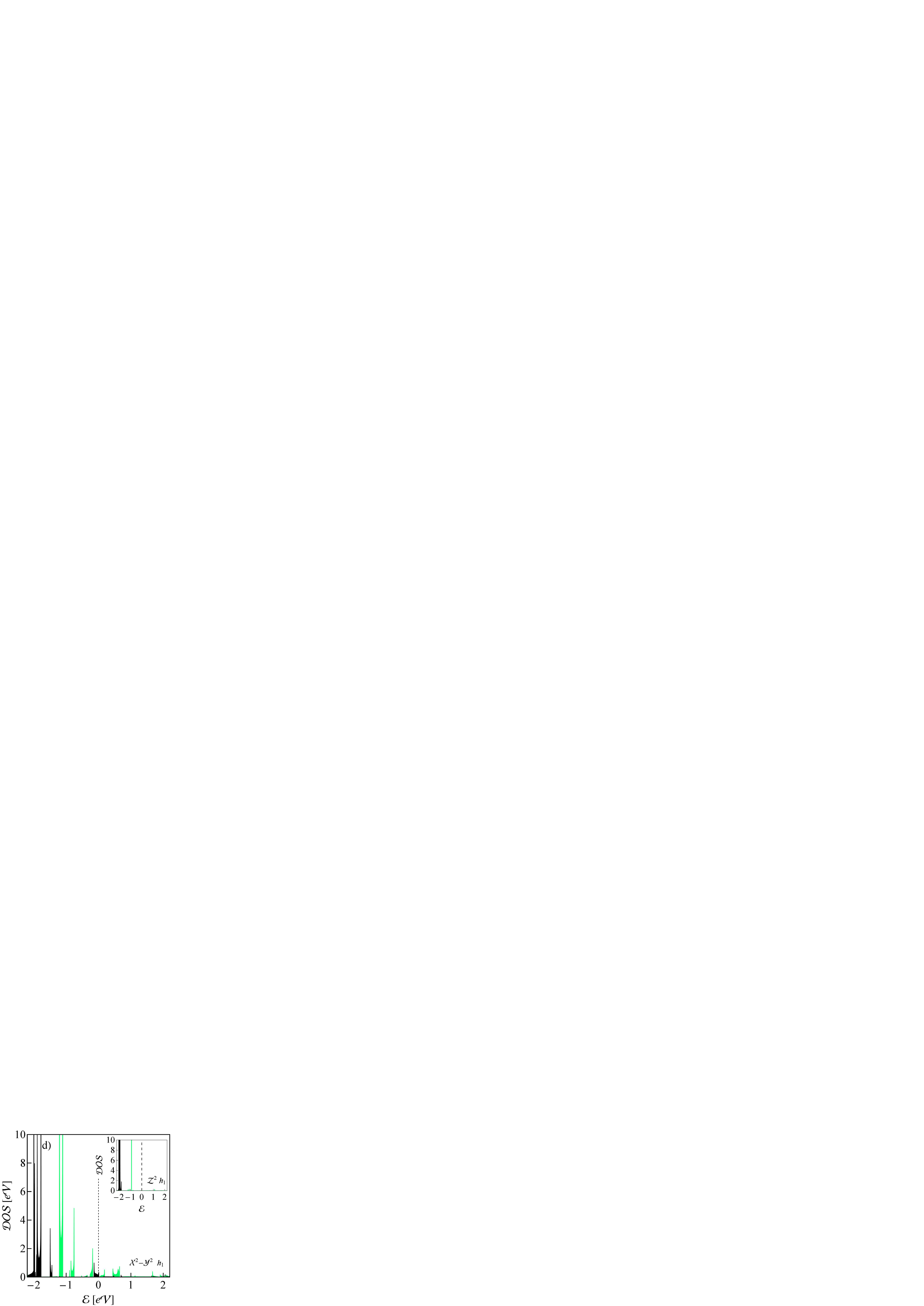}
\caption{\label{fig:epsart7}(Color online) Spin and orbital resolved densities of states of Co adatom at $h_{1}$ position in 4ZGNR. The black denote spin up contributions and green/light grey the spin down contributions. Inset of Figs. b, c present $yz$ and $xy$ partial DOS of Co adatom at $h_{0}$ site. }
\end{figure}
Fig. 7 displays  the corresponding spin and orbital resolved  Co adatom densities of states. In pure graphene, in consequence of  $C_{6v}$ point symmetry the  $4s$ and $3d_{z}^{2}$  orbitals do not hybridize with graphene orbitals close to the Dirac points.  In nanoribbon this symmetry is broken, but still hybridization of these orbitals is only very weak.  The rest of Co  $d$ orbitals hybridize strongly, what  results in covalent interactions.    The bonding, almost completely occupied orbitals  $d_{x^{2}-y^{2}}$ , $d_{xy}$ lie lower in energy than $d_{xz}$ and $d_{yz}$, the latter are partially filled and they  play an active role in formation of magnetic moment. Depending on the position of adatom  some orbitals may swap the roles. This can be seen comparing for example  the $d_{xy}$ , $d_{yz}$ , partial DOS for $h_{1}$ and $h_{0}$ sites (Figures 7b, c).
\begin{table}
\caption{\label{tab:table2}Illustration of the impact of Coulomb interaction and exchange interaction on the values of adsorption energies of Co adatom in hollow positions of 4ZGNR (GGA+U) \cite{Liechtenstein}.}
\begin{ruledtabular}
\begin{tabular}{lcr}
$$&${\cal{E}}_{ads}^{h_{1}} [eV]$&${\cal{E}}_{ads}^{h_{2}} [eV]$\\
\hline
${\cal{U}}=0$ ${\cal{J}}=0$ & $-1.60$ & $-1.37$\\
${\cal{U}}=2$ ${\cal{J}}=0$ & $-1.06$ & $-0.76$\\
${\cal{U}}=4$ ${\cal{J}}=0$ & $-0.93$ & $-0.84$\\
${\cal{U}}=2$  ${\cal{J}}=0.9$ & $-1.20$ & $-0.87$\\
${\cal{U}}=4$  ${\cal{J}}=0.9$ & $-1.14$ & $-0.87$\\
\end{tabular}
\end{ruledtabular}
\end{table}
For $h_{0}$ position  $d_{yz}$ orbital becomes fully occupied whereas $d_{xy}$ shifts closer to ${\cal{E}}_{{\cal{F}}}$  and takes over the role of magnetic orbital. The   reversal of  the roles is a consequence of the change of symmetry and reduced coordination, what alters hybridization amplitudes (see hybridization Tab. IV) and consequently modifies the  widths and effective orbital splittings of $4d$ levels.
\begin{figure}
\includegraphics[width=7.2 cm,bb=0 0 660 630,clip]{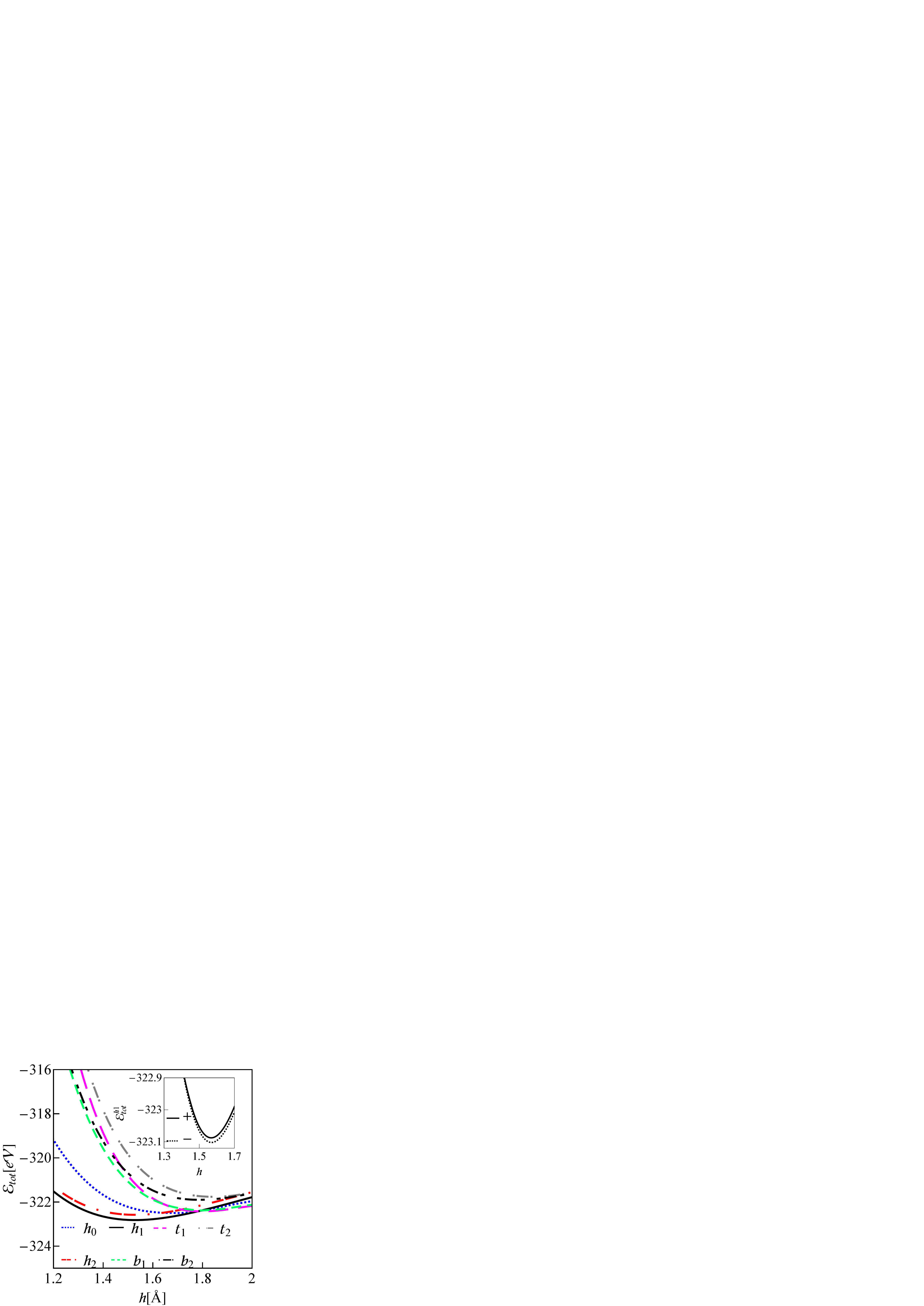}
\caption{\label{fig:epsart8}(Color online) Total energy curves for different positions of Co adatom on 4ZGNR. Inset is the zoom-view of spin resolved energy curves for $h_{1}$ position presented close to minimum.}
\end{figure}
To understand the energetics of Co adsorption on graphene nanoribbon we performed a series of calculations for different vertical distances  of Co and nanoribbon plane.  The calculated equilibrium heights of adatom together with adsorption energies are summarized in Tab. III and Figure 8 presents  selected adsorption energies curves.
\begin{figure}
\includegraphics[width=7.3 cm,bb=0 0 438 448,clip]{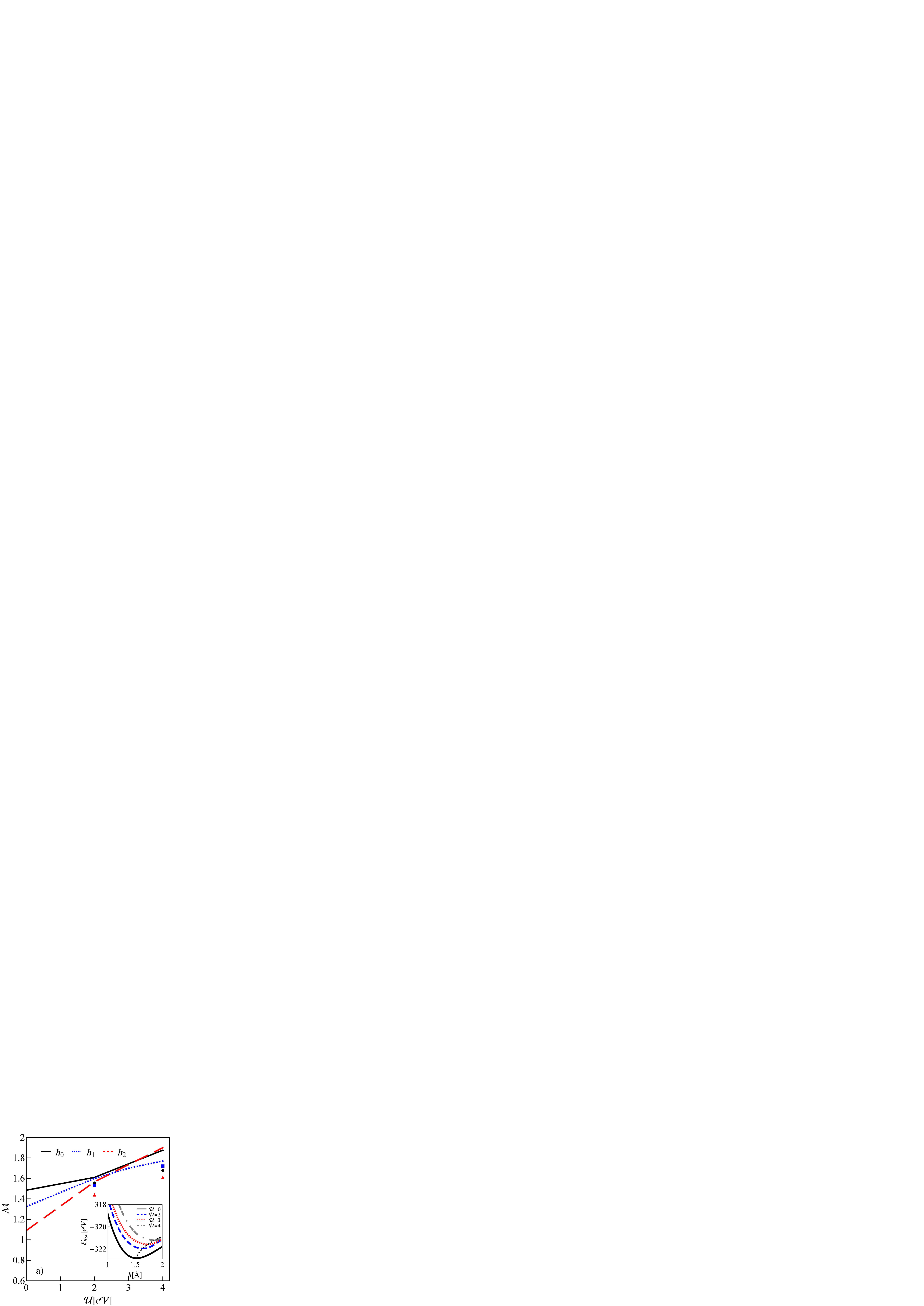}
\includegraphics[width=7.2 cm,bb=0 0 660 658,clip]{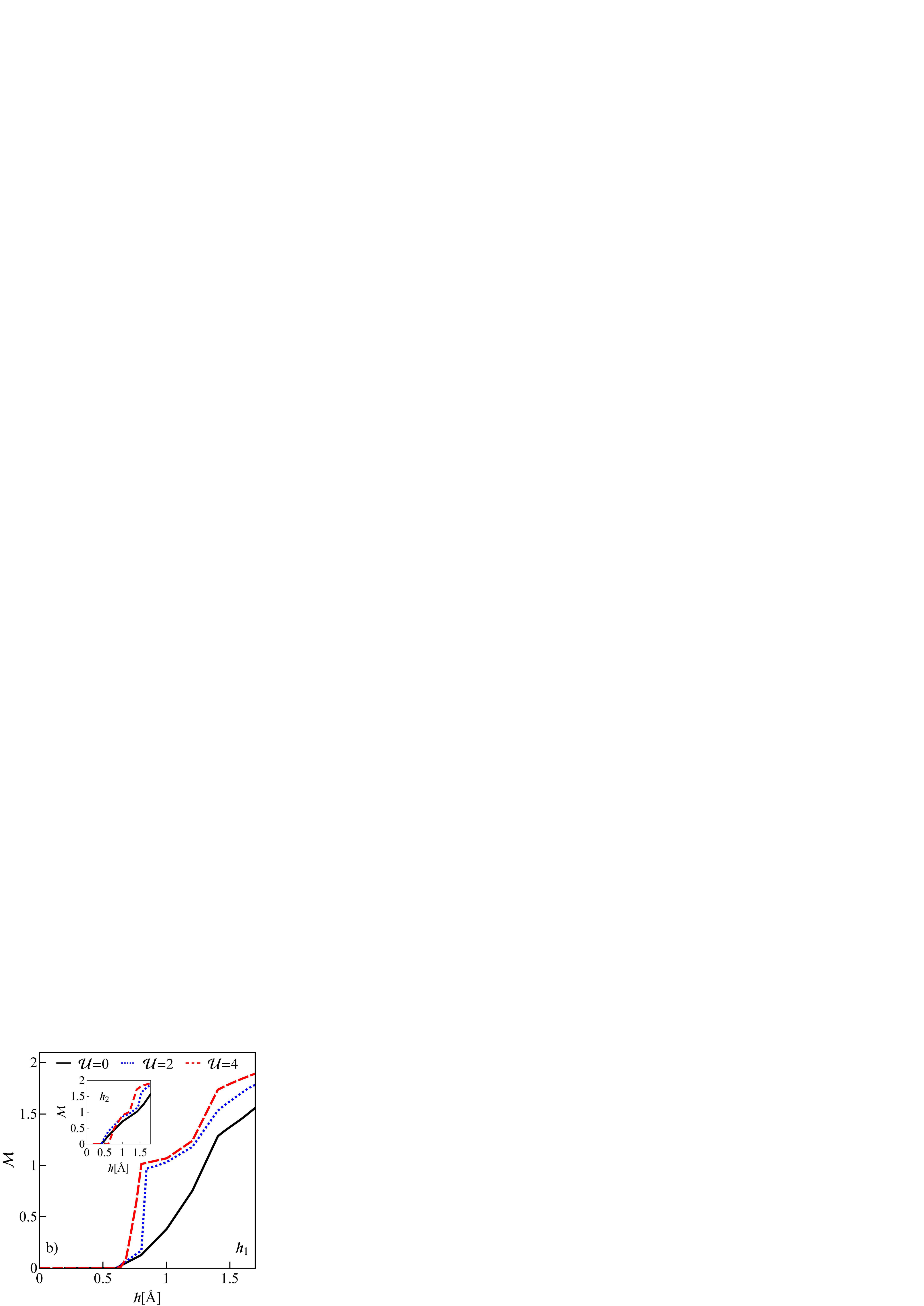}
\caption{\label{fig:epsart9}(Color online) a) GGA+U estimations of total magnetic moment of Co impurity in 4ZGNR for diffrent values of Coulomb interaction parameter. The dots represent additional effects of exchange (${\cal{J}}=0.9$ eV) for $h_{0}$ (circle), $h_{1}$ (square) and $h_{2}$ (triangle). Inset shows the total energy curves for $h_{1}$ position for selected values of ${\cal{U}}$. b) Adatom-graphene nanoribbon distance dependence of total magnetic moment of Co impurity for $h_{1}$ and $h_{2}$ (inset) positions.}
\end{figure}
The formed covalent bonds are directional and the bond strength depends on the adatom coordination. Therefore it is unsurprising  that the adsorption energy is strongly dependent on the adsorption site. The presented energies are the minimal values obtained after relaxation.
\begin{table}
\caption{\label{tab:table3}Equilibrium heights above nanoribbon plane  and adsorption energies of Co adatom at different positions of 4ZGNR (VASP with GGA-PBE).}
\begin{ruledtabular}
\begin{tabular}{lcr}
$-$&$h_{eq}[$\AA$]$&${\cal{E}}_{ads} [eV]$\\
\hline
$h_{0}$ & $1.7$ & $-1.261$\\
$h_{1}$ & $1.53$ & $-1.396$\\
$h_{2}$ & $1.53$ & $-1.158$\\
$t_{1}$ & $1.83$ & $-0.341$\\
$t_{2}$ & $1.85$ & $-0.407$\\
$b_{1}$ & $1.79$ & $-0.946$\\
$b_{2}$ & $1.85$ & $-0.477$\\
\end{tabular}
\end{ruledtabular}
\end{table}
Since nanoribbon polarization is nonuniform, adsorbsion energy depends on Co spin polarization, but as shown in the example of $h_{1}$ position (inset of Fig. 8),  the adsorption curves for different spin orientations do not differ significantly.  Nevertheless the predicted spin orientations of Co adatoms deposited at $h_{1}$ and $h_{3}$ positions are opposite, whereas at the ribbon center, where no net magnetic polarization of ZGNR occurs  the energies for both Co spin orientations are degenerate. Co adatom  prefers hollow positions, where  the impurity is not associated with a particular sublattice, but instead binds three carbon atoms from each.
Following Power et al. \cite{Power} we have also checked that at the edge a new type of metastable adsorption site is realized ($h_{0}$, Fig. 1), where impurity connects to three edge zigzag atoms. This position is however only reached after relaxation along the path starting from adatom originally siting at $t_{1}$ or $t_{2}$  positions.
\begin{table*}
\caption{\label{tab:table4}Nearest neighbors hybridization amplitudes of Co adatom located in hollow positions of 4ZGNR.}
\begin{ruledtabular}
\begin{tabular}{c|cc|cccc|cccc}
 &\multicolumn{2}{c|}{$h_{0}$}&\multicolumn{4}{c|}{$h_{1}$}&\multicolumn{4}{c}{$h_{2}$}\\
 $d$&$|{\cal{V}}^{{\cal{A}}_{1}}_{d}|$&$|{\cal{V}}^{{\cal{B}}_{1}}_{d}|$&$|{\cal{V}}^{{\cal{A}}_{1}}_{d}|$&$|{\cal{V}}^{{\cal{A}}_{2}}_{d}|$
 &$|{\cal{V}}^{{\cal{B}}_{1}}_{d}|$&$|{\cal{V}}^{{\cal{B}}_{2}}_{d}|$&$|{\cal{V}}^{{\cal{A}}_{2}}_{d}|$&$|{\cal{V}}^{{\cal{A}}_{3}}_{d}|$
 &$|{\cal{V}}^{{\cal{B}}_{2}}_{d}|$&$|{\cal{V}}^{{\cal{B}}_{3}}_{d}|$\\\hline
 $z^{2}$&$0.13$&$0.64$&$0.001$&$0.001$&$0.001$&$0.001$&$0.001$&$0.001$&$0.001$&$0.001$\\
 $xz$&$0.47$&$0$&$0$&$0.68$&$0.66$&$0$&$0$&$0.67$&$0.67$&$0$\\
 $yz$&$0.13$&$0.65$&$1.92$&$0.59$&$0.59$&$1.87$&$1.81$&$0.55$&$0.55$&$1.81$\\
 $xy$&$0.47$&$0$&$0$&$0.28$&$0.28$&$0$&$0$&$0.28$&$0.28$&$0$\\
 $x^{2}-y^{2}$&$0.3$&$0.52$&$0.79$&$0.25$&$0.21$&$0.76$&$0.78$&$0.25$&$0.25$&$0.78$\\
\end{tabular}
\end{ruledtabular}
\end{table*}

Taking into account electron correlations is crucial for the description of adsorption of transition metal atoms.  Here we  present in Tab. II some testing results obtained within GGA + U type approach. The estimated adsorption energy significantly lowers with the increase of ${\cal{U}}$  whereas ${\cal{J}}$ enhances it.
Inset of Fig. 9a  illustrates that for higher values of Coulomb interaction  the equilibrium distance of Co adatom increases.
Table VI presents GGA-PBE  orbital occupations and total and orbital contributions to magnetic moments of Co for $h_{0}$, $h_{1}$  and  $h_{2}$ positions, which correspond to the earlier presented local densities of states (Fig. 7). Magnetic moments of Co at the ZGNR depend on the adsorption site, but their absolute value  in all cases are  much   smaller than the moment of  the free atom (${\cal{M}}_{Co}\approx3\mu_{B}$ \cite{Jacob}).  The decrease of magnetic moment is  dictated by electron transfer from $4s$ to $3d$ states and corresponding change of occupancy of the unpaired $3d$ orbitals.  The spin-down component of the hybrid states is almost entirely below ${\cal{E}}_{{\cal{F}}}$, while a large amount of the spin-up component lies above ${\cal{E}}_{{\cal{F}}}$. As aforementioned, in the case of  hollow sites moments come mainly from $d_{xz}$ and $d_{yz}$ orbitals and their contributions are  $1.103$  and $0.941$ for  $h_{1}$ and $h_{2}$ respectively,  i.e. they do not differ much from unity. We have checked that  the trend of decrease of moment with moving with the adsorption site to the centre of the ribbon  persists in wider ribbons and for ${\cal{N}} = 10$ total magnetic moment reaches  at the center   ${\cal{M}}=1.094$  what is close to the value for  Co adatom on graphene (${\cal{M}}=1.083$). Similarly binding energy in the center of a wide ribbon  converges to the value for graphene ${\cal{E}}_{ads}=-1.23$ eV, already for ${\cal{N}}=8$ amounts   value ${\cal{E}}^{h_{4}}_{ads}=-1.37$ eV.

It is known that pure LDA approach inaccurately estimates magnetic moments and therefore we present in Fig. 9   how the results are modified by inclusion of correlations  within  GGA+U scheme. The finite ${\cal{U}}$ solutions give larger values of  estimated   magnetic moments  and    Hund's coupling reduces the moment and diminishes the difference of $d_{xz}$ and $d_{yz}$ contributions.
\begin{figure}
\includegraphics[width=7 cm,bb=18 0 640 1144,clip]{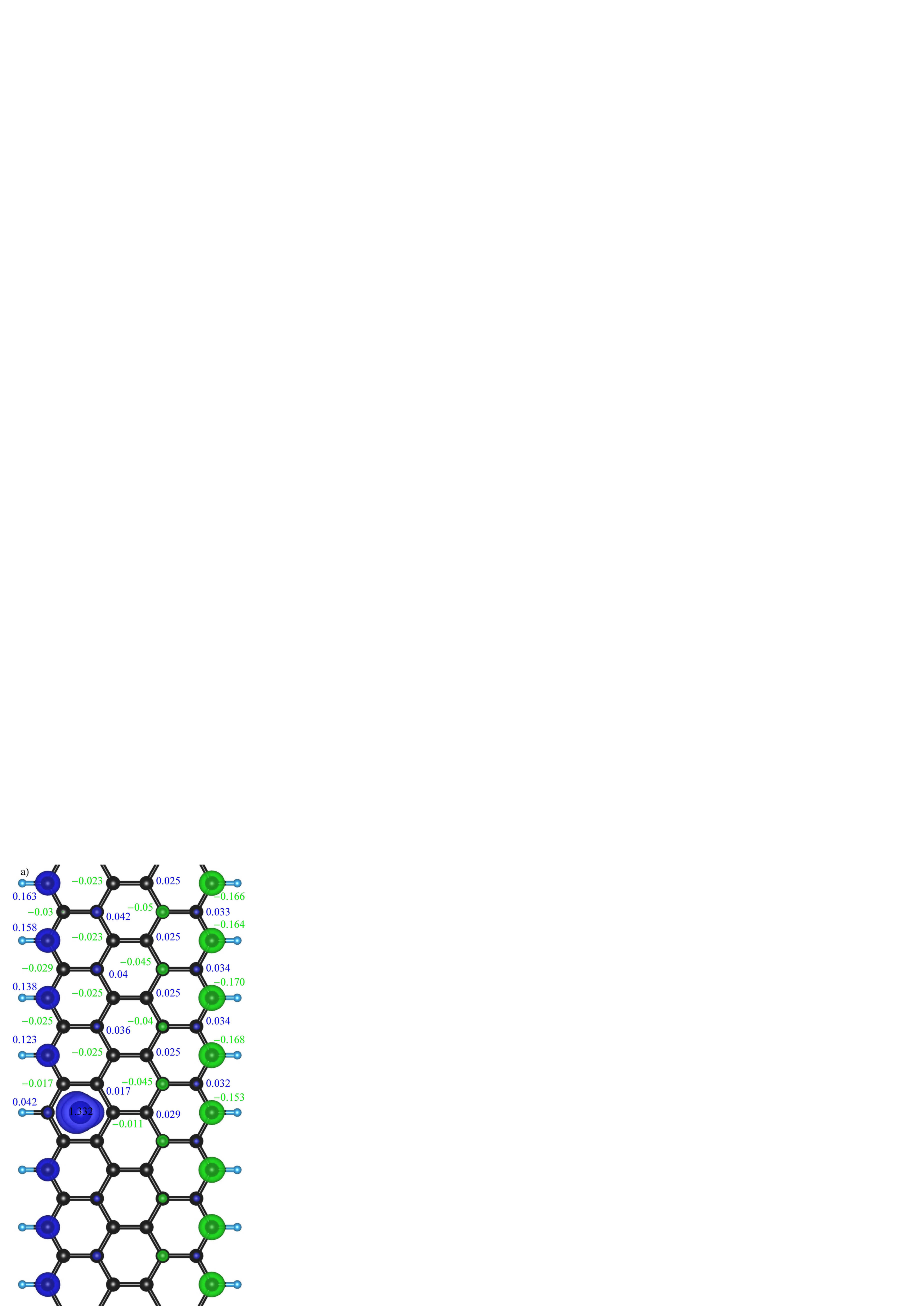}
\includegraphics[width=7 cm,bb=18 0 640 1144,clip]{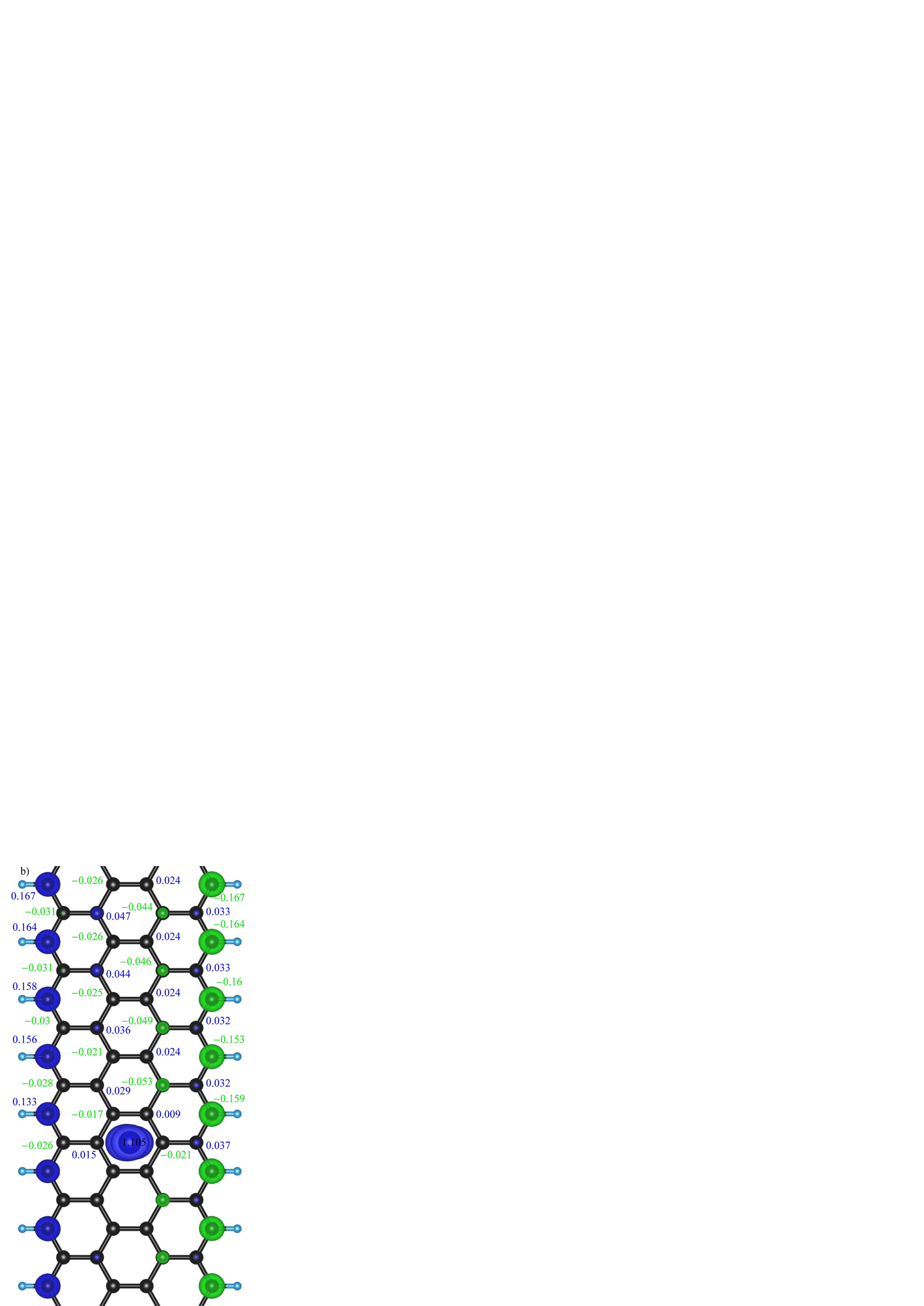}
\caption{\label{fig:epsart10}(Color online) Spin density plots of Co adatom and 4ZGNR matrix ($8\times1$ supercell). The blue/dark grey densities correspond to spin up and green/light grey to spin down components. a) Co impurity in $h_{1}$ position b) Co in $h_{2}$ position.}
\end{figure}
We also present in Fig. 9 magnetic moments dependencies of Co adatom on the height  for three values of Coulomb interaction parameter and Fig. 9a, b compares distance dependencies for $h_{1}$ and $h_{2}$ positions. Increase of the  overlap of impurity to ribbon states with decrease of the distance results in a reduction of magnetic   moment. Close to nanoribbon surface the occupation of weakly hybridized $4s$ state is  less favorable  than these of strongly hybridizing $3d$ orbitals. Hybridization lowers the energy of the orbitals and due to increased delocalization the Coulomb repulsion is reduced. In consequence of $4s-3d$   charge transfer the  decrease of magnetic moment results.
\begin{table}
\caption{\label{tab:table5}Total magnetic moments of Co atoms located in hollow positions of NZGNR
($N = 4,6,8$) compared with magnetic moment of Co at  hollow site of   graphene (VASP with GGA-PBE).}
\begin{ruledtabular}
\begin{tabular}{cccccc}
 &${\cal{M}}$ &${\cal{M}}_{h_{1}}$ &${\cal{M}}_{h_{2}}$ &${\cal{M}}_{h_{3}}$ &${\cal{M}}_{h_{4}}$\\
\hline
${\cal{G}}+{\cal{C}}o$& $1.083$ & $-$ & $-$ & $-$ & $-$\\
$4{\cal{ZGNR}}+{\cal{C}}o$& $-$ & $1.413$ & $1.178$ & $-$ & $-$\\
$6{\cal{ZGNR}}+{\cal{C}}o$& $-$ & $1.383$ & $1.209$ & $1.122$ & $-$\\
$8{\cal{ZGNR}}+{\cal{C}}o$& $-$ & $1.376$ & $1.258$ & $1.180$ & $1.115$\\
\end{tabular}
\end{ruledtabular}
\end{table}
Figures 10a, b. show the spin polarization patterns induced by the presence of Co adatom at $h_{1}$ and $h_{2}$ sites calculated for $8\times 1$ supercell. For the twice reduced supercell ($4\times 1$) the local polarizations around impurity are almost identical to $8\times 1$ case, the differences are only seen at the border of supercells. In the case of $4\times 1$ cell the edge magnetic moments at a grater distance from impurity do not approach the values for pure nanoribbon, this is achieved for $8\times 1$ cell. As it is seen from Fig. 10 the polarization effect is strongest for Co located in  $h_{1}$ position and nearest edge atoms are most sensitive to perturbation. Our calculations indicate the charge transfer from Co to the $\pi$ bands of nanoribbon and no transfer is observed to the $\sigma$ bands.  The occupation  of the neighboring carbon $p_{z}$ orbitals is increased with adsorption and the spin polarization of  the nearby atoms at the edge is locally suppressed.

\section{KONDO EFFECT OF CO ATOM ON HOLLOW SITE}
\subsection{Model}
Static mean field DFT description of the electronic structure of Co adsorbed on the nanoribbon does not capture the effects of dynamic correlations of  strongly interacting  $3d$ electrons. To complement the  missing   local correlations of adatom electrons  we complete the model by Hubbard type term and exchange \cite{Pruschke1, Makarovski}.  The description of nanoribbon substrate and its coupling to impurity is  maintained within DFT formalism.  The Kohn-Sham Hamiltonian thereby serves as the non-interacting reference frame onto which we add local intra-atomic interactions. As we have presented in the preceding chapter, in the case of the considered  hollow location of  Co  atom the  $d_{xz}$ and $d_{yz}$ orbitals  are responsible for formation of magnetic moment,  their fluctuations in occupations and spins are essential for low energy physics.   We discuss therefore double orbital Anderson-like Hamiltonian in the form:
\begin{eqnarray}
&&{\mathcal{H}} ={\mathcal{H}}_{d}+{\mathcal{H}}_{v}+{\mathcal{H}}_{{\cal{ZGNR}}},
\end{eqnarray}
where impurity is described by
\begin{eqnarray}
&&{\mathcal{H}}_{d}=\sum_{m\sigma}\varepsilon_{0}n_{m\sigma}+{\cal{U}}\sum_{m}n_{m+}n_{m-}
+({\cal{U}}-{\cal{J}}/2)n_{m}n_{m'}\nonumber\\
&&-2{\cal{J}}\overrightarrow{{\cal{S}}}_{m}\cdot\overrightarrow{{\cal{S}}}_{m'}
\end{eqnarray}
and  $\varepsilon_{0}$ is the bare energy of local levels,  assumed to be equal for both orbitals,  ${\cal{U}}$ is the energy of  intra or interorbital Coulomb interaction   and ${\cal{J}}$ is Hund's exchange coupling.
Nanoribbon Hamiltonian reads:
\begin{eqnarray}
&&{\mathcal{H}}_{{\cal{ZGNR}}}=\sum_{kn\sigma}{\epsilon}_{kn\sigma}c^{\dag}_{kn\sigma}c_{kn\sigma},
\end{eqnarray}
\begin{table*}
\caption{\label{tab:table6}Magnetic moments and spin occupancies of different Co orbitals for hollow positions of 4ZGNR (VASP).}
\begin{ruledtabular}
\begin{tabular}{c|ccc|ccc|ccc}
 &\multicolumn{3}{c|}{$h_{0}$}&\multicolumn{3}{c|}{$h_{1}$}&\multicolumn{3}{c}{$h_{2}$}\\
 $d$&${\cal{N}}_{+}$&${\cal{N}}_{-}$&${\cal{M}}$&${\cal{N}}_{+}$&${\cal{N}}_{-}$&${\cal{M}}$
 &${\cal{N}}_{+}$&${\cal{N}}_{-}$&${\cal{M}}$\\\hline
 $z^{2}$&$0.966$&$0.933$&$0.033$&$0.997$&$0.994$&$0.003$&$0.996$&$0.989$&$0.007$\\
 $xz$&$0.963$&$0.456$&$0.507$&$0.963$&$0.503$&$0.46$&$0.937$&$0.729$&$0.208$\\
 $yz$&$0.955$&$0.889$&$0.066$&$0.979$&$0.336$&$0.643$&$0.978$&$0.245$&$0.733$\\
 $xy$&$0.997$&$0.062$&$0.935$&$0.934$&$0.837$&$0.097$&$0.905$&$0.845$&$0.06$\\
 $x^{2}-y^{2}$&$0.965$&$0.926$&$0.039$&$0.892$&$0.763$&$0.129$&$0.879$&$0.782$&$0.097$\\
 $tot$&$-$&$-$&$1.579$&$-$&$-$&$1.332$&$-$&$-$&$1.105$\\
 \end{tabular}
\end{ruledtabular}
\end{table*}
with $\varepsilon_{kn\sigma}$ denoting DFT ribbon eigenvalues and corresponding eigenfunctions $|kn\sigma\rangle$. The interaction between nanoribbon electrons and local levels is described by hybridization term:
\begin{eqnarray}
&&{\mathcal{H}}_{v}=\sum_{kn\sigma m}({\cal{V}}_{kn\sigma m}c^{\dag}_{kn\sigma}d_{m\sigma}+h.c.),
\end{eqnarray}
where hybridization amplitudes $V_{kn\sigma m}$ are hopping matrix elements between nanoribbon DFT eigenstates and $d$ orbitals. In this work, the realistic  ab initio hybridization is taken from  GGA-PBE calculations based on VASP code.
Hybridization strengths we use are not strictly single impurity couplings due to the periodicity of the adopted  first principles computations schemes, but for large enough supercells  they can approximately play this role. The nearest neighbor impurity-nanoribbon hopping integrals are extracted from DFT data $V^{i}_{m}=\sum_{kn}\langle d_{m}|\overline{kn}\rangle\overline{\varepsilon_{kn}}\langle \overline{kn}|p^{i}_{z}\rangle$,where $|m\rangle$ denotes adatom orbital, $|\overline{kn}\rangle$ and $\overline{\varepsilon_{kn}}$ are DFT eigenstates and energies of $Co+ZGNR$ system.

Table IV presents   hybridization amplitudes in real space with restriction to the dominant n.n. contributions.
For comparison we present amplitudes for all $3d$ orbitals. Note the smallness of the amplitudes to $A_{1}$ and $B_{2}$ for $d_{xz}$ in $h_{1}$ position and large amplitudes to these atoms for $d_{yz}$. Pictorially this difference  can be understood recalling  the shapes of these  orbitals. Remembering that the edge states dominate the energy window near the gap,  one can expect  distinctly different roles of  $d_{xz}$ and $d_{yz}$ orbitals  in Kondo physics for $h_{1}$ position.  For $h_{2}$ the role of edge states is diminished. Comparison of the amplitudes for  $h_{0}$ and hollow sites   helps to understand  the earlier mentioned  reversal of the role between $d_{yz}$ and $d_{xy}$ when positions of the adatom interchange.

To describe orbital degrees of freedom on the same footing as spin it is useful to introduce orbital pseudospin ${\cal{T}}$ defined by ${\cal{T}}=\Psi_{d}^{\dag}\tau\Psi_{d}$, where $\tau$ is Pauli matrix in orbital space $\{d_{xz},d_{yz}\}$ and $\Psi_{d}^{\dag}$ represents spin-orbital field operator $\Psi_{d}^{\dag}=[d^{\dag}_{xz+},d^{\dag}_{xz-},d^{\dag}_{yz+},d^{\dag}_{yz-}]$.

\subsection{Hybridization function}
Hybridization function describes coupling of impurity to nanoribbon. Hereafter we restrict to nearest neighbors of impurity and consider only $d_{xz}$ and $d_{yz}$ orbitals. Hybridization then reads:
\begin{eqnarray}
&&\Sigma^{h_{1(2)}}_{mm'\sigma}=\sum_{kn}\frac{V^{*h_{1(2)}}_{kn\sigma m}V^{h_{1(2)}}_{kn\sigma m'}}{z-\varepsilon_{kn\sigma}},
\end{eqnarray}
where $\varepsilon_{kn\sigma}$ and $|kn\sigma\rangle$ are DFT eigenenergies and eigenstates of bare graphene nanoribbon and $V_{kn\sigma m}=(1/\sqrt{N_{x}})\sum_{l_{h}}\sum_{a_{h}}V^{a_{h}}_{m}e^{ika_{h}}\langle kl_{h}\sigma|kn\sigma\rangle$, where $a_{h}$ are n. n. vectors connecting adatom with carbon sites from surrounding hexagon, $l_{h}$ labels four carbon chains along infinite $x$ direction crossing the hexagon, $N_{x}$ is the number of sites in carbon chain in $x$ direction and $|kl\rangle=(1/\sqrt{N_{x}})\sum e^{iki_{l}}|p^{i_{l}}_{z}\rangle$. All hybridization functions presented below have been calculated using VASP code.
\begin{figure}
\includegraphics[width=7 cm,bb=0 0 660 652,clip]{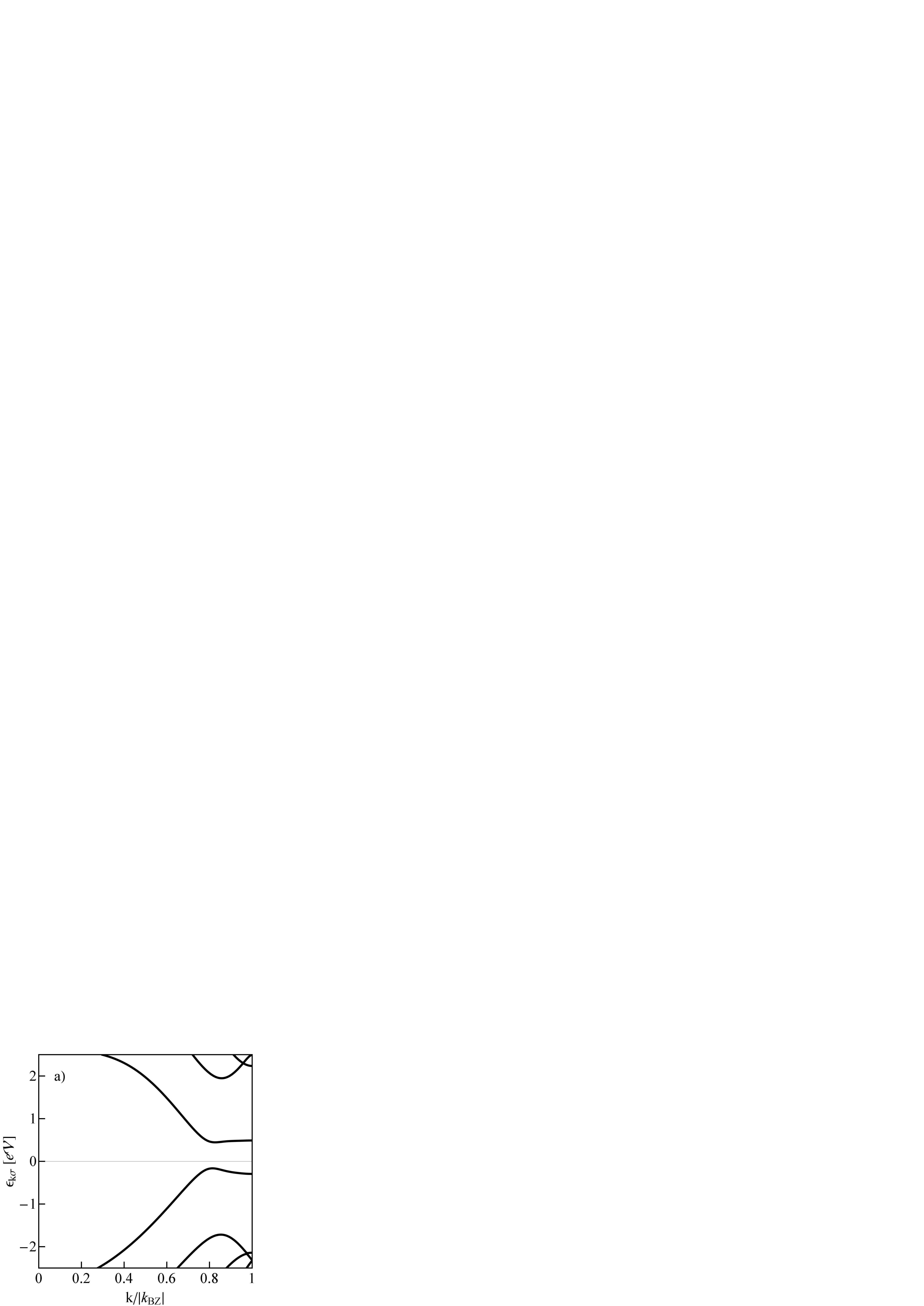}
\includegraphics[width=6.6 cm,bb=0 0 660 704,clip]{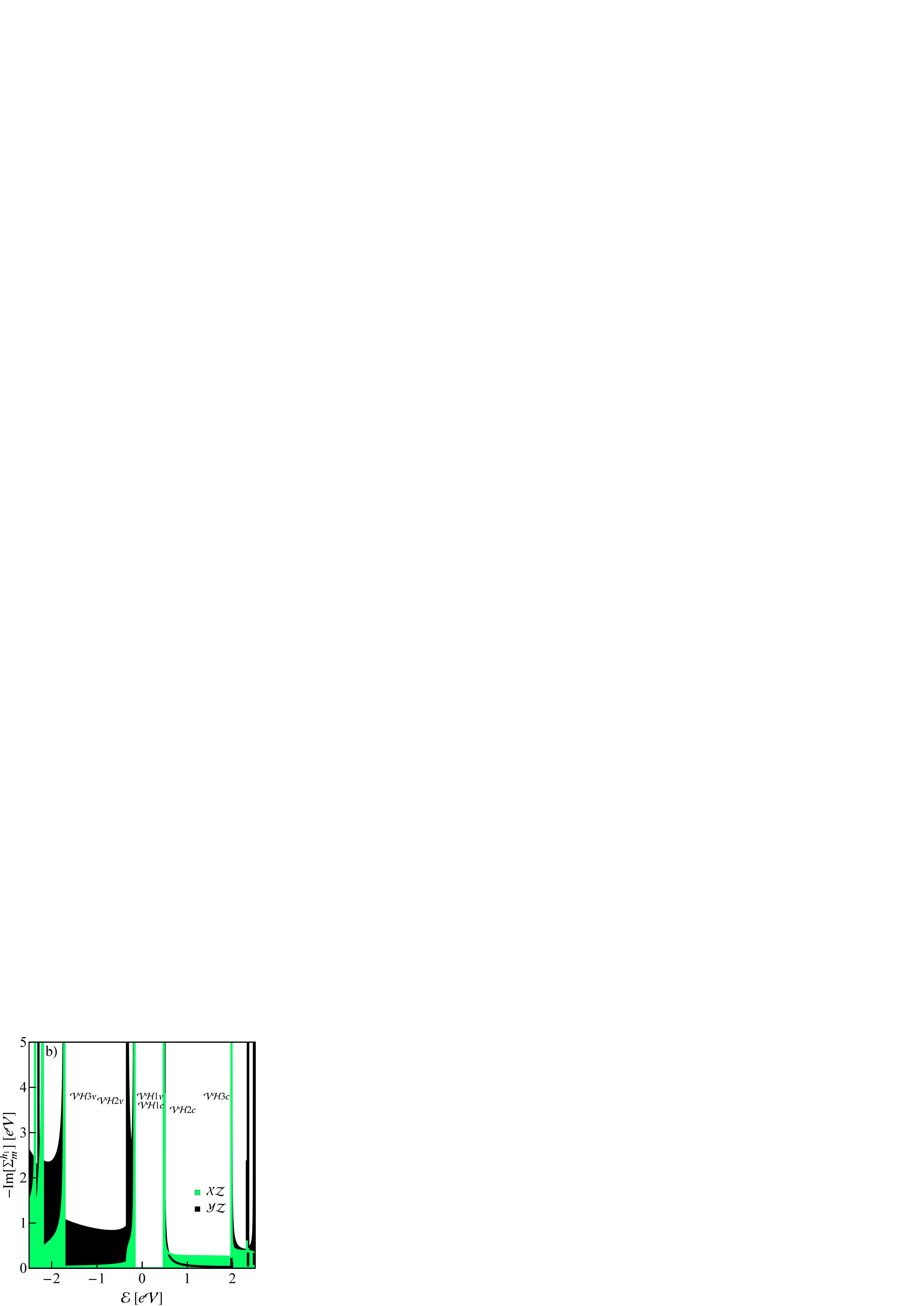}\\
\includegraphics[width=6.8 cm,bb=0 0 660 704,clip]{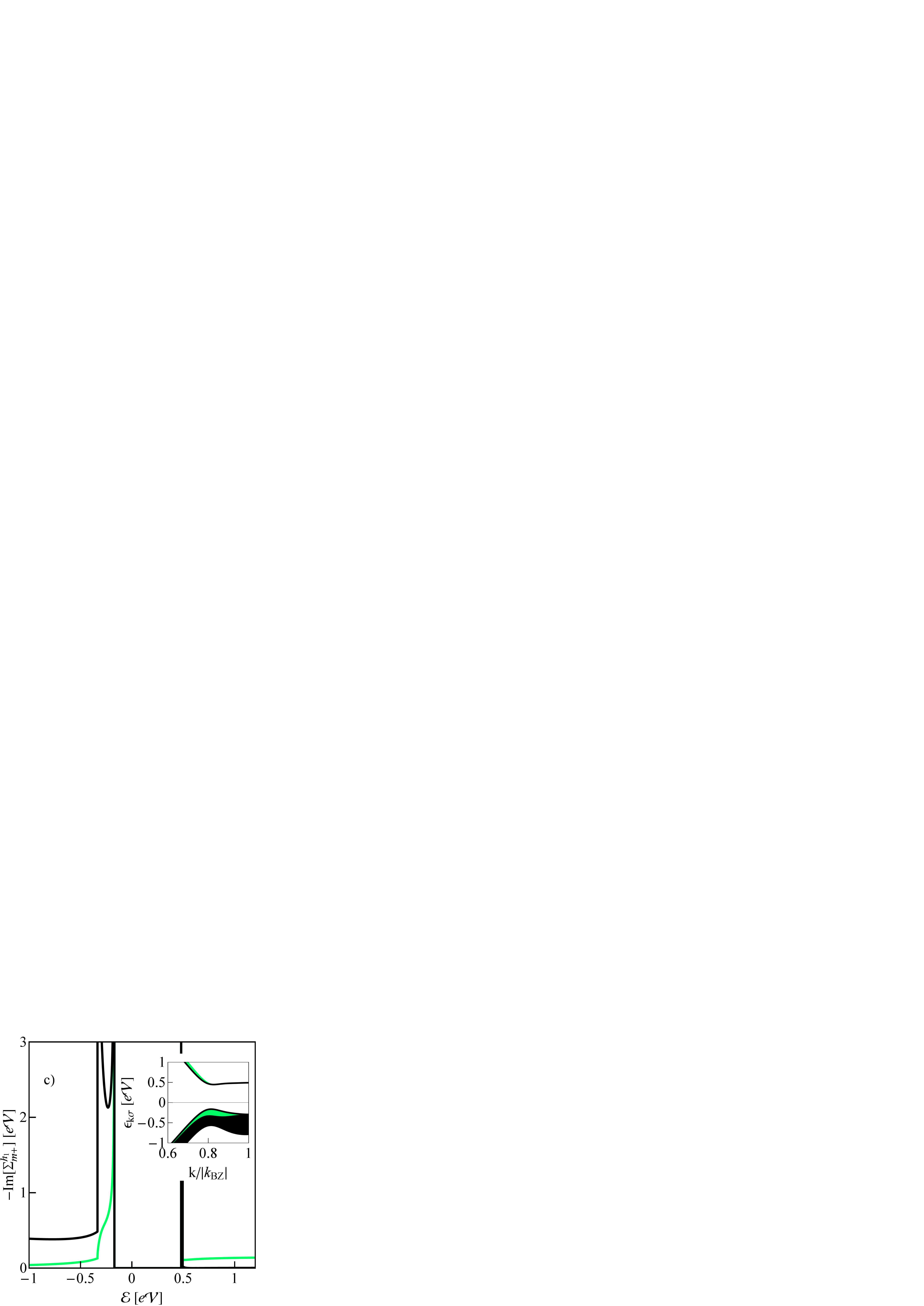}
\includegraphics[width=6.8 cm,bb=0 0 660 704,clip]{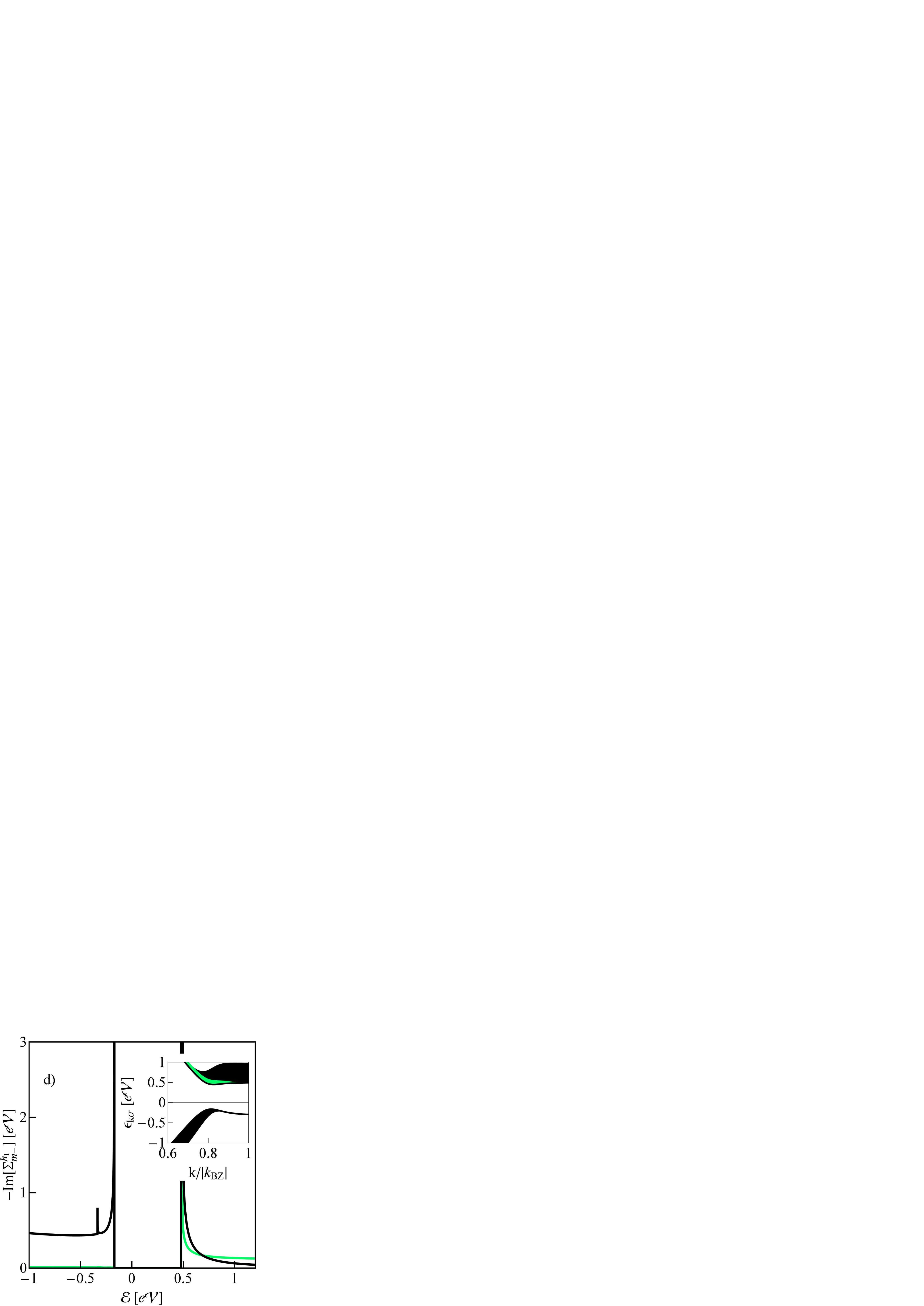}
\caption{\label{fig:epsart11}(Color online) Low energy band structure of 4ZGNR (a) together with orbitally resolved hybridization functions of Co adatom in $h_{1}$ position (b, c, d). Black color denotes $yz$ contribution and geen/light grey $xz$. In Fig. b labeling of Van Hove singularities used further in the text is introduced. Figures c, d present spin up and spin down parts of hybridization respectively. Insets of Figs. c, d  are zoom views of lowest dispersion curves decorated with $xz$ and $yz$ contributions.}
\end{figure}
For pure graphene $\Sigma_{mm'}\sim\delta_{mm'}$ due to $C_{6v}$ symmetry of hollow sites. In nanoribbon, where this symmetry is broken also off-diagonal terms occur, but as we have checked due to rapid oscillations in k-space they are much smaller than diagonal elements and additionally they affect the impurity states in forth power in hybridization, whereas the diagonal in second power. Based on these arguments we neglect in the following, for simplicity of calculations, the off-diagonal self energies.
Hybridization function plays the role of  embedding self-energy. The real parts of self-energies are associated with the shift of the local energies, while the imaginary parts give the broadening of  impurity levels. Fig. 11   shows imaginary part of the low energy hybridization functions for $h_{1}$ position together with the corresponding VASP nanoribbon bands from this range. In the following we refer to the presented singularities and therefore we introduce their labeling in Fig. 11b.  More detailed pictures of spin and orbital resolved hybridization with  both real and imaginary parts are presented in Fig. 12.  In general  the hybridization functions are spin dependent, what is mainly dictated by spin dependence of local nanoribbon Green's functions. The  opposite local polarizations at $h_{1}$ and $h_{3}$ (Fig. 1) reflects in the change of roles of spins in  hybridization function $\Sigma^{h_{1}}_{m\sigma}=\Sigma^{h_{3}}_{m\overline{\sigma}}$. At $h_{2}$, where   polarization  contributions from the opposite edges compensate   hybridizations are equal for both spin orientations.  The hybridization functions are rich in structure, of special importance for Kondo effect are observed Van Hove singularities (VHS's) occurring in position of minima, maxima or saddle points of the bands.
Vanishing of derivatives of dispersion curves indicates energies, where singularities are expected, but whether singularity clearly reflects in orbital resolved hybridization  depends on the weight of   contribution of a given  symmetry to the bands  in the considered energy range. This fact is illustrated in Figs. 11 c, d, where highest conduction and lowest valence bands are decorated by amplitudes specifying   projection of the eigenfunctions onto the symmetry of a given local orbital.
\begin{figure}
\includegraphics[width=7 cm,bb=0 0 660 630,clip]{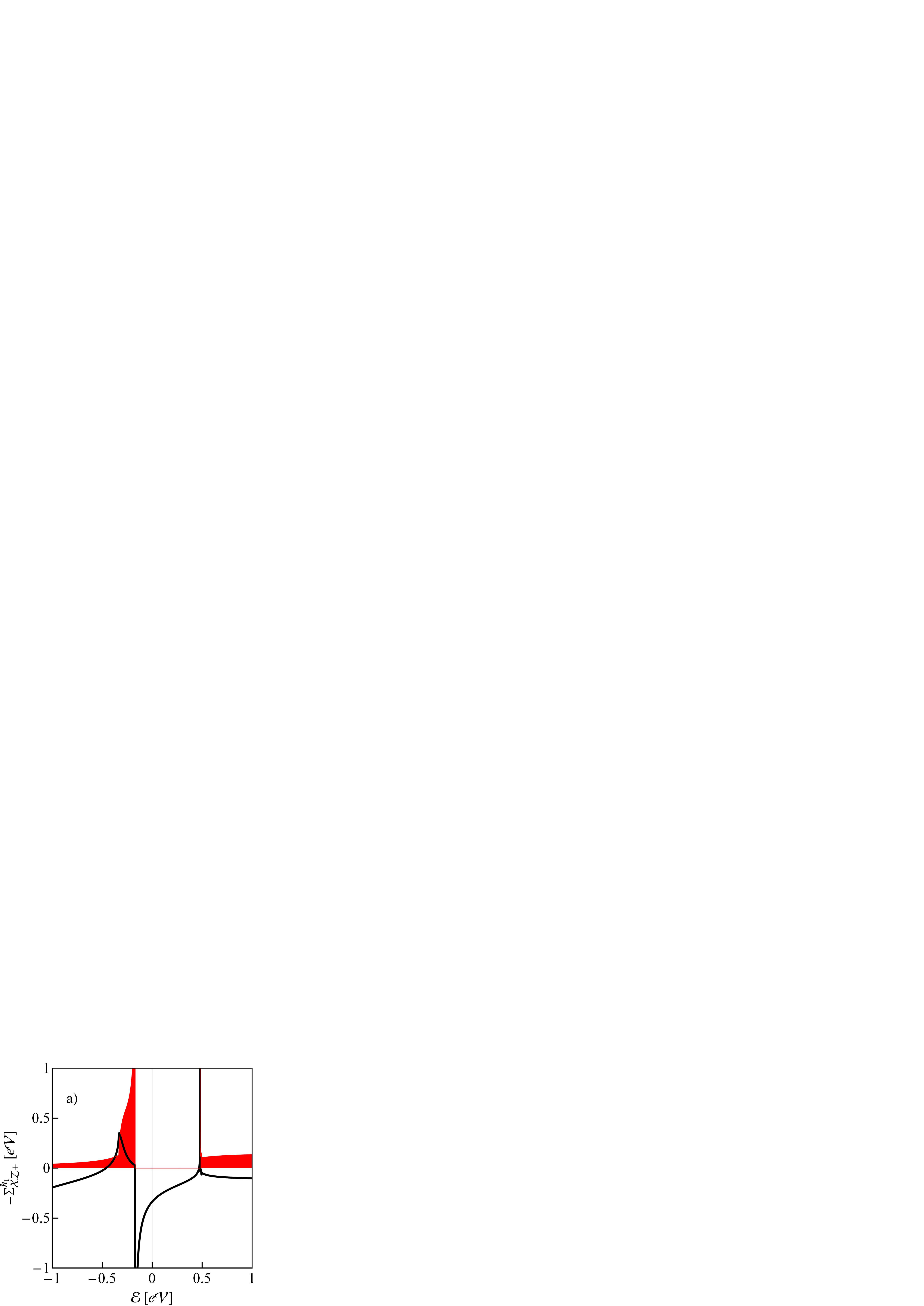}
\includegraphics[width=7 cm,bb=0 0 660 630,clip]{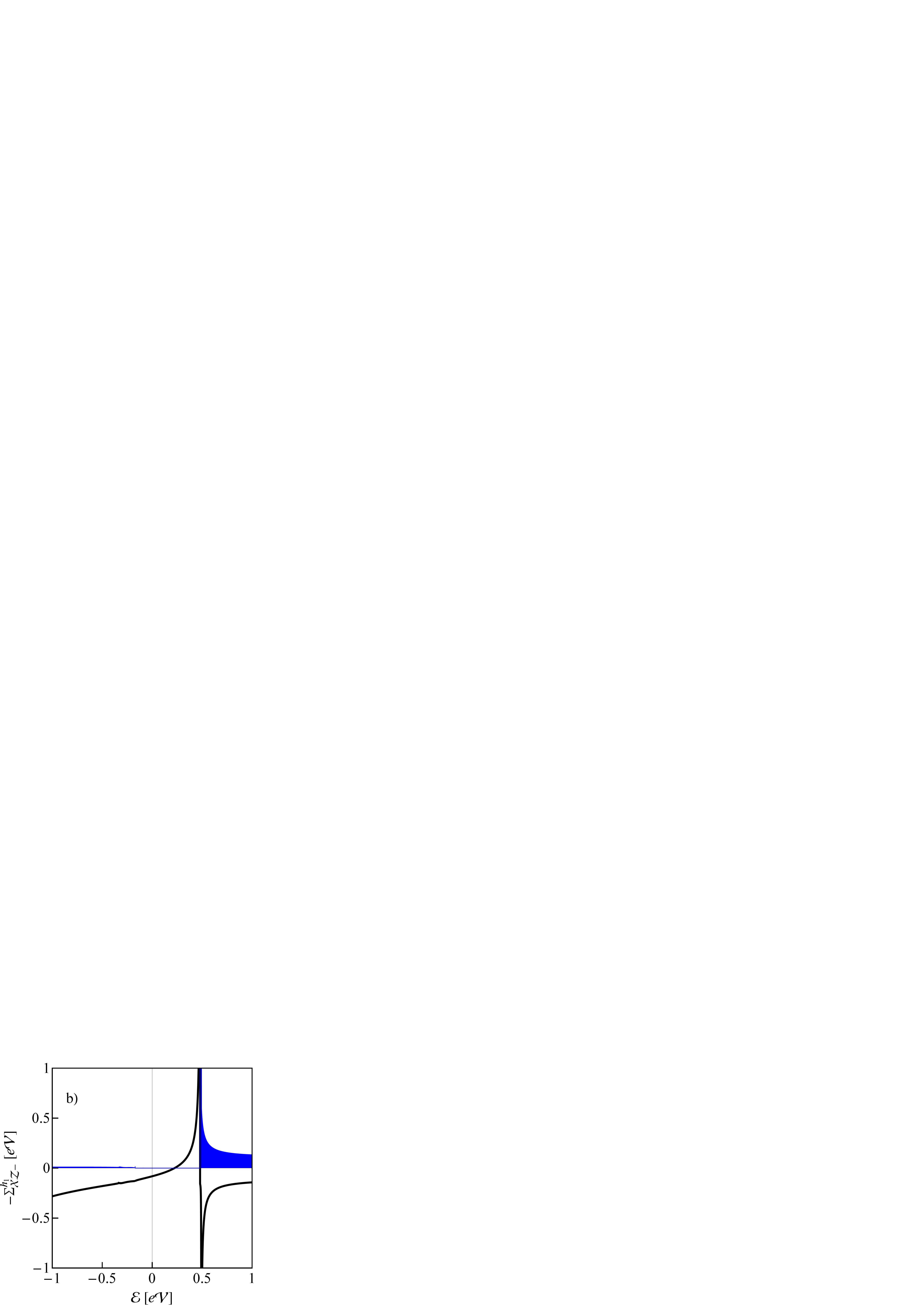}\\
\includegraphics[width=7 cm,bb=0 0 660 630,clip]{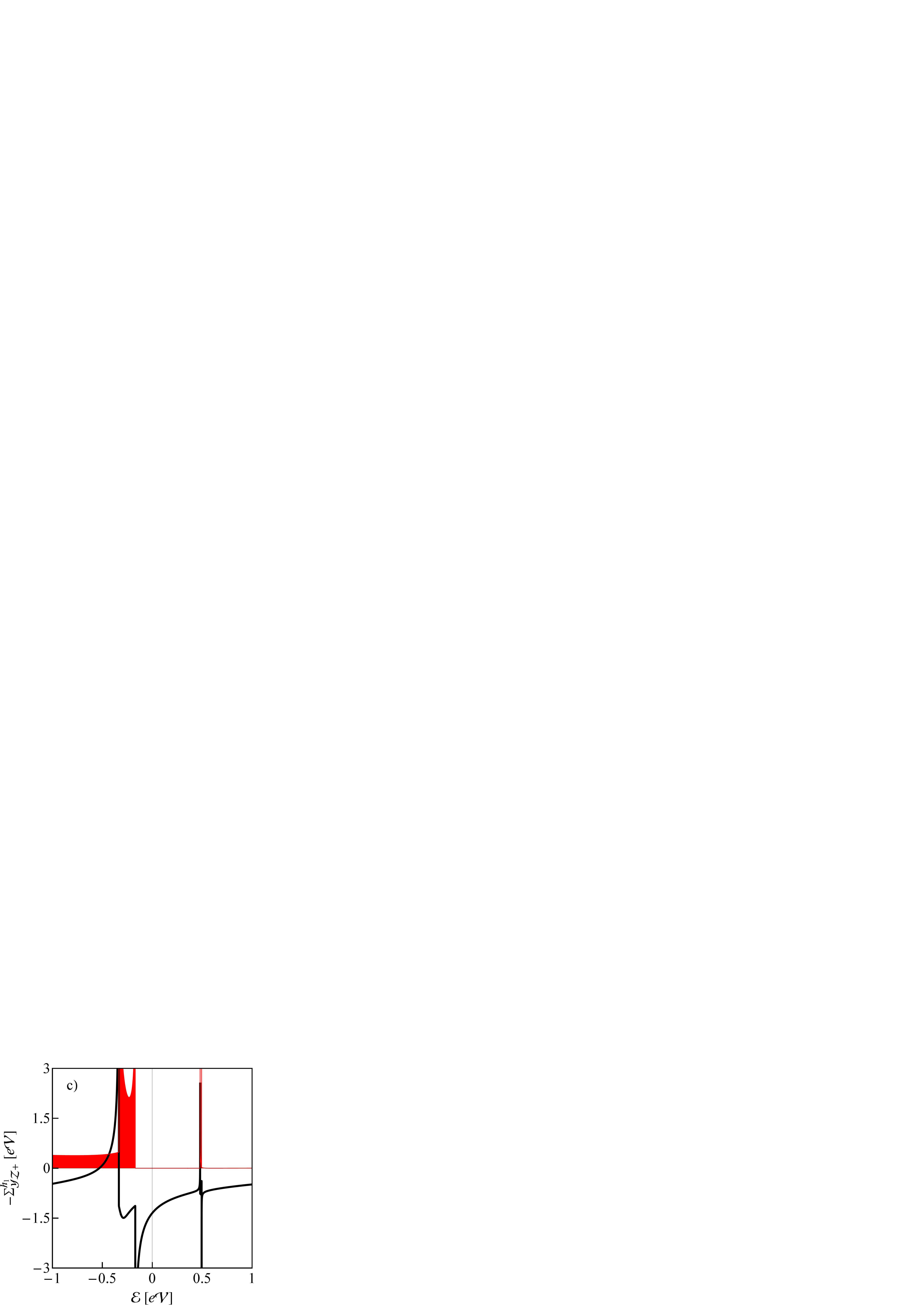}
\includegraphics[width=7 cm,bb=0 0 660 630,clip]{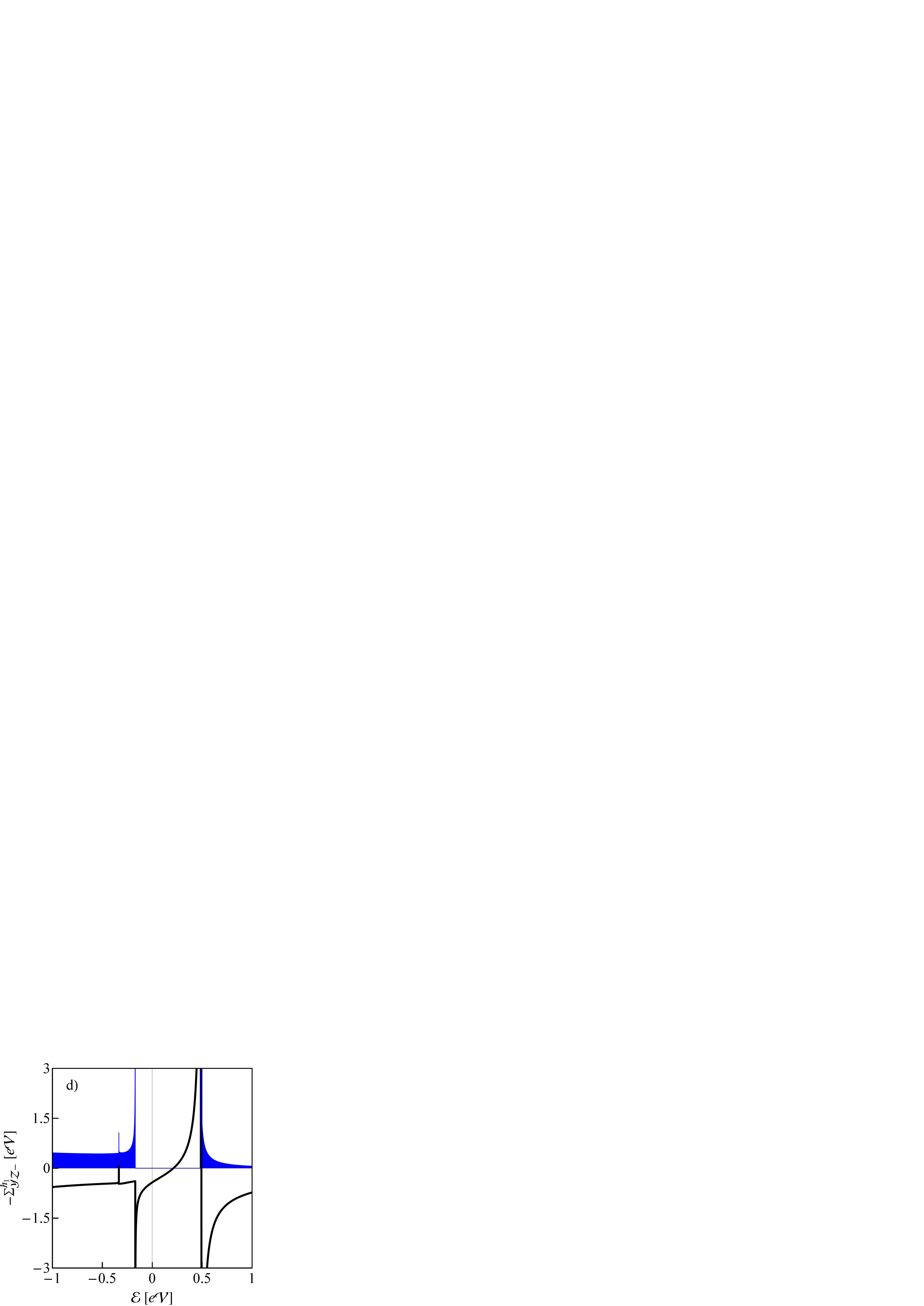}
\caption{\label{fig:epsart12}(Color online) Orbital and spin resolved hybridization functions for Co at $h_{1}$ position in 4ZGNR. Solid black lines represent real parts ($Re[\Sigma^{h_{1}}_{m\sigma}]$) and imaginary parts ($Im[\Sigma^{h_{1}}_{m\sigma}]$) are shown by green/light grey filled curves.}
\end{figure}
For example in the energy window presented in Figs. 11c, d   two pronounced singularities are observed (${\cal{VH}}1\nu$ and ${\cal{VH}}2\nu$) below the gap for $yz$ symmetry, characterized by  peaks in imaginary parts of hybridizations and  discontinuities in real parts. For $xz$ symmetry on the other hand, similar behavior is seen only  close to the gap.
\begin{figure}
\includegraphics[width=7 cm,bb=0 0 660 636,clip]{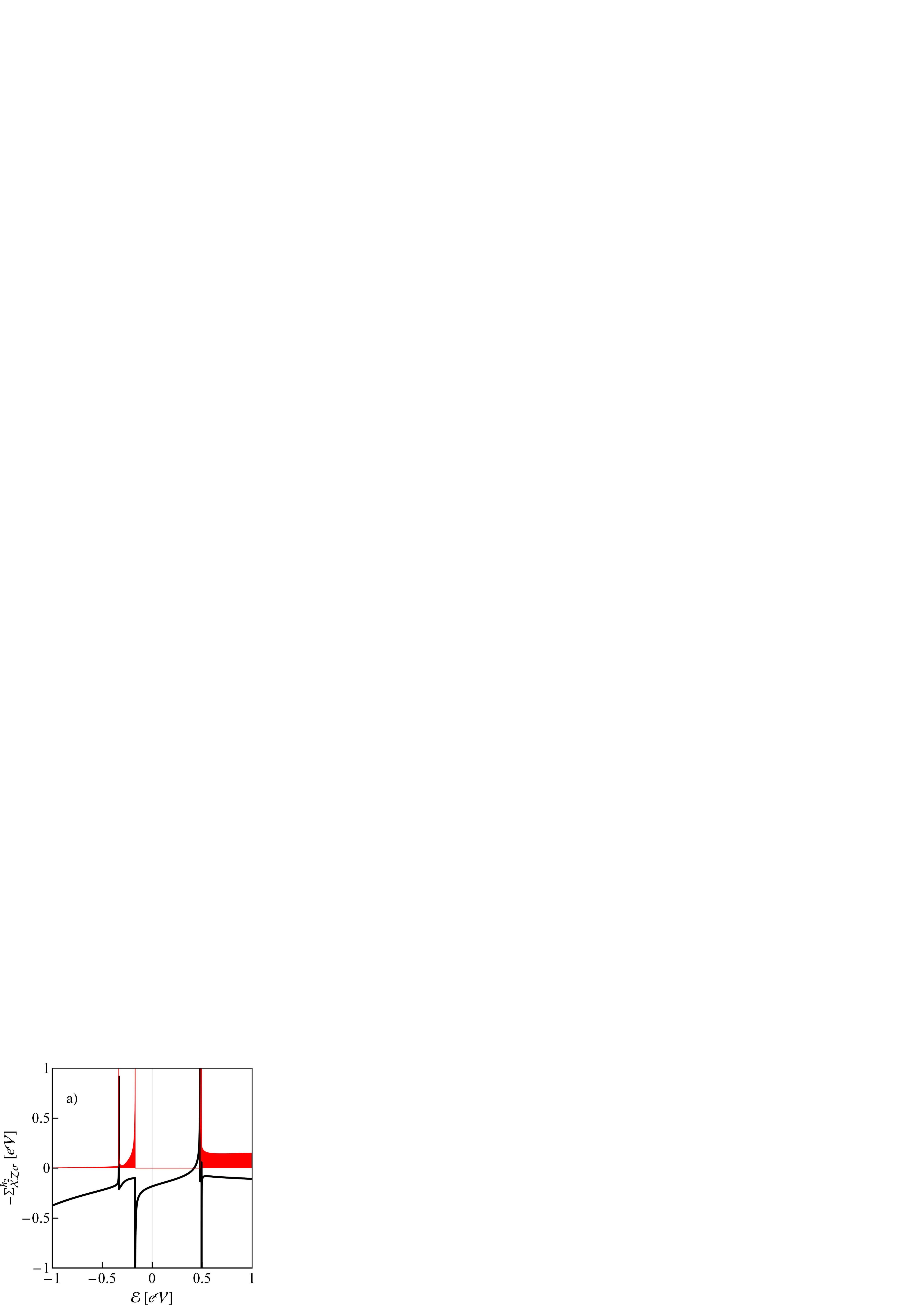}
\includegraphics[width=7 cm,bb=0 0 660 636,clip]{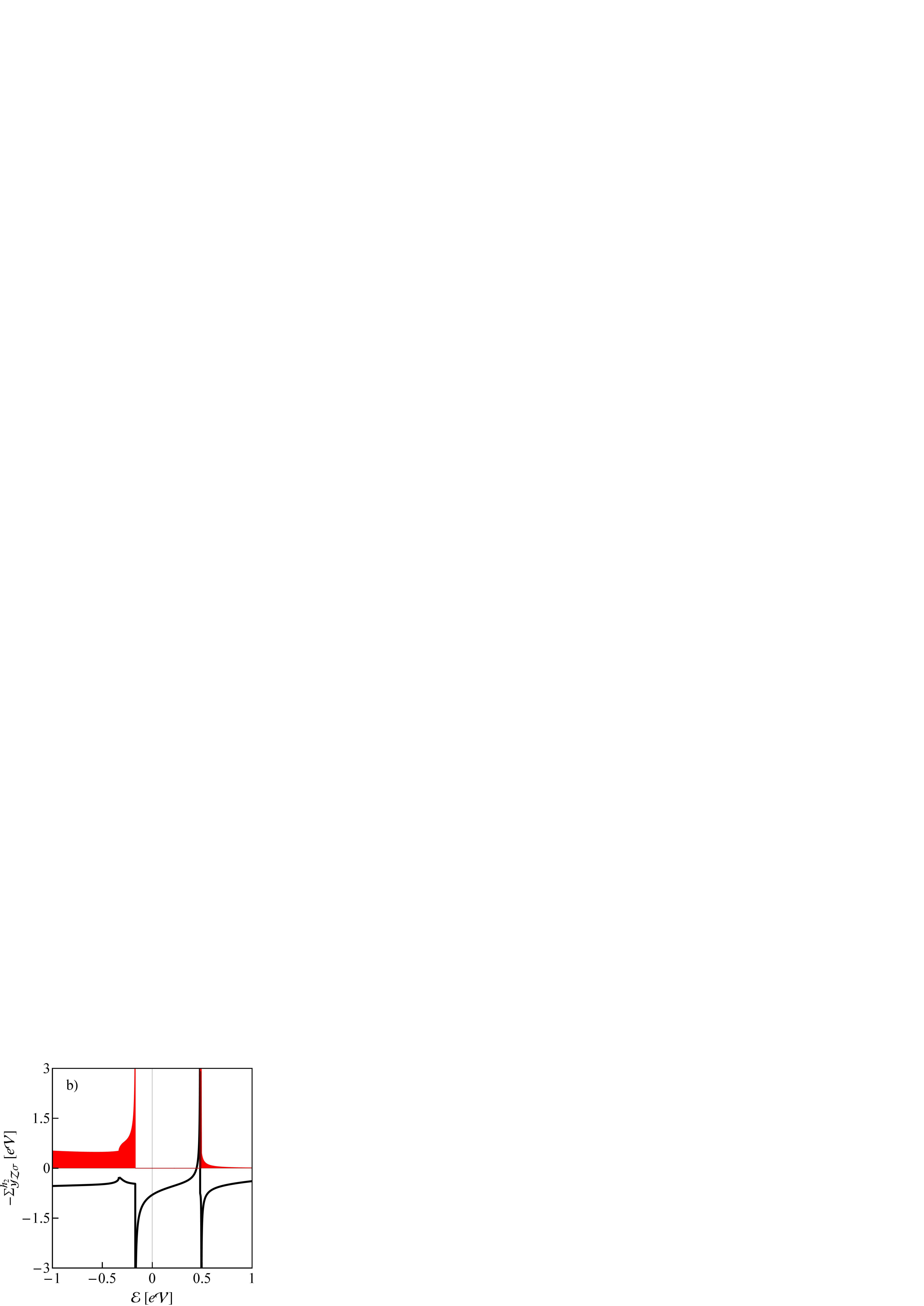}
\caption{\label{fig:epsart13}(Color online) Orbital resolved hybridization function of Co in $h_{2}$ position of 4ZGNR. Solid black lines represent real part and green/light grey filled curves show the imaginary parts of hybridization.}
\end{figure}
For ${\cal{E}} =-0.3348$ (${\cal{VH}}2\nu$) the $xz$  contribution to the bands is small (see the insets of Fig. 11c, d). The character of many-body resonances  are determined by the deepness of the local level with respect to the Fermi energy and hybridization strength, both of these values  dramatically change  near the singularity and therefore it is expected, that  an interesting physics emerges   in the vicinity  of these energy points.

\subsection{Slave boson mean field approach}
The described modeling of single adatom embedded onto graphene nanoribbon by Anderson like Hamiltonian allows us to examine the strong correlations by the well elaborated  techniques with known applicability regimes.  Our main interest  focuses on the impact of the details of nanoribbon electronic and magnetic  structure on  the single impurity Kondo effect.
\begin{figure}
\includegraphics[width=6.8 cm,bb=0 0 660 660,clip]{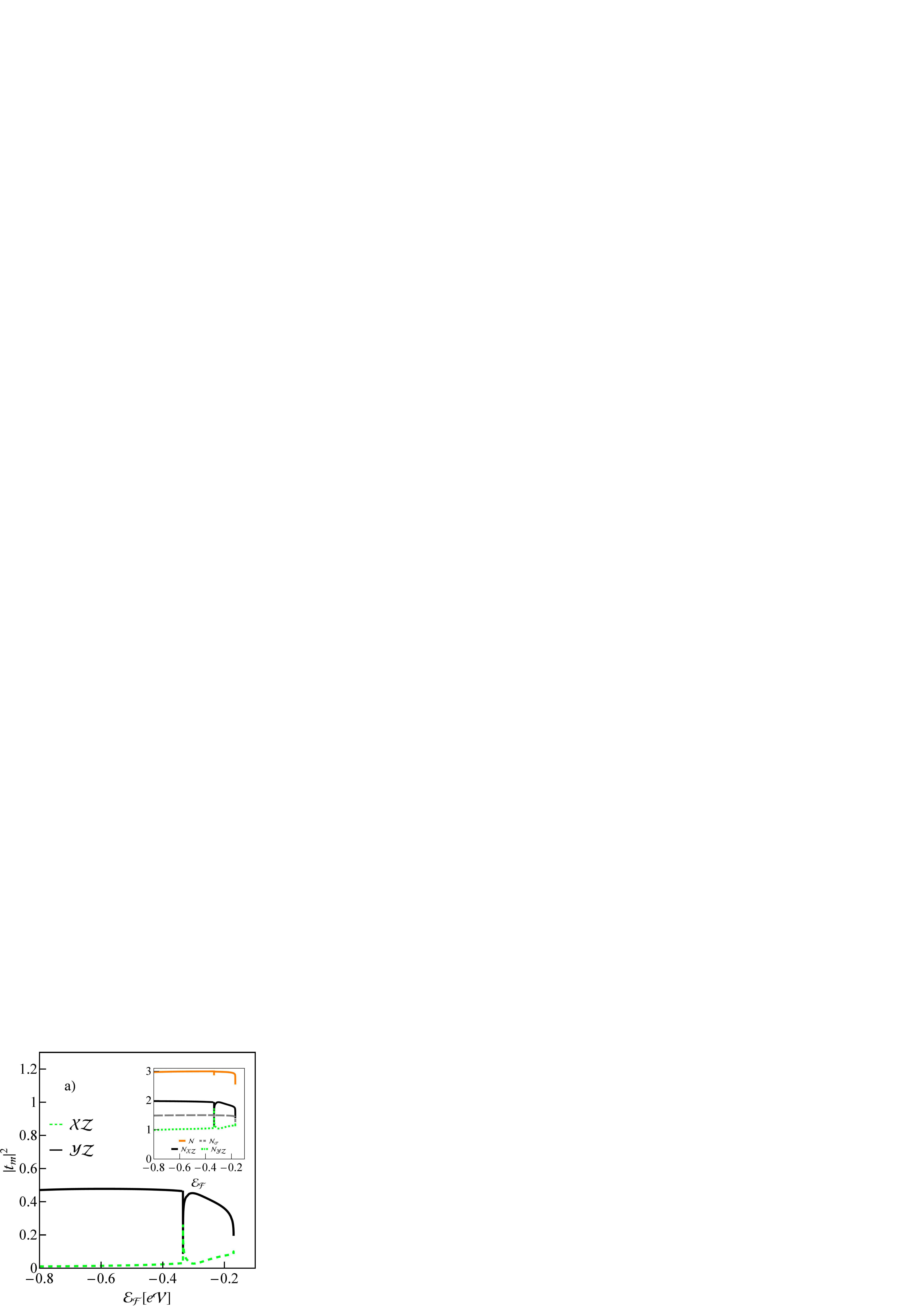}
\includegraphics[width=7.1 cm,bb=0 0 660 636,clip]{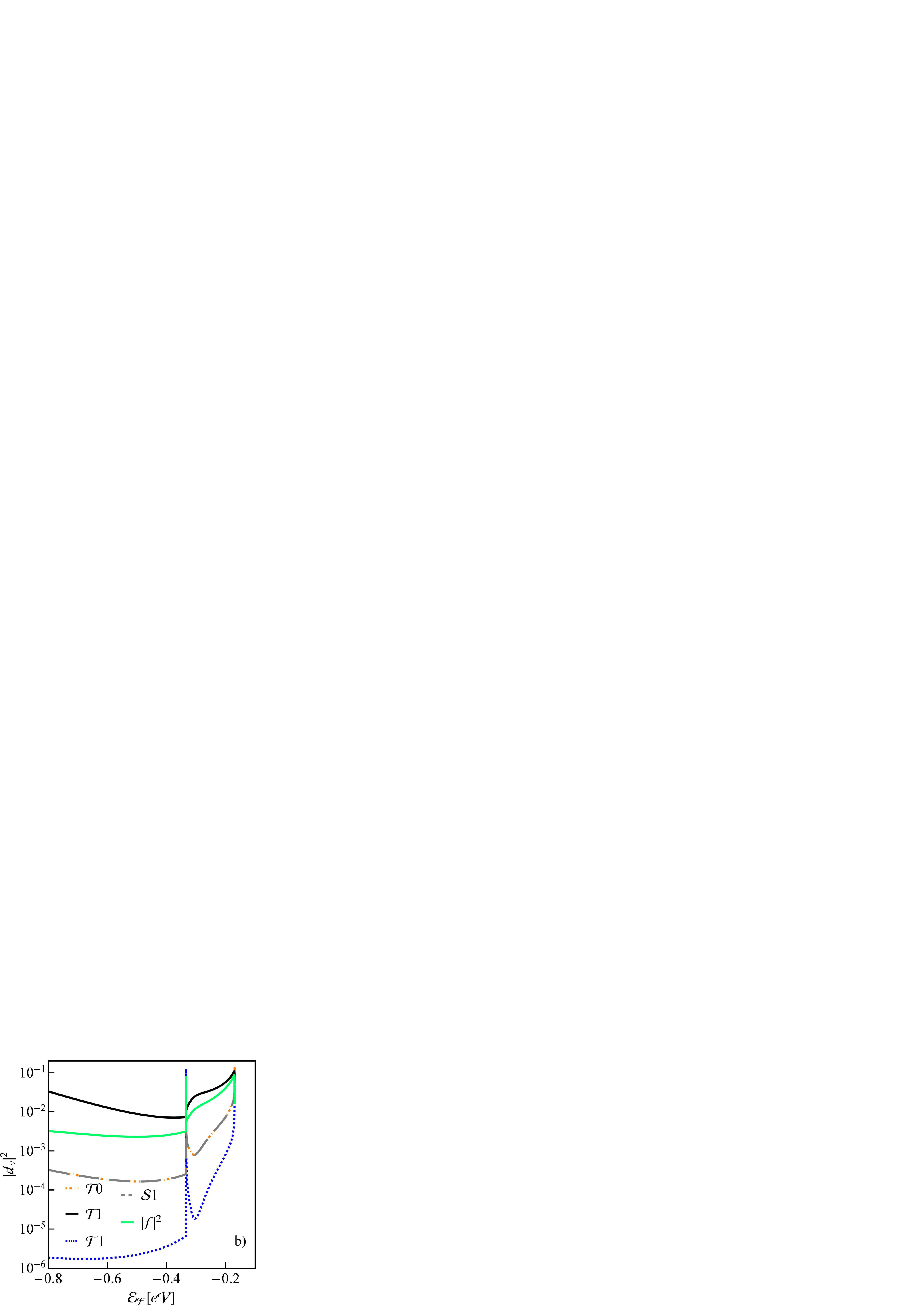}\\
\includegraphics[width=6.8 cm,bb=0 0 660 702,clip]{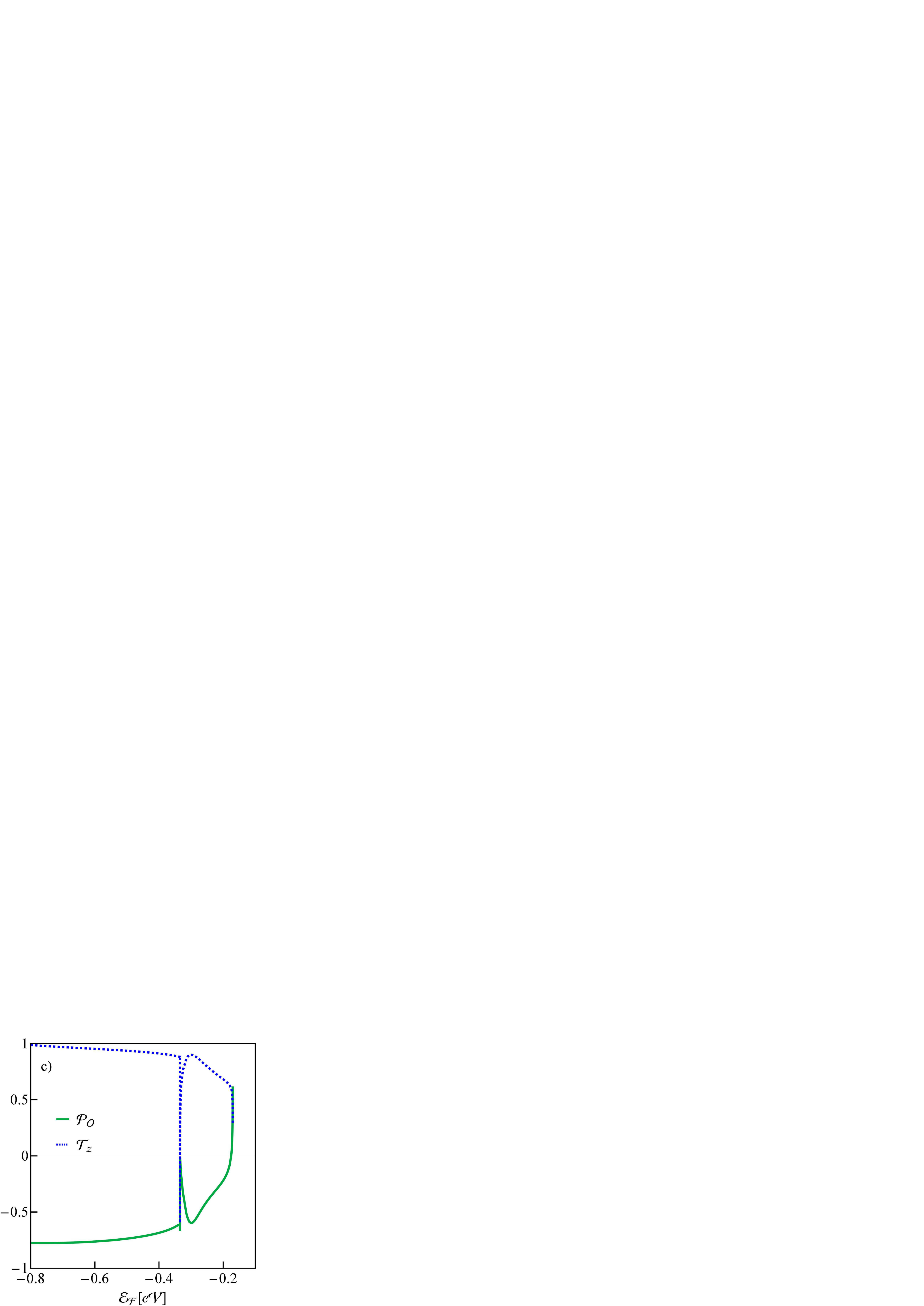}
\caption{\label{fig:epsart14}(Color online) Expectation values of slave boson operators, polarizations and expectation values of orbital pseudospin of Co impurity at $h_{2}$ position of 4ZGNR a) triple occupation slave boson operators $|t_{m}|^{2}=|t_{m+}|^{2}=|t_{m-}|^{2}$ b) double and full occupation slave boson operators $|d_{{\cal{T}}0}|^{2}$, $|d_{{\cal{T}}1}|^{2}$, $|d_{{\cal{T}}\overline{1}}|^{2}$, $|d_{{\cal{S}}1}|^{2}=|d_{{\cal{S}}\overline{1}}|^{2}$, $|d_{{\cal{S}}0}|^{2}=|d_{{\cal{T}}0}|^{2}$ and $|f|^{2}$ c) orbital polarization of Co at $h_{2}$ site (${\cal{P}}_{{\cal{O}}}=\frac{\varrho_{xz}({\cal{E}}_{\cal{F}})-\varrho_{yz}({\cal{E}}_{\cal{F}})}{\varrho_{xz}({\cal{E}}_{\cal{F}})+\varrho_{yz}({\cal{E}}_{\cal{F}})}$, where $\varrho_{m}({\cal{E}})=\sum_{\sigma}\varrho_{m\sigma}({\cal{E}})$ and $\varrho_{m\sigma}$ denotes spin-orbital partial density of states) and expectation value of Co orbital pseudospin (${\cal{T}}_{z}$). Inset of Fig. a shows total, spin and orbital ${d_{xz},d_{yz}}$ occupancies.}
\end{figure}
The basic analysis of  variation of  many-body correlations with tuning the chemical potential  is based on  mean field  slave boson approach of Kotlar and Ruckenstein \cite{Ruckenstein}.  This approximation  concentrates exclusively on many-body resonances taking into account spin and orbital fluctuations, but neglecting charge fluctuations.  In principle SBMFA strictly  applies close to  the unitary Kondo limit, but  due to its simplicity this method is also often used for systems with broken symmetry \cite{Bulka, Lim, Lipinski, Trocha}.  It is believed that  it   captures  the essential features of the examined problem also in this case. SMBFA is unreliable for higher temperatures.
\begin{table}
\caption{\label{tab:table7}Two electron eigenstates of hamiltonian (3) with the assigned slave bosons.}
\begin{ruledtabular}
\begin{tabular}{lr}
$eigenstate$&$slave-boson$\\
\hline
$|\uparrow,\uparrow\rangle$ & $d_{{\cal{S}}=1{\cal{S}}_{z}=1}=d_{{\cal{S}}1}$\\
$(1/\sqrt{2})(|\uparrow,\downarrow\rangle+|\downarrow,\uparrow\rangle)$ & $d_{{\cal{S}}=1{\cal{S}}_{z}=0}=d_{{\cal{S}}0}$\\
$|\downarrow,\downarrow\rangle$ & $d_{{\cal{S}}=1{\cal{S}}_{z}=\overline{1}}=d_{{\cal{S}}\overline{1}}$\\
$|\uparrow\downarrow,0\rangle$ & $d_{{\cal{T}}=1{\cal{T}}_{z}=1}=d_{{\cal{T}}1}$\\
$(1/\sqrt{2})(|\uparrow,\downarrow\rangle-|\downarrow,\uparrow\rangle)$ & $d_{{\cal{T}}=1{\cal{T}}_{z}=0}=d_{{\cal{T}}0}$\\
$|0,\uparrow\downarrow\rangle$ & $d_{{\cal{T}}=1{\cal{T}}_{z}=\overline{1}}=d_{{\cal{T}}\overline{1}}$\\\end{tabular}
\end{ruledtabular}
\end{table}
This is a consequence of break of  the required gauge invariance which is associated with charge conservation, what leads to artificial sharp transition to the state with vanishing expectation value of boson fields \cite{Hewson}.
\begin{figure}
\includegraphics[width=7 cm,bb=0 0 660 654,clip]{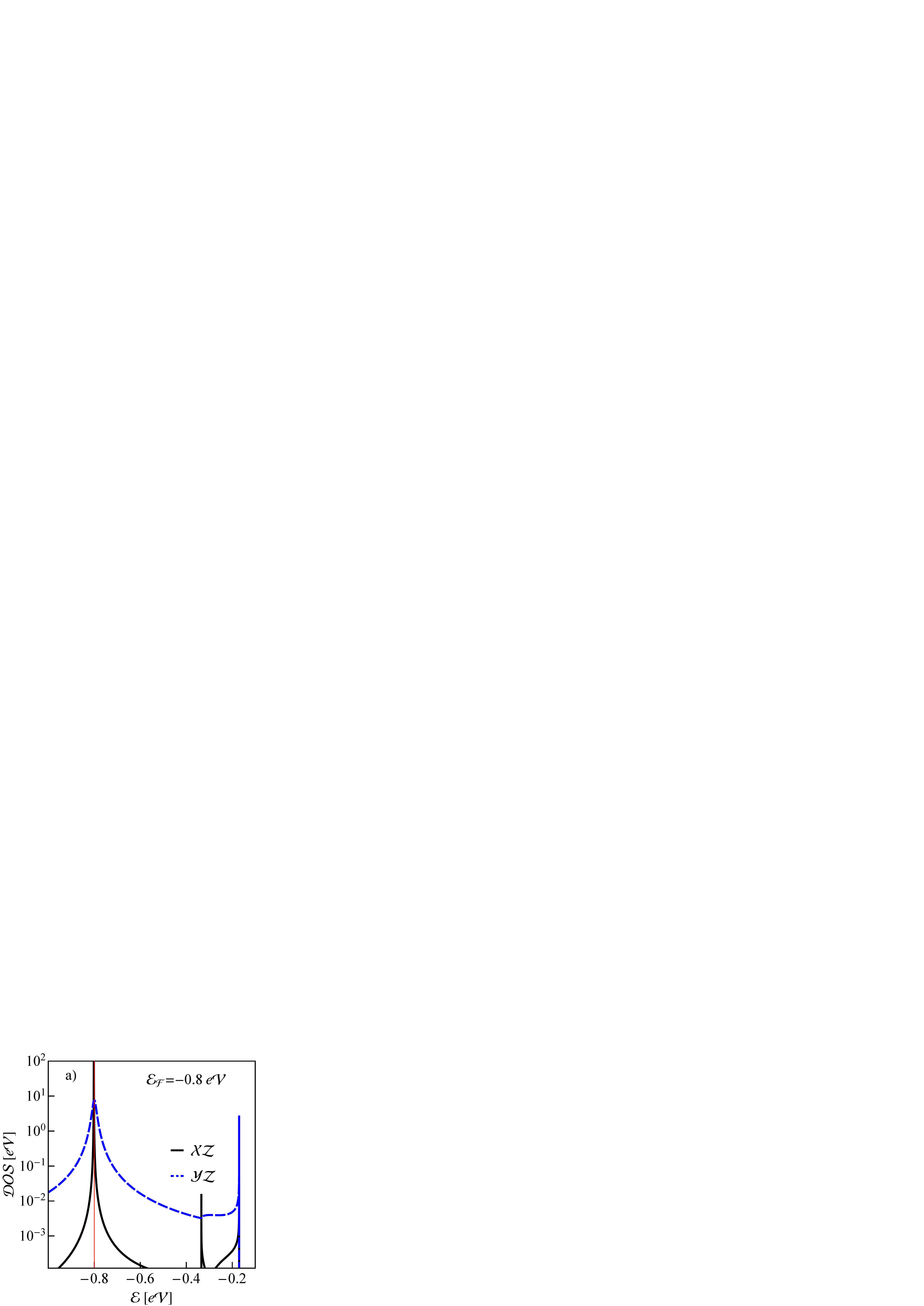}
\includegraphics[width=7 cm,bb=0 0 660 654,clip]{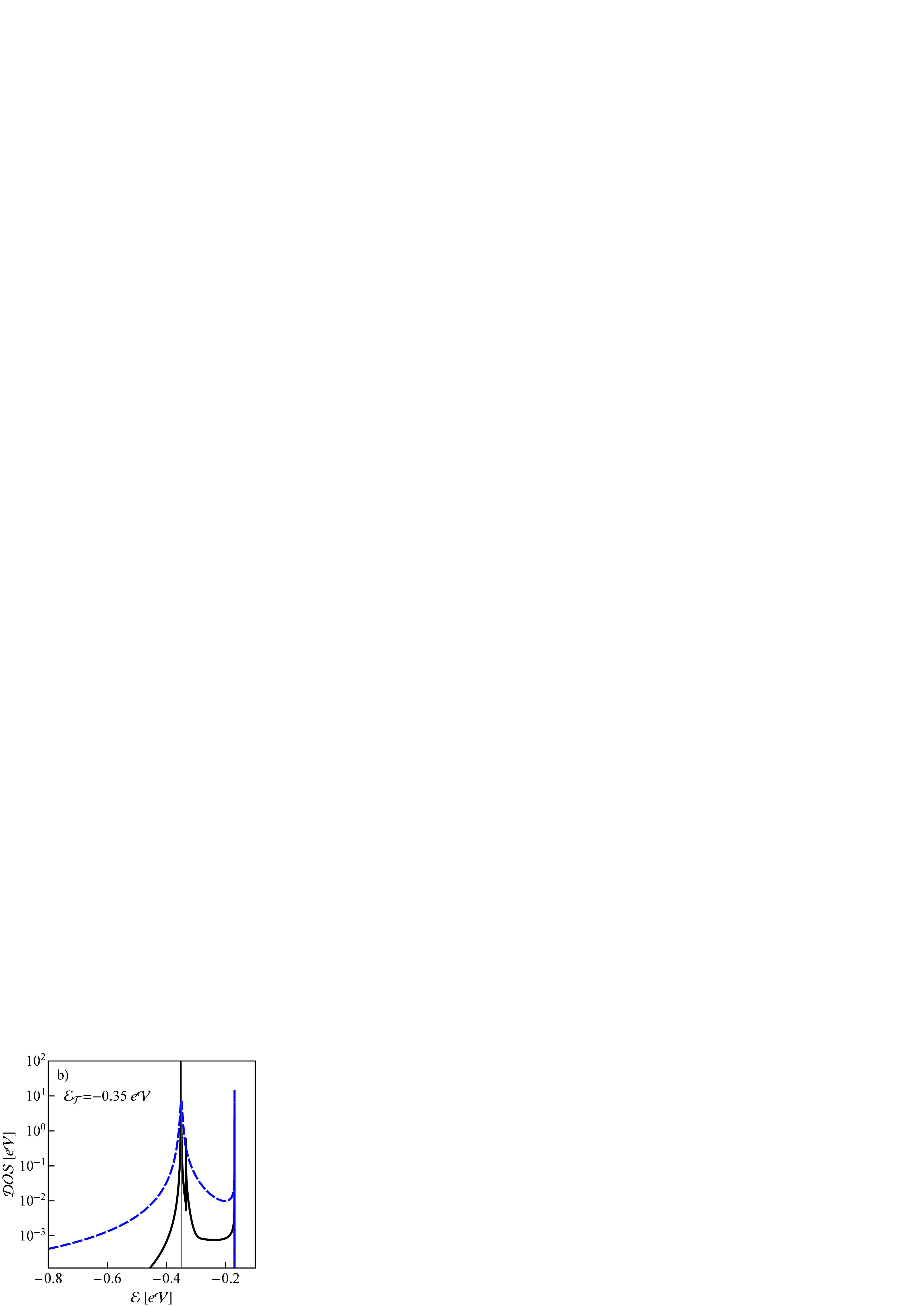}\\
\includegraphics[width=7 cm,bb=0 0 660 654,clip]{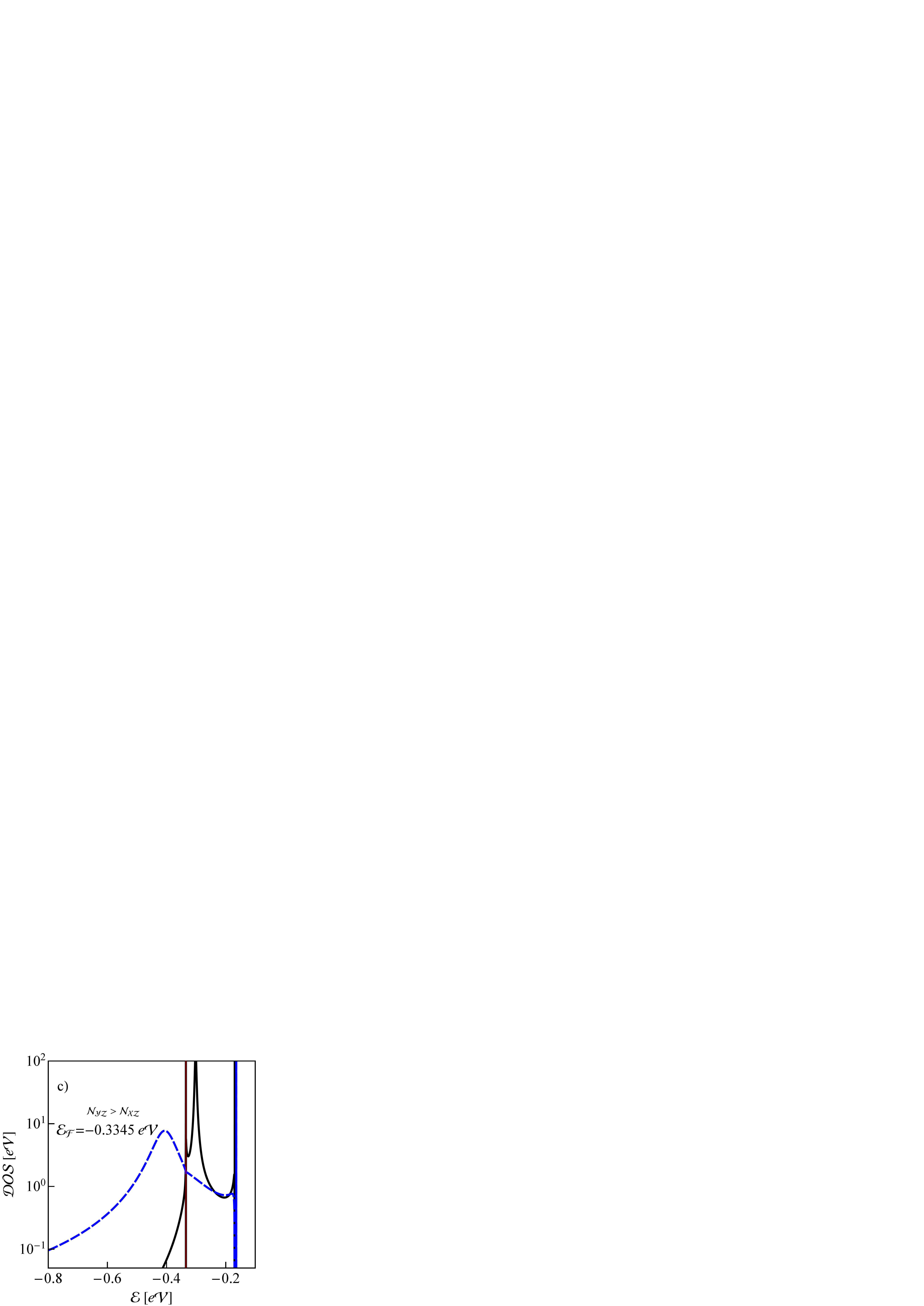}
\includegraphics[width=7 cm,bb=0 0 660 654,clip]{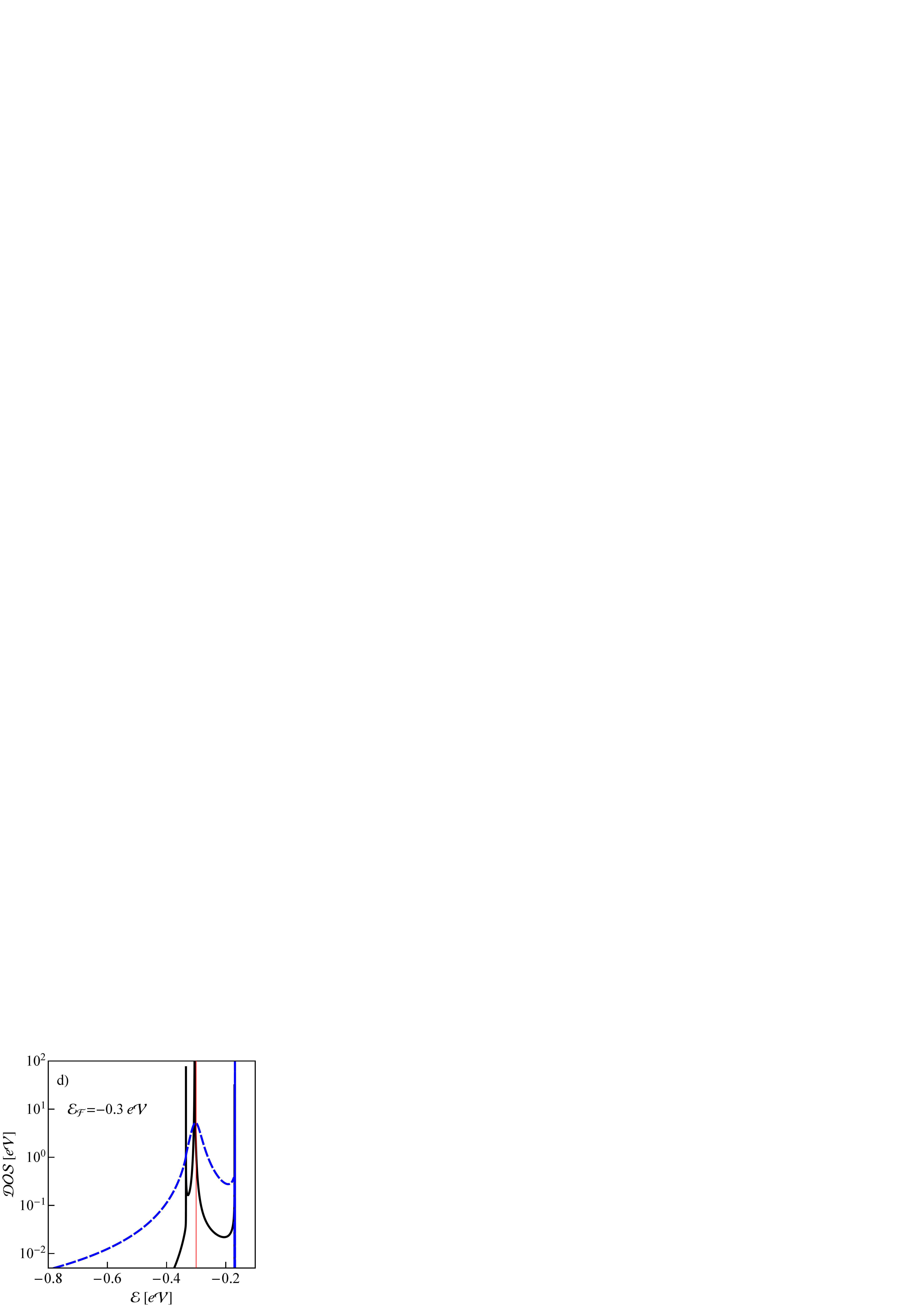}
\caption{\label{fig:epsart15}(Color online) Selected partial orbital DOS of Co at $h_{2}$ site in 4ZGNR. The vertical red/thin lines indicate positions of Fermi levels.}
\end{figure}
To get some insight into the higher temperatures regime and to see the influence of charge fluctuations we complement the analysis in the next section  by presentation of  some NCA results (fluctuation of boson fields) and EOM calculations.  For brevity the discussion of the latter results is restricted only  to a single  value of chemical potential.
	
In Kotliar and Ruckenstein (KR) formalism one introduces a set of boson operators for each of  electronic configuration of the impurity. For the considered two orbital impurity there are $16$ auxiliary Bose fields projecting onto the  empty ($e$), single occupied ($p_{m\sigma}$), doubly occupied ($d_{\nu}$, with $\nu={\cal{S}}{\cal{S}}_{z}$ or $\nu={\cal{T}}{\cal{T}}_{z}$ i.e. $\nu={\cal{S}}1,{\cal{S}}0,{\cal{S}}\overline{1},{\cal{T}}1,{\cal{T}}0,{\cal{T}}\overline{1}$), triple occupied ($t_{m\sigma}$) and fully (quadruple) occupied ($f$)  states  \cite{Dong}. For $e$ and $p$ operators the assignment of eigenstates is clear, for $t$ operator we use the notation $t_{m\sigma}\leftrightarrow|m\sigma, \overline{m}\uparrow \overline{m}\downarrow\rangle$ and the eigenstates corresponding to $d_{\nu}$ are listed in Tab. VII. In order to eliminate unphysical states the completeness relation for these operators ${\cal{I}}=e^{\dag}e+\sum_{m\sigma}p^{\dag}_{m\sigma}p_{m\sigma}+\sum_{\nu}d^{\dag}_{\nu}d_{\nu}+\sum_{m\sigma}t^{\dag}_{m\sigma}t_{m\sigma}+f^{\dag}f$,and the correspondence between fermions and bosons (${\cal{Q}}_{m\sigma}=p^{\dag}_{m\sigma}p_{m\sigma}+d^{\dag}_{{\cal{S}}{\cal{S}}_{z}(\sigma)}d_{{\cal{S}}{\cal{S}}_{z}(\sigma)}
+(\frac{1}{2})(d^{\dag}_{{\cal{S}}0}d_{{\cal{S}}0}+d^{\dag}_{{\cal{T}}0}d_{{\cal{T}}0})+d^{\dag}_{{\cal{T}}{\cal{T}}_{z}(m)}d_{{\cal{T}}{\cal{T}}_{z}(m)}
+t^{\dag}_{m\sigma}t_{m\sigma}
+\sum_{\sigma'}t^{\dag}_{\overline{m}\sigma'}t_{\overline{m}\sigma'}$) have to be imposed (${\cal{S}}_{z}(\pm)=1(\overline{1})$ and ${\cal{T}}_{z}[xz(yz)]=1(\overline{1})$). These constraints can be enforced by introducing Lagrange multipliers  $\lambda$,$\lambda_{m\sigma}$ and the effective SB Hamiltonian then reads:
\begin{eqnarray}
&&{\mathcal{H}^{{\cal{K}}-{\cal{R}}}}=\sum_{m\sigma}({\cal{E}}_{0}+\lambda_{m\sigma})n^{f}_{m\sigma}+\lambda({\cal{I}}-1)
+\sum_{m\sigma}\lambda_{m\sigma}({\cal{Q}}_{m\sigma}-n^{f}_{m\sigma})+\nonumber\\&&
+({\cal{U}}-{\cal{J}})\sum_{{\cal{S}}_{z}}d^{\dag}_{{\cal{S}}{\cal{S}}_{z}}d_{{\cal{S}}{\cal{S}}_{z}}
+\sum_{{\cal{T}}_{z}}[{\cal{U}}+(1-|{\cal{T}}_{z}|){\cal{J}}]d^{\dag}_{{\cal{T}}{\cal{T}}_{z}}d_{{\cal{T}}{\cal{T}}_{z}}
+\nonumber\\&&+({3\cal{U}-\cal{J}})\sum_{m\sigma}t^{\dag}_{m\sigma}t_{m\sigma}
+({6\cal{U}}-2{\cal{J}})f^{\dag}f+\sum_{kn\sigma m}({\cal{V}}^{h_{1(2)}}_{kn\sigma m}c^{\dag}_{kn\sigma}z_{m\sigma}f_{m\sigma}+h.c.)+{\mathcal{H}}_{{\cal{ZGNR}}}
\end{eqnarray}
The effective hopping in Eq. (7) is expressed by $z^{+}_{m\sigma}f^{+}_{m\sigma}$  ($z_{m\sigma}f_{m\sigma})$ with $z_{m\sigma}=[e^{+}p_{m\sigma}+p^{+}_{m\overline{\sigma}}d_{{\cal{S}}{\cal{S}}_{z}(\sigma)}
+p^{+}_{\overline{m}\overline{\sigma}}(\frac{1}{2})(d_{{\cal{S}}0}+d_{{\cal{T}}0})
+p^{+}_{\overline{m}\sigma}d_{{\cal{T}}{\cal{T}}_{z}(m)}+d^{+}_{{\cal{T}}\overline{{\cal{T}}_{z}(m)}}t_{m\sigma}
+(\frac{1}{2})(d^{+}_{{\cal{S}}0}+d^{+}_{{\cal{T}}0})t_{\overline{m}\sigma}
+d^{+}_{{\cal{S}}\overline{{\cal{S}}_{z}(\sigma)}}t_{\overline{m}\overline{\sigma}}
+t^{+}_{m\overline{\sigma}}f]/(\sqrt{{\cal{Q}}_{m\sigma}}\sqrt{1-{\cal{Q}}_{m\sigma}})$.

The stable mean field  solutions are found from the saddle point of partition function of (7), i.e. from the minimum of the free energy with respect to the slave boson parameters and Lagrange multipliers. The results for $h_{1}$ and $h_{2}$  positions are presented in Figures 14-19. According to our earlier DFT discussion  we restrict to the two orbital subspace ($d_{xz}$, $d_{yz}$) considering  the case of triple electron occupancy (single hole) and choosing a typical for Co on graphene nanostructure Coulomb interaction parameter ${\cal{U}} = 3 eV$ \cite{Wehling, Jacob, Rudenko} and  the bare orbital level energy ${\cal{E}} =  -(2{\cal{U}}+{\cal{U}})/2 = -7.5 eV$. This choice of parameters yields within Hartree-Fock approximation the required triple occupancy ($N = 3$) and reproduces the DFT magnetic moments.
\begin{figure}
\includegraphics[width=6.6 cm,bb=0 0 660 660,clip]{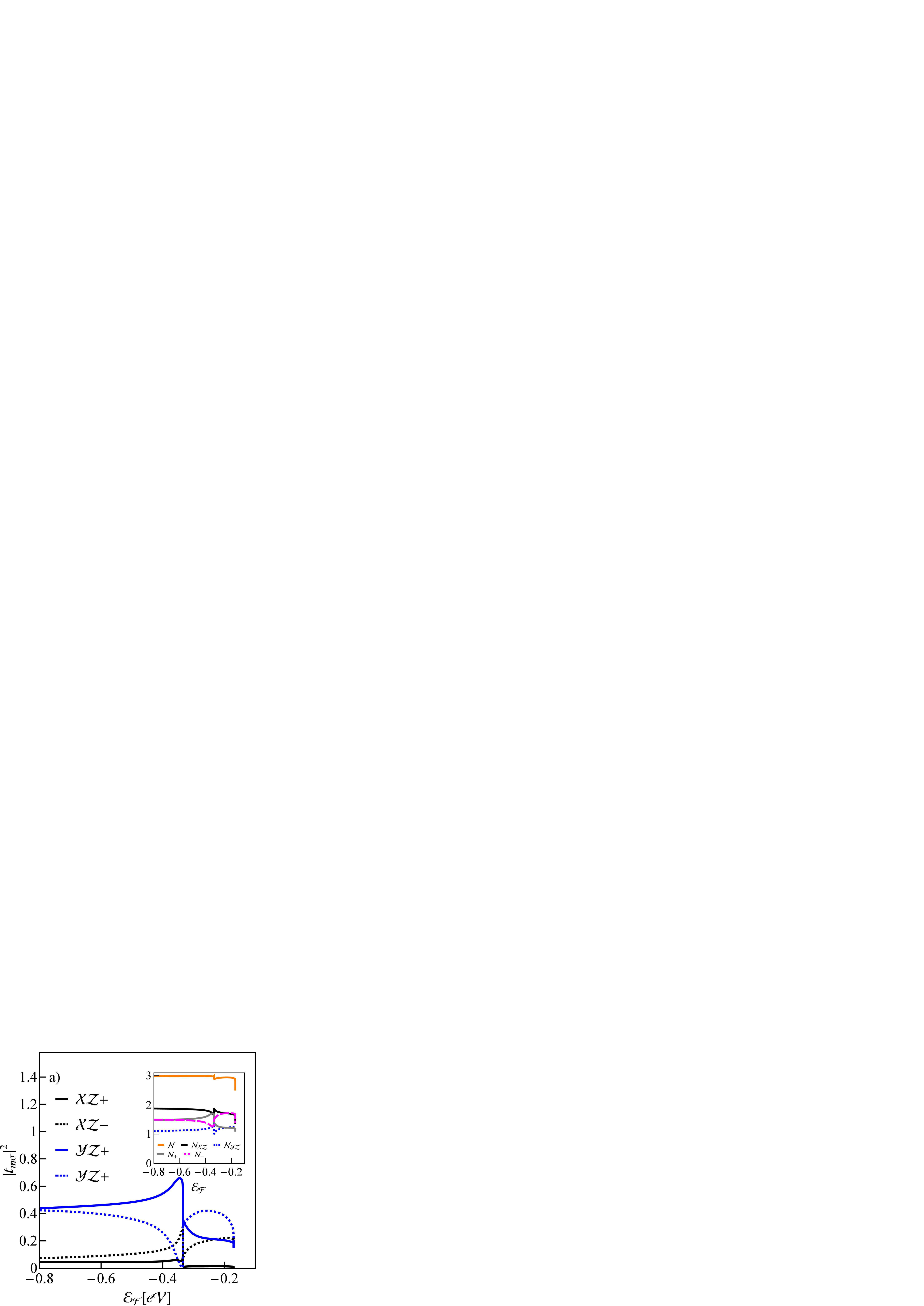}
\includegraphics[width=6.6 cm,bb=0 0 660 636,clip]{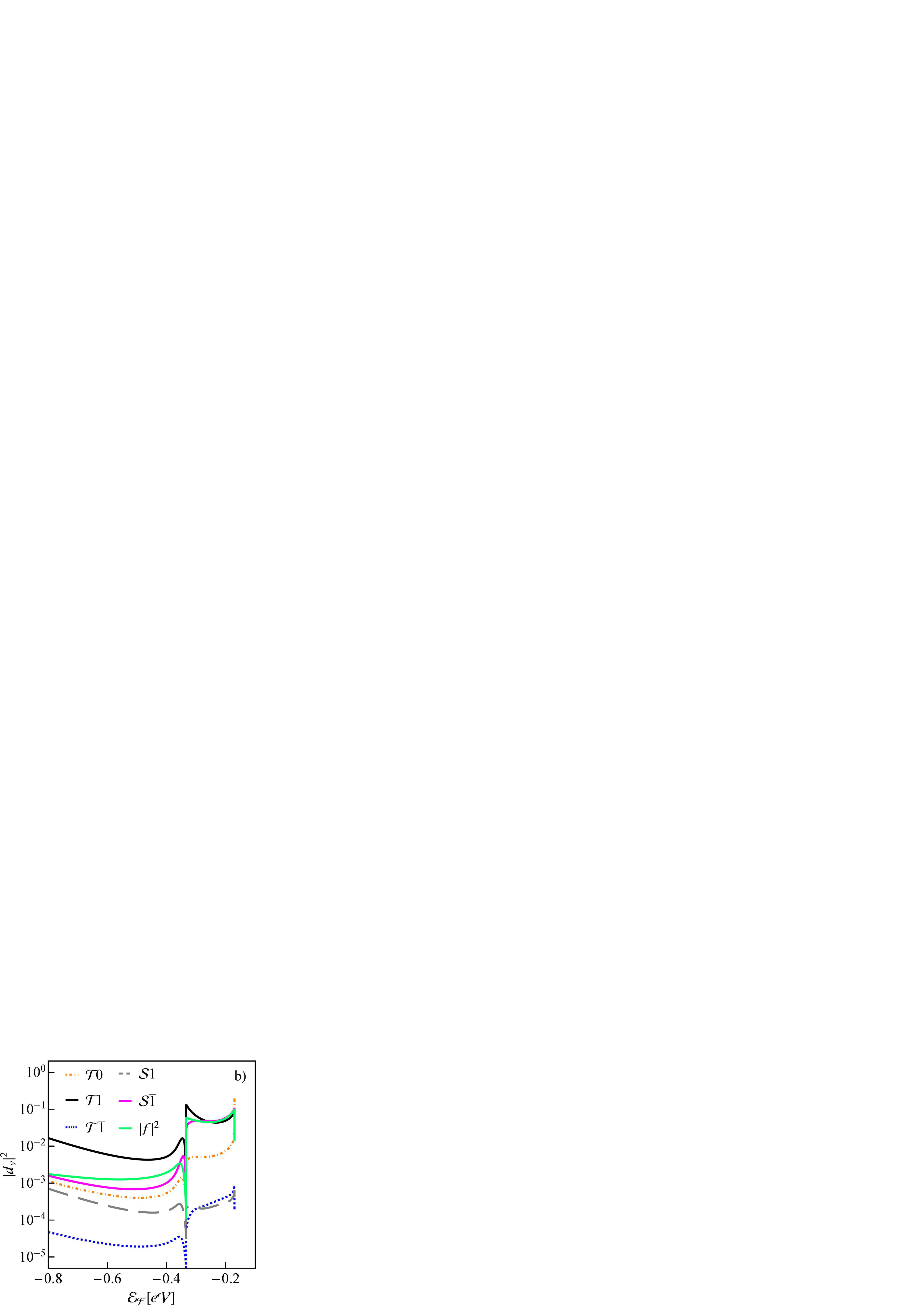}\\
\includegraphics[width=6.6 cm,bb=0 0 660 660,clip]{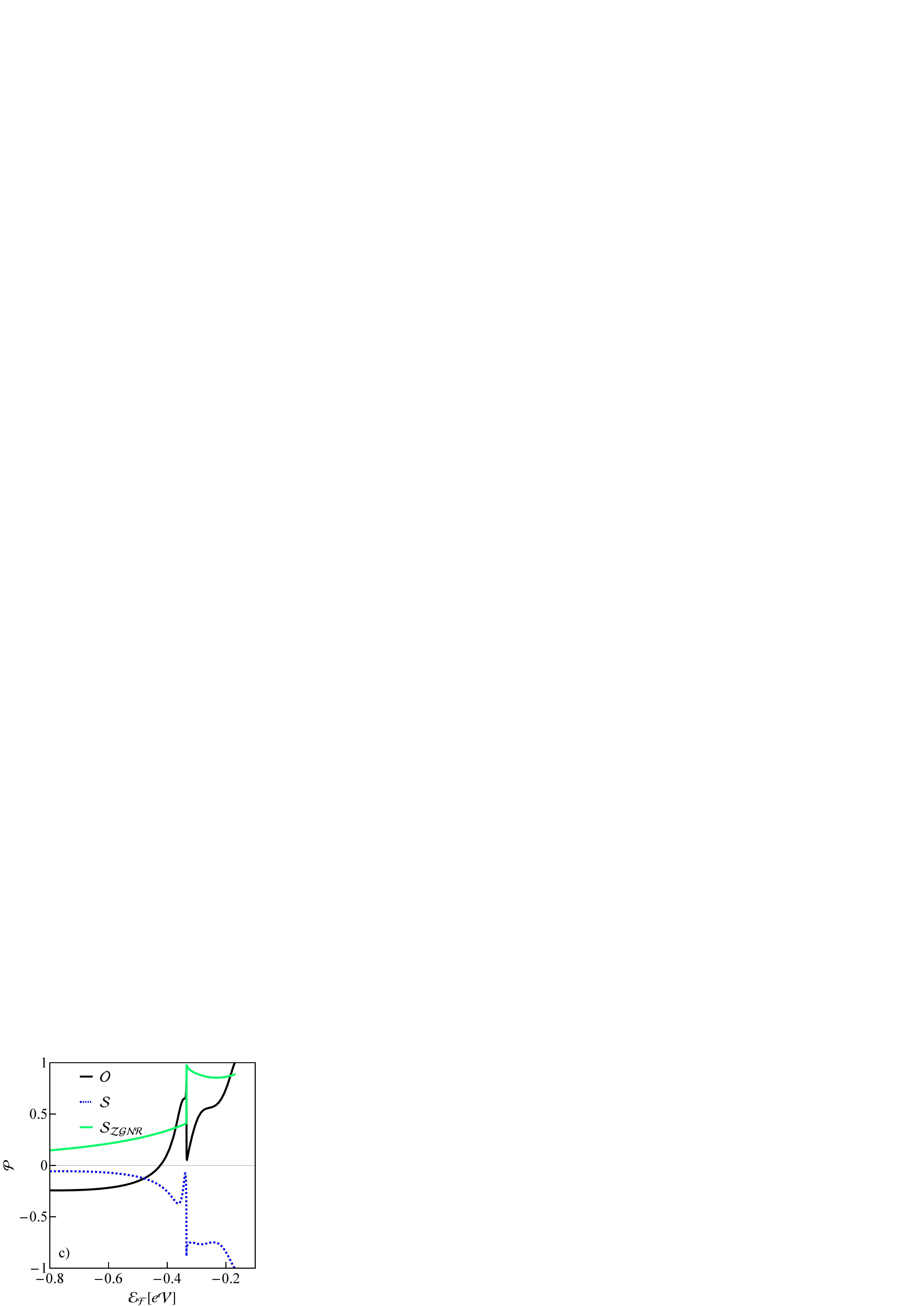}
\includegraphics[width=6.6 cm,bb=0 0 660 702,clip]{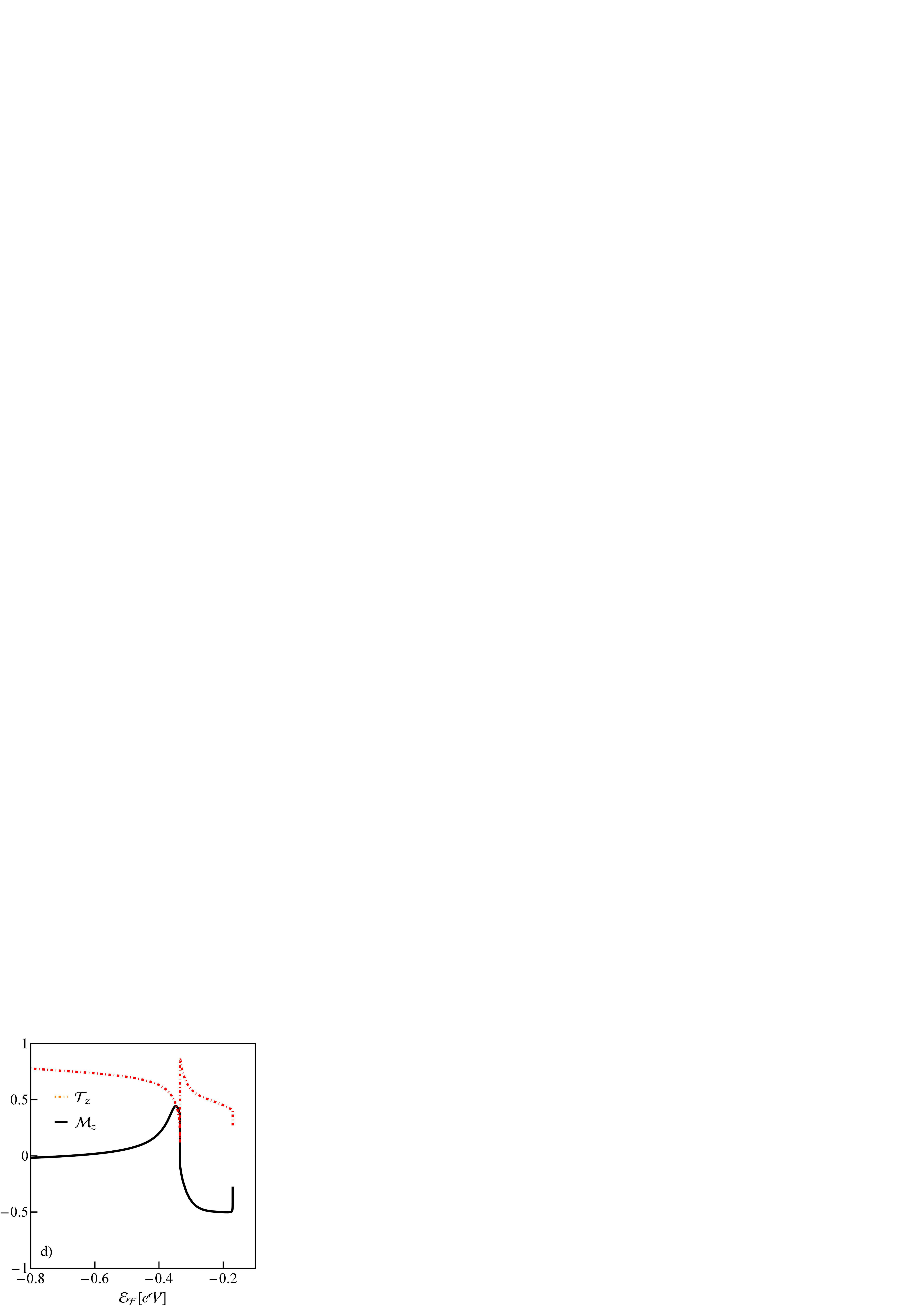}
\caption{\label{fig:epsart16}(Color online) Expectation values of slave boson operators, polarizations and expectation values of orbital pseudospin and magnetic moment of Co impurity at $h_{1}$ position of 4ZGNR a) triple occupation slave boson operators $|t_{m\sigma}|^{2}$ b) double and full occupation SB operators $|d_{{\cal{T}}0}|^{2}$, $|d_{{\cal{T}}1}|^{2}$, $|d_{{\cal{T}}\overline{1}}|^{2}$, $|d_{{\cal{S}}1}|^{2}$, $|d_{{\cal{S}}\overline{1}}|^{2}$, $|d_{{\cal{S}}0}|^{2}=|d_{{\cal{T}}0}|^{2}$ and $|f|^{2}$ c) local spin polarization of the nanoribbon around $h_{1}$ (${\cal{P}}_{{\cal{S}}_{{\cal{ZGNR}}}}=\sum_{i}\frac{\varrho_{i+}({\cal{E}}_{\cal{F}})-\varrho_{i-}({\cal{E}}_{\cal{F}})}{\varrho_{i+}({\cal{E}}_{\cal{F}})+\varrho_{i-}({\cal{E}}_{\cal{F}})}$, where $\varrho_{i\sigma}$ denotes local ZGNR density of states at the n. n. carbon sites around hollow position), orbital (${\cal{P}}_{{\cal{O}}}$) and spin polarization of Co adatom (${\cal{P}}_{{\cal{S}}}$) d) magnetic moment and orbital pseudospin of Co impurity. Inset of Fig. a presents spin and orbital resolved contributions to the occupancies of Co.}
\end{figure}

Let us first  discuss  the $h_{2}$ case, where local nanoribbon environment is unpolarized. Figures 14 present expectation values of slave boson operators, orbital and spin occupations, orbital and spin moments, orbital and spin polarizations, all quantities  plotted as a function of chemical potential. To interpret the results  it is worth to refer to the energy dependence of the corresponding hybridization functions (Fig. 13).
Outside the singularities  ($-0.8<{\cal{E}}<-0.4$) , where hybridization function of $yz$ symmetry dominates over $xz$ hybridization, Kondo physics is governed mainly by spin fluctuations in $yz$ sector (${\cal{N}}_{yz}\approx1$) , orbital  $xz$ is almost completely filled  (${\cal{N}}_{xz}\approx2$). We have checked that there are no SBMFA  solutions for $xz$ channel when the interorbital fluctuation path is closed (i.e. when the two last terms in (3) are neglected). When interorbital path opens the coupled spin-orbital fluctuations create resonances in both orbital sectors. Very crudely one can visualize these processes  as virtual complete  filling or emptying of $yz$  orbital by hoppings  resulting in  fast SU(2) type spin fluctuations  in  $yz$ channel (broad peak). These fluctuations are however not completely decoupled from $xz$ channel.  Orbital $xz$ is much  weaker coupled to nanoribbon  and hoppings are less frequent. Virtual creation of a hole on $xz$ orbital increases the probability of double occupancy of $yz$ orbital. Temporary the reverse of roles of orbitals is possible. Such orbital fluctuations  enable weak effective spin fluctuations in $xz$ sector despite its high occupancy.  The average time of such fluctuations is however relatively  long, what reflects in an observed narrow $xz$ quasiparticle resonance .
The representative  density of states of $h_{2}$ impurity in this range (${\cal{E}}_{{\cal{F}}} = -0.8$) is shown in Figure 15.
\begin{figure}
\includegraphics[width=7.2 cm,bb=0 0 660 620,clip]{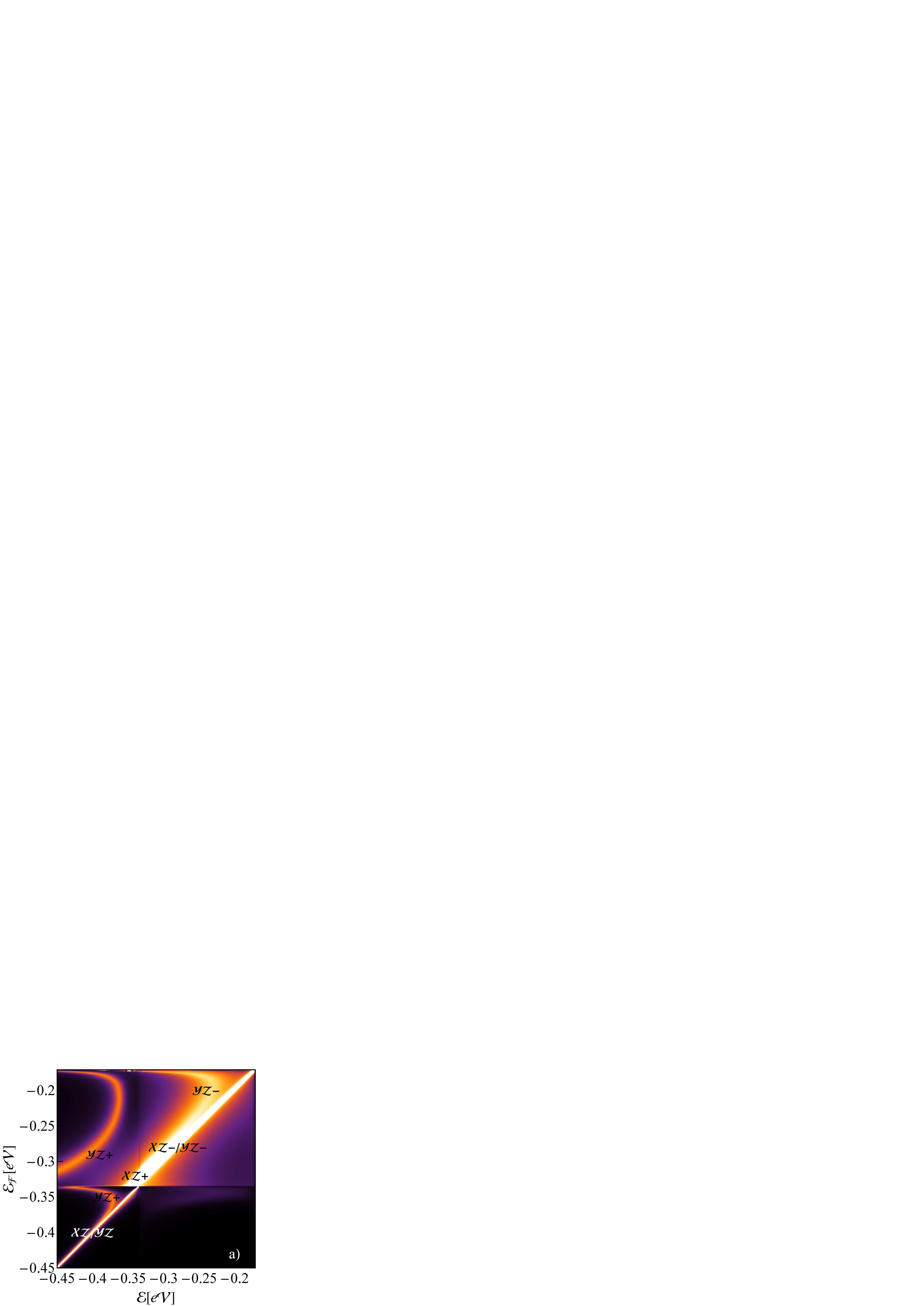}
\includegraphics[width=6.8 cm,bb=0 0 660 654,clip]{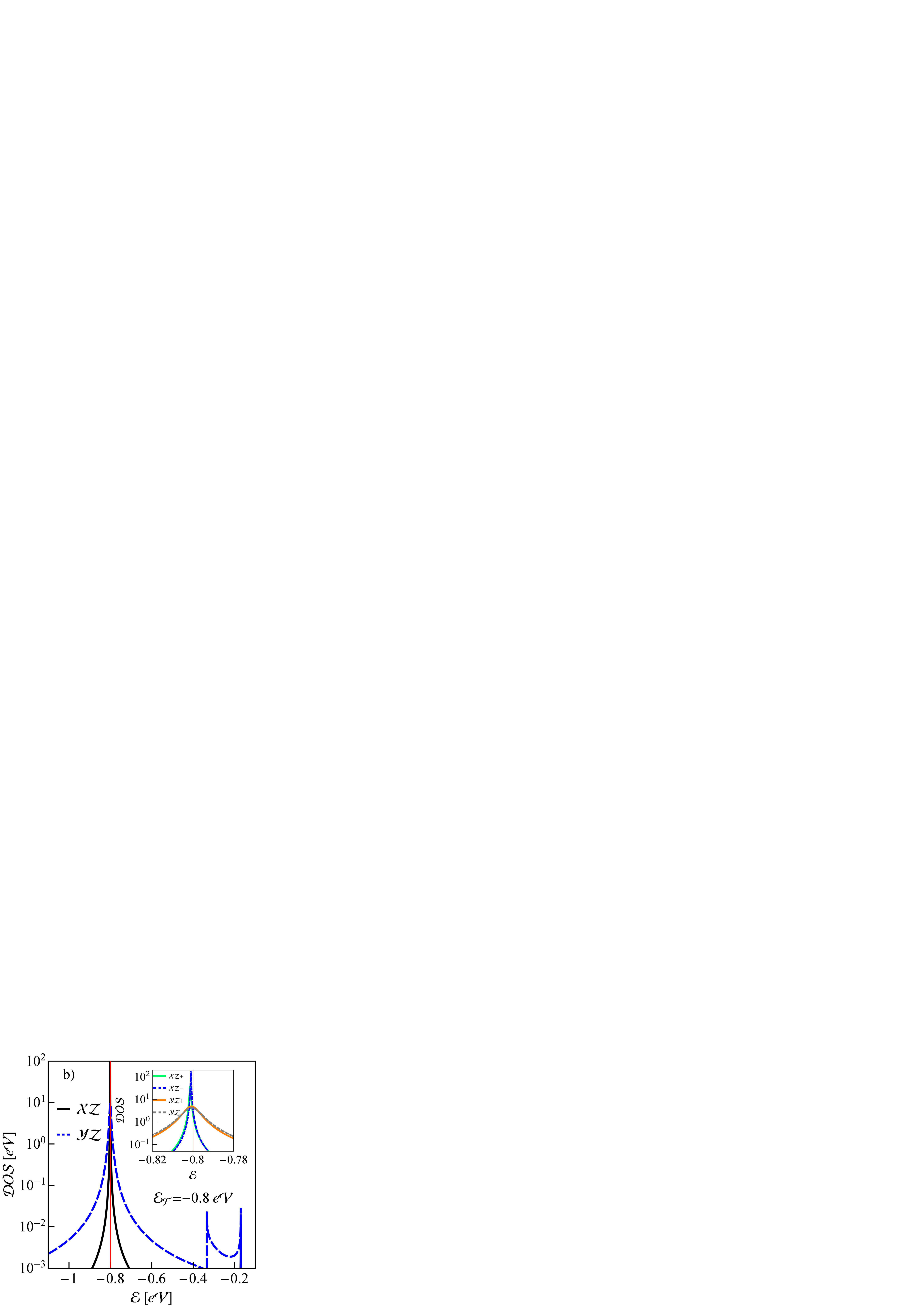}\\
\includegraphics[width=7 cm,bb=0 0 660 654,clip]{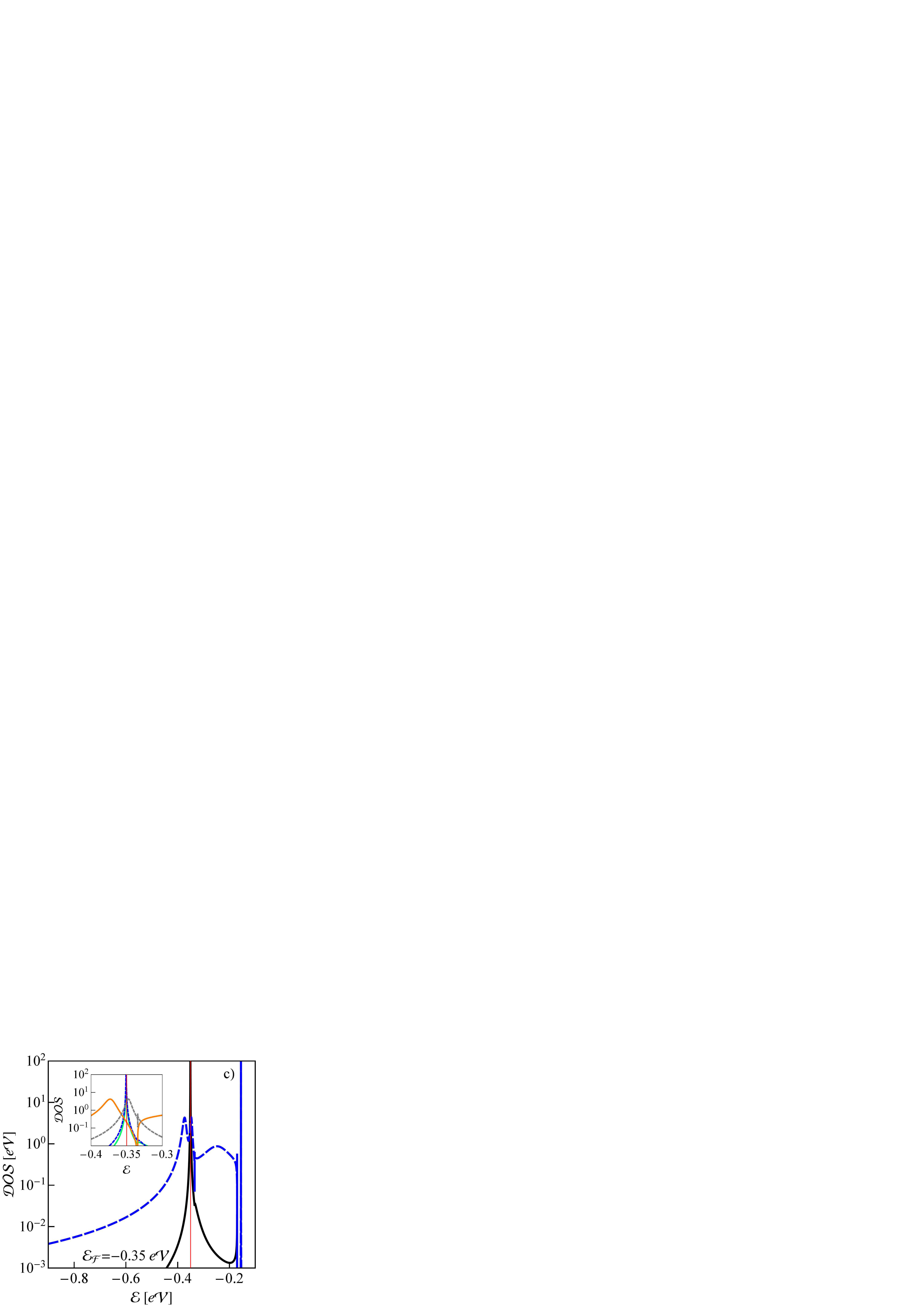}
\includegraphics[width=6.8 cm,bb=0 0 660 672,clip]{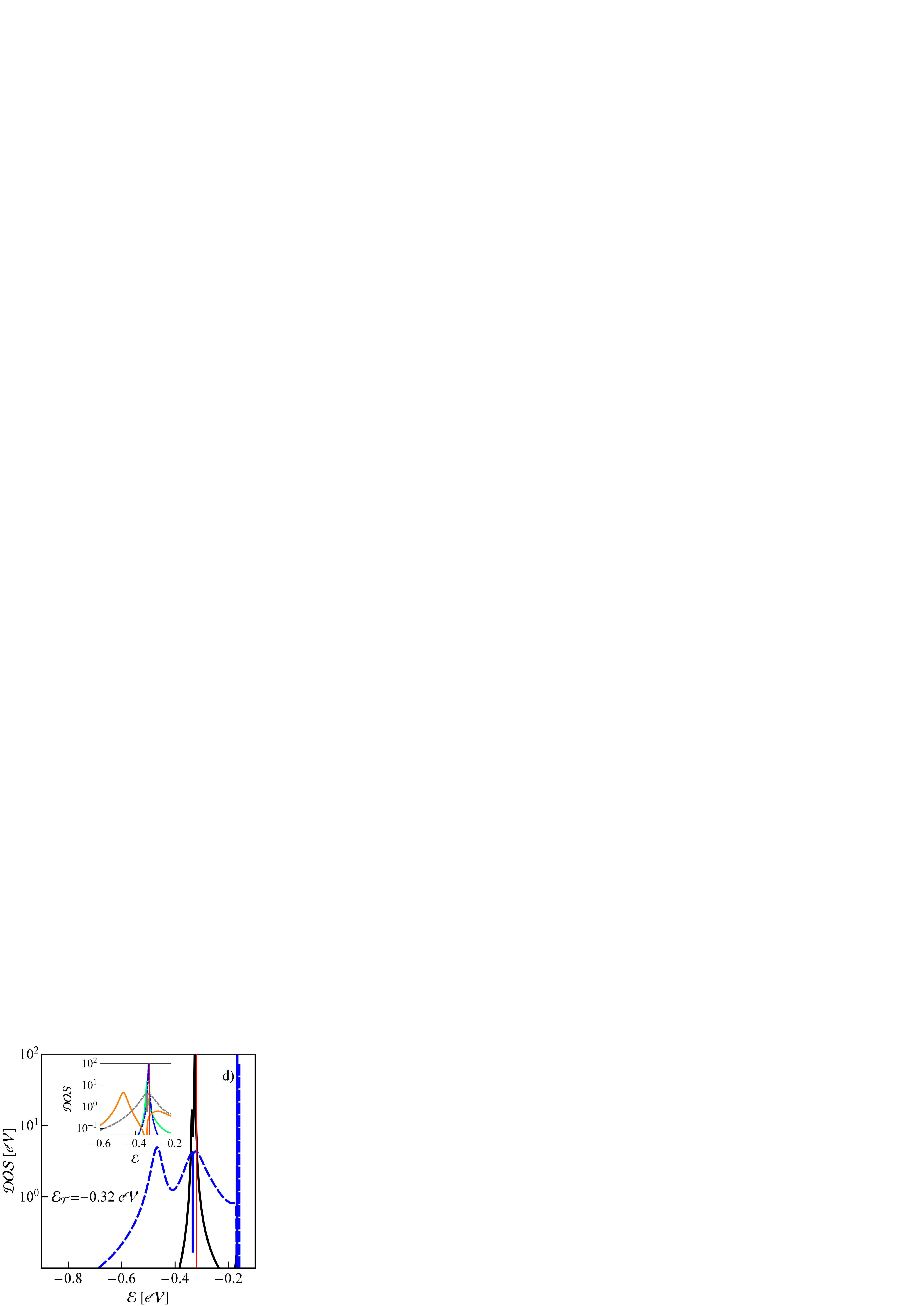}
\caption{\label{fig:epsart17}(Color online) a) Density of states map of Co impurity in $h_{1}$ position of 4ZGNR b, c, d) partial orbital densities of states for the selected values of the Fermi levels. Inset are the zoom-views of the spin and orbital resolved DOS.}
\end{figure}
When ${\cal{E}}_{{\cal{F}}}$ moves closer to singularity  ${\cal{VH}}2\nu$, $yz$  hybridization does not change considerably, but in $xz$ sector  Van Hove singularity  manifests strongly. Close to ${\cal{VH}}2\nu$    the expectation values of  slave boson operators $t$ and orbital occupations approach each other in consequence of strong enhancement of $xz$ hybridization, but  symmetric SU(4) case is not realized for any energy because this would require the equality of both real and imaginary parts of hybridizations functions. As it is seen in Fig. 15b   the resonances in this region for both orbitals  (${\cal{E}} = -0.35$ eV) are distinctively different. Interestingly, moving still closer to singularity around (${\cal{E}} =-0.3345$), in an extremely narrow energy range, orbital $xz$ even takes over the dominant role in many-body processes, what reflects in a change of sign of orbital pseudospin. In the region of singularity strong deviations of orbital occupancies from integer values are observed what indicates, that system is driven out from Kondo state into mixed valence state.
\begin{figure}
\includegraphics[width=7.2 cm,bb=0 0 660 632,clip]{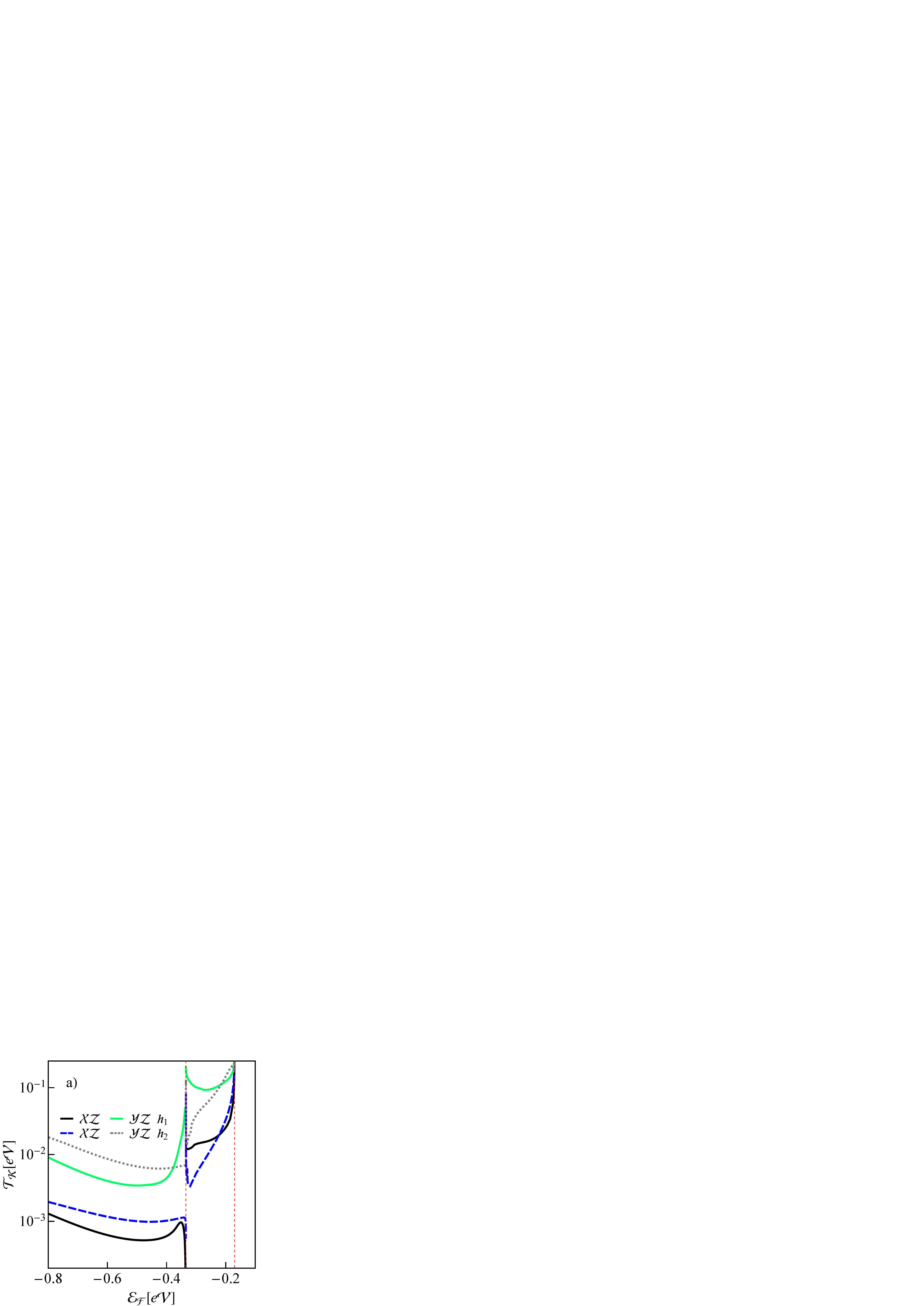}
\includegraphics[width=7.2 cm,bb=0 0 660 632,clip]{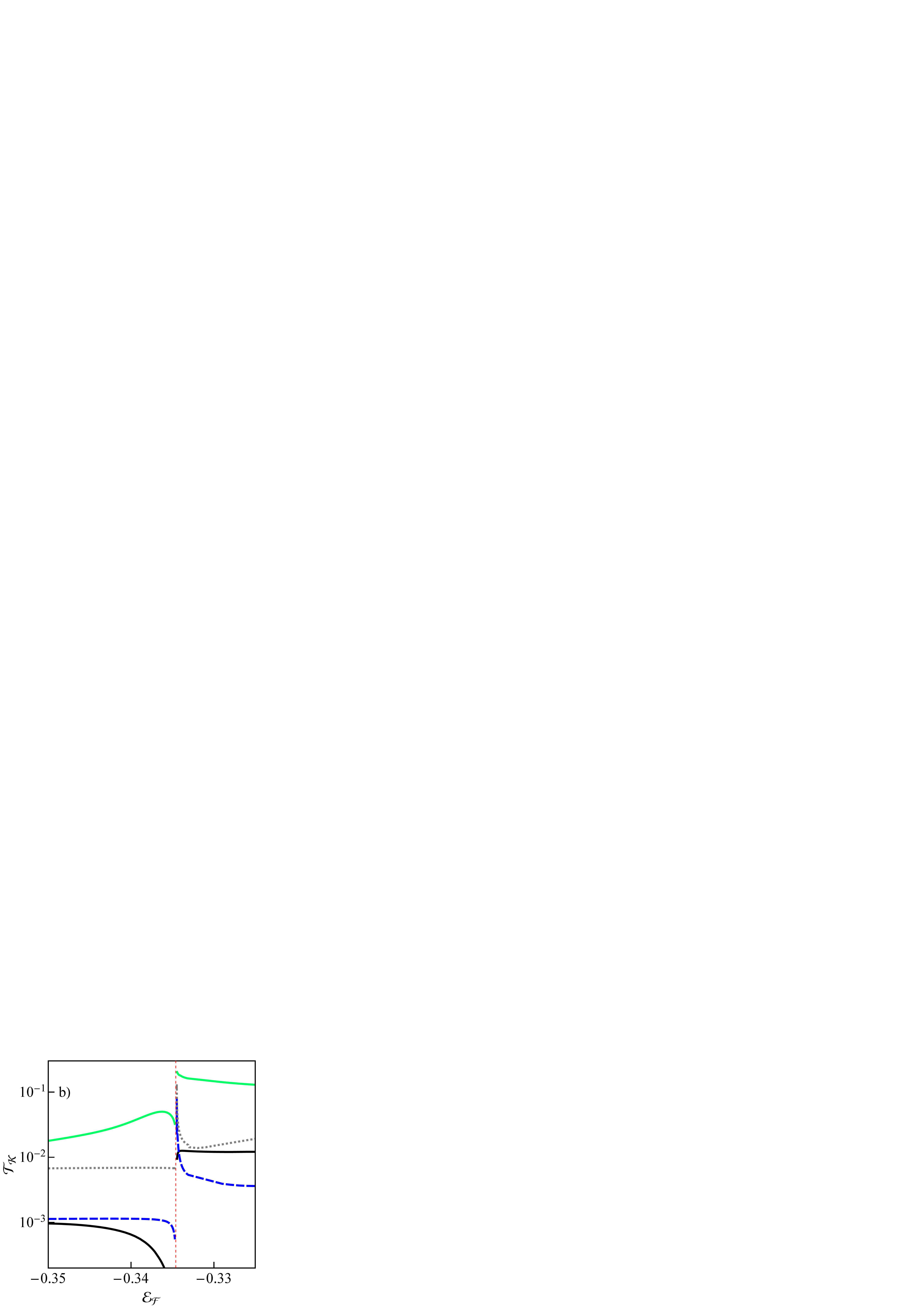}
\caption{\label{fig:epsart18}(Color online) Characteristic quasiparticles temperatures $T_{{\cal{K}}}$ for $xz$ and $yz$ channels vs. position of the Fermi level. Fig. b presents details of dependencies close to ${\cal{VH}}2\nu$ singularity.}
\end{figure}
For chemical potential above ${\cal{VH}}2\nu$ again the  dominance of $yz$ orbital is restored and system moves into Kondo state again. One should remember however, that the presented picture in vicinity of singularity should be treated with caution, only as a crude visualization of tendencies.  Around singularity the system is pushed into non-Fermi liquid regime due to the observed divergences of self energies, and in principle for a discussion of this range  summation of higher order corrections to  MFA  is  indispensable \cite{Irkhin, Irkhin2}.  When ${\cal{E}}_{{\cal{F}}}$ moves closer to the edge and both real and imaginary parts of hybridization are strongly enhanced for both symmetries  broadening of many-body resonances results and   delta like structures are observed at the band edges, which  extend into the gap  for  ${\cal{E}}_{{\cal{F}}}$ moving very close to the edge (Fig. 15c, d). They reflect the new poles of impurity Green's function and these structures are essential in order to satisfy the sum rules. Of interest are also the dips occurring for energies where singularities occur. They emerge due to an interplay of  correlations effect and singular  substrate electron density of states.  When Fermi level crosses the singularity the dip sits at the the Fermi level, but  singularities  also reflect in spectral function when chemical potential is not in  close proximity to VHS (Fig. 15).

Let us now turn to $h_{1}$ case.  The  spin polarization of  nanoribbon  breaks the spin degeneracy. The  number of independent slave boson operators increases and the many body resonances become  spin dependent. Again of special interest  are the regions around  singularities of  DOS.  In addition to the earlier described effects,   also  new phenomena associated with polarization are observed. The  sharp change of local nanoribbon spin polarization in the vicinity of ${\cal{VH}}2\nu$  reveals  in a drastic, but opposite change of impurity polarization  and suppression of  screening processes  of Co magnetic moment. Singularity   most strongly reflects in the abrupt increase  of spin distinction in $yz$ orbital channel what is a consequence of clearly exhibited singularity in the  corresponding hybridization function for one spin direction and only very weak trace of it  for the opposite spin.  Remarkable is the  resulting   jump and  change of sign of magnetic moment and fall and next jump of orbital pseudospin when ${\cal{E}}_{{\cal{F}}}$ crosses singularity.  All the anomalies are the consequence of dramatically enhanced imaginary part of hybridization and a jump from negative to positive values of the real part of hybridization.
The dramatic changes of spin or orbital characteristics when Fermi level crosses the singularities is of potential interest for spintronics (orbitronics) since these changes can be induced by gate voltage.
\begin{figure}
\includegraphics[width=6.8 cm,bb=0 0 660 660,clip]{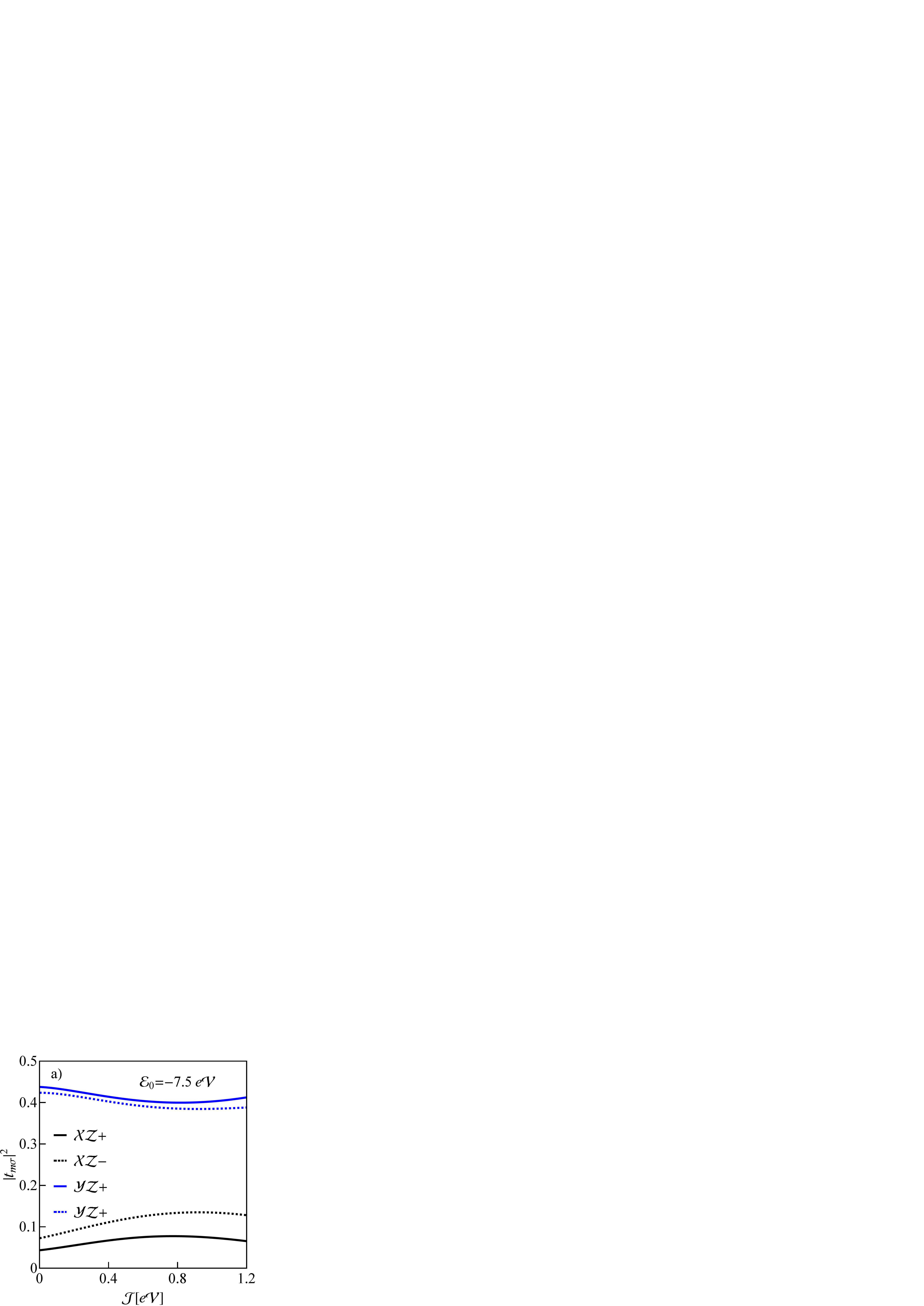}
\includegraphics[width=6.8 cm,bb=0 0 660 660,clip]{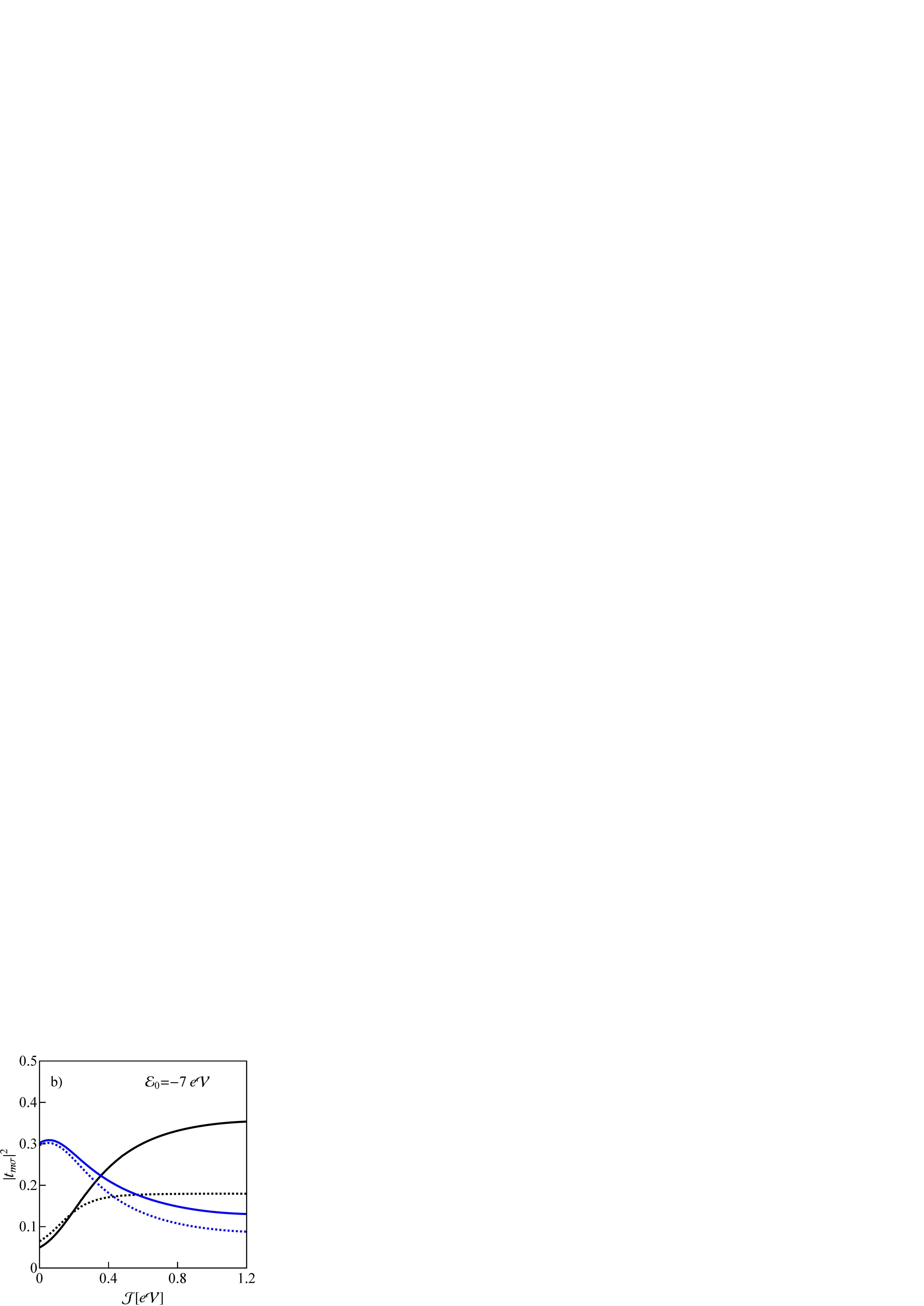}\\
\includegraphics[width=6.6 cm,bb=0 0 660 616,clip]{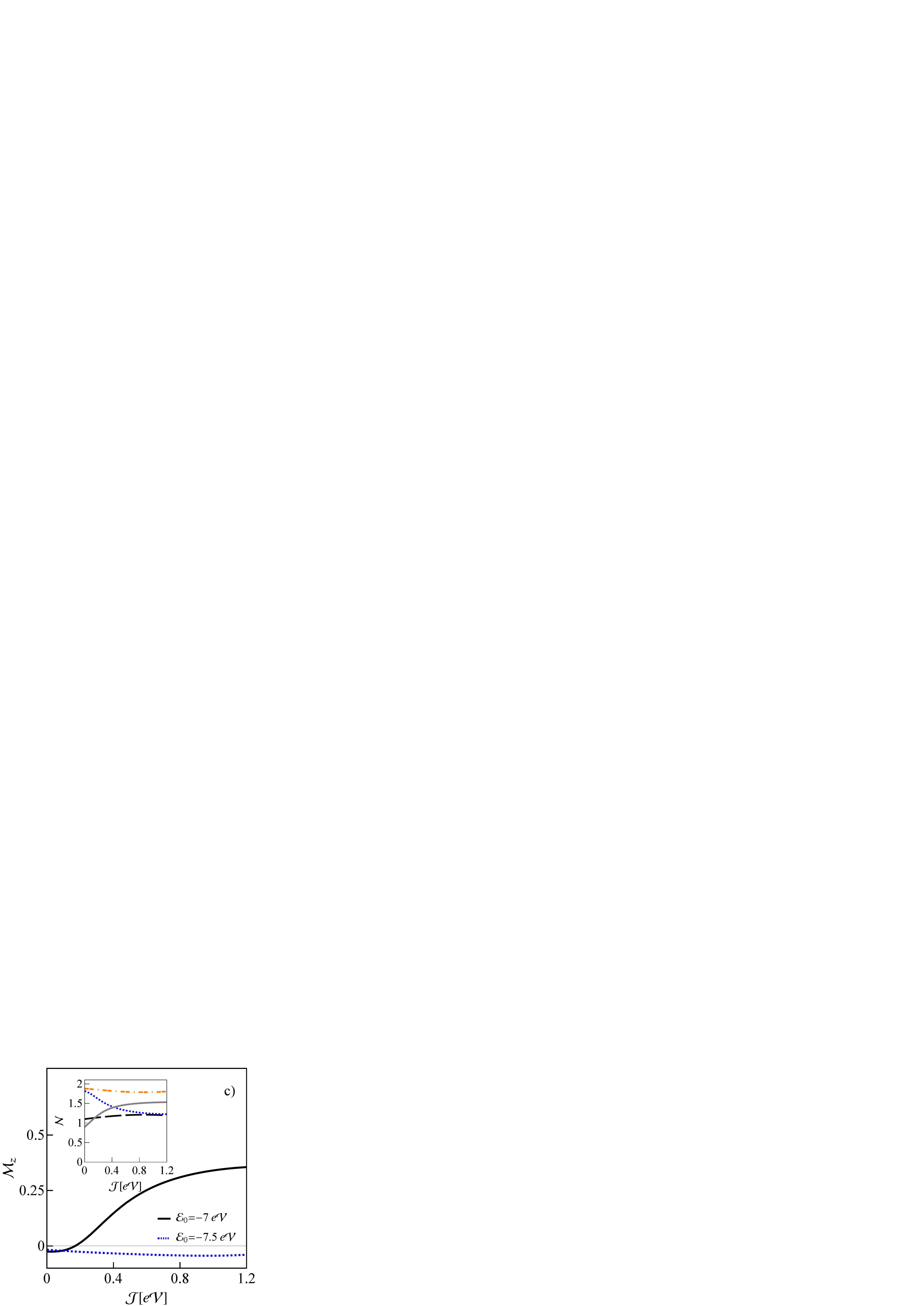}
\includegraphics[width=7 cm,bb=0 0 660 636,clip]{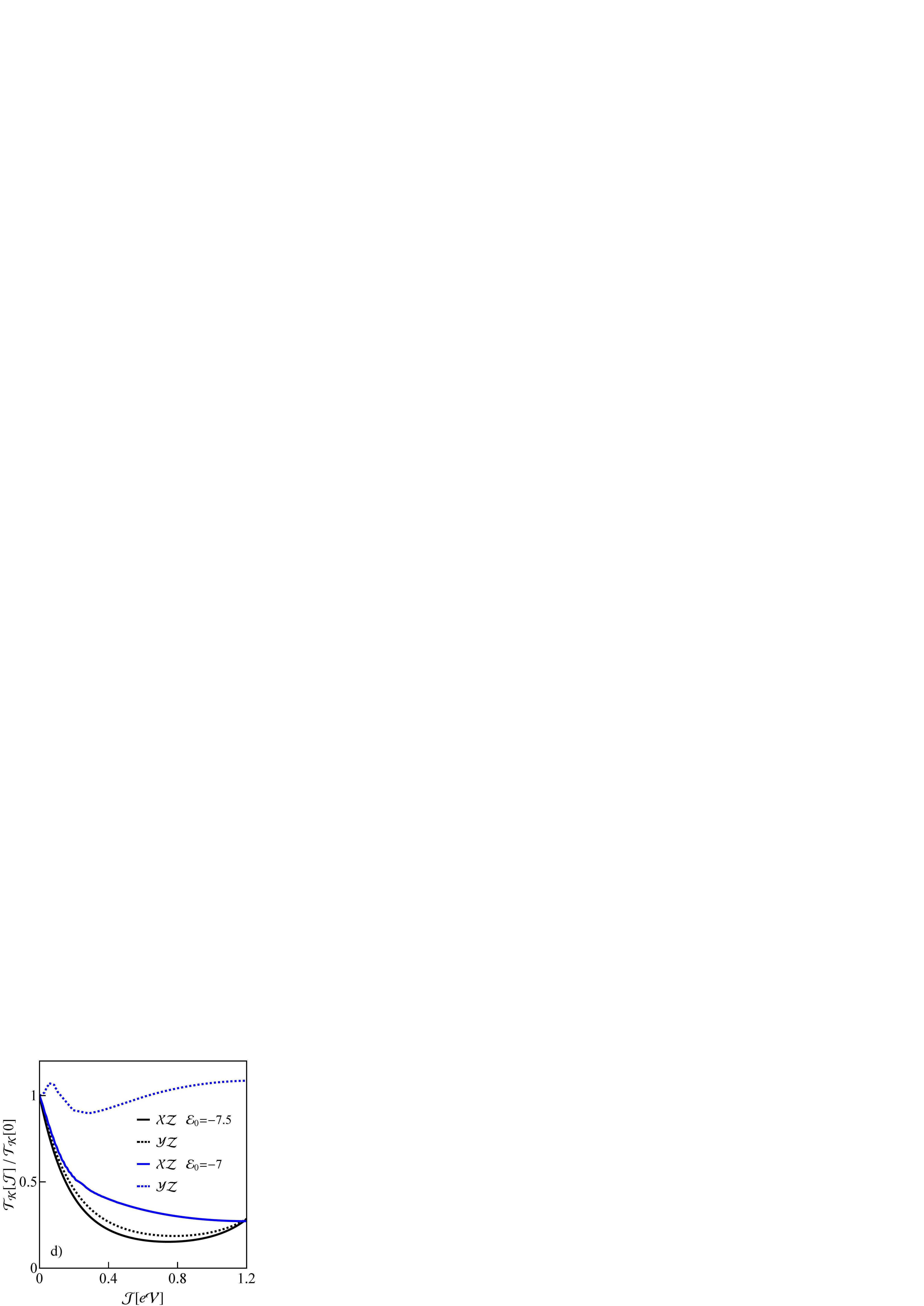}
\caption{\label{fig:epsart19}(Color online) Illustration of characteristics of the Kondo state on the value of exchange parameter a, b) slave boson expectation values $|t_{m\sigma}|^{2}$ for ${\cal{N}}\approx3$ ($\varepsilon_{0}=-7.5$ eV) and ${\cal{N}}\approx2.7$ ($\varepsilon_{0}=-7$ eV) occupancies (${\cal{E}}_{{\cal{F}}}=-0.8$). c) Magnetic moments of Co impurity. Inset shows orbital occupancies: for $\varepsilon_{0}=-7.5$ dashed line (${\cal{N}}_{yz}$) and dashed-dotted line (${\cal{N}}_{xz}$), $\varepsilon_{0}=-7$ solid line (${\cal{N}}_{yz}$) and dotted line (${\cal{N}}_{xz}$). d) Relative Kondo temperature.}
\end{figure}
Fig. 17a illustrates the evolution of density of states with the shift of the Fermi level. We have  also marked  the lines of maxima of DOS. The clearly seen horizontal (${\cal{E}}_{{\cal{F}}}=-0.3348 eV$) and vertical (${\cal{E}}=-0.3348 eV$)  straight lines of reduced intensity reflect the position of ${\cal{VH}}2\nu$ singularity.   The representative spin and orbital resolved densities of states, which correspond to the selected horizontal cross sections (${\cal{E}}_{{\cal{F}}} = const$) of the map (Fig. 17a) are displayed in Figure 17b, c, d.  Since in the considered energy range  $yz$ hybridization  is stronger than for $xz$ symmetry the corresponding many body $yz$ resonances are in general broader. Although spin distinction in $xz$ hybridization is remarkable (Fig. 12) it does not reflect in clear distinction of corresponding many-body peaks (Fig. 17). This observation is in accordance with our earlier interpretation of $xz$ resonance as a repercussion of spin fluctuations in $yz$ shell transferred to $xz$ by orbital fluctuations. In contrast to the $h_{2}$  case, for $h_{1}$ position ${\cal{VH}}2\nu$ singularity plays  an important role  for $yz$ symmetry, especially for majority spins.   The peak splitting of $yz$ resonance  is a combined effect of spin dependence and singularity induced dips in the corresponding densities of states. Fig. 18 presents Kondo temperatures for $h_{1}$ and $h_{2}$ positions. We define $T_{{\cal{K}}}$ through widths and position of quasiparticle resonance \cite{Hewson}. $T_{{\cal{K}}m}=(1/2)\sum_{\sigma}\sqrt{\widetilde{\Delta}^{2}_{m\sigma}+\widetilde{{\cal{E}}}^{2}_{m\sigma}}$, where  $\widetilde{{\cal{E}}}_{m\sigma}$ is the distance between Fermi energy and quasiparticle resonance and $\widetilde{\Delta}_{m\sigma}$ is the width at half maximum. Since the characteristic quasiparticles energies are distinctively different for both orbital channels we show the corresponding characteristic temperatures separately.
The estimated characteristic temperatures are of order of $20$ K  and $200$ K for $xz$ and $yz$ channels respectively, and they are   strongly enhanced or suppressed in the region of singularity depending on which side the chemical potential approaches the singularity. This tendency reflects the opposite  shift of effective orbital energies on both sides caused by real parts of self energy which are discontinuous and change sign in the singularity point.
Fig. 19 illustrates an impact of  Hund's coupling on  Kondo physics. We show two examples ${\cal{N}}\approx3$ and  ${\cal{N}}\approx2.7$. Insight on  the slave boson dependencies and orbital occupancies highlights the stronger impact of Hund's coupling for ${\cal{N}}\approx2.7$.  In this case  a remarkable weakening of Kondo screening is observed for high values of exchange coupling (increase of magnetic moment). In general one can expect that an increase of magnetic correlations with the increase of exchange coupling should result, as a consequence of competitiveness of different correlations, in a decrease of Kondo temperature. The spin and orbital degrees of freedom fluctuate less freely in this case. This general tendency is really observed in most presented cases. For reduced occupancy however, ${\cal{N}}\approx2.7$ characteristic temperature $T^{yz}_{{\cal{K}}}$  changes nonmonotonically,  what reflects change of partial occupancy form slightly below half filling to values above. For ${\cal{N}}_{yz}=1$ a maximum of Kondo temperature is observed.

\subsection{Charge fluctuation effects - NCA and EOM approaches}
The SBMFA results get  less good with increasing temperature due to fluctuations. Some account of fluctuations is achieved by systematic corrections to MFA approach using e.g. hybridization expansion or applying the equation of motion method.
In this section we briefly  analyze the role of charge fluctuations in the considered many-body processes. For  transparency of considerations we discuss only the case when the  chemical potential is located not to close to any VH singularity.  We limit to the lowest-order in hybridization  self-consistent  approximation NCA  and EOM with Lacroix's decoupling approximation \cite{Lacroix}.
These methods  apply for  higher temperatures, but  they  give reliable results also down to a fraction of $T_{\cal{K}}$  \cite{Bickers, Hewson, Kashcheheyevs}. They   fail however for $T\ll T_{\cal{K}}$, but in this range  in turn    SBMFA is valid.
\begin{figure}
\includegraphics[width=6.6 cm,bb=0 0 660 694,clip]{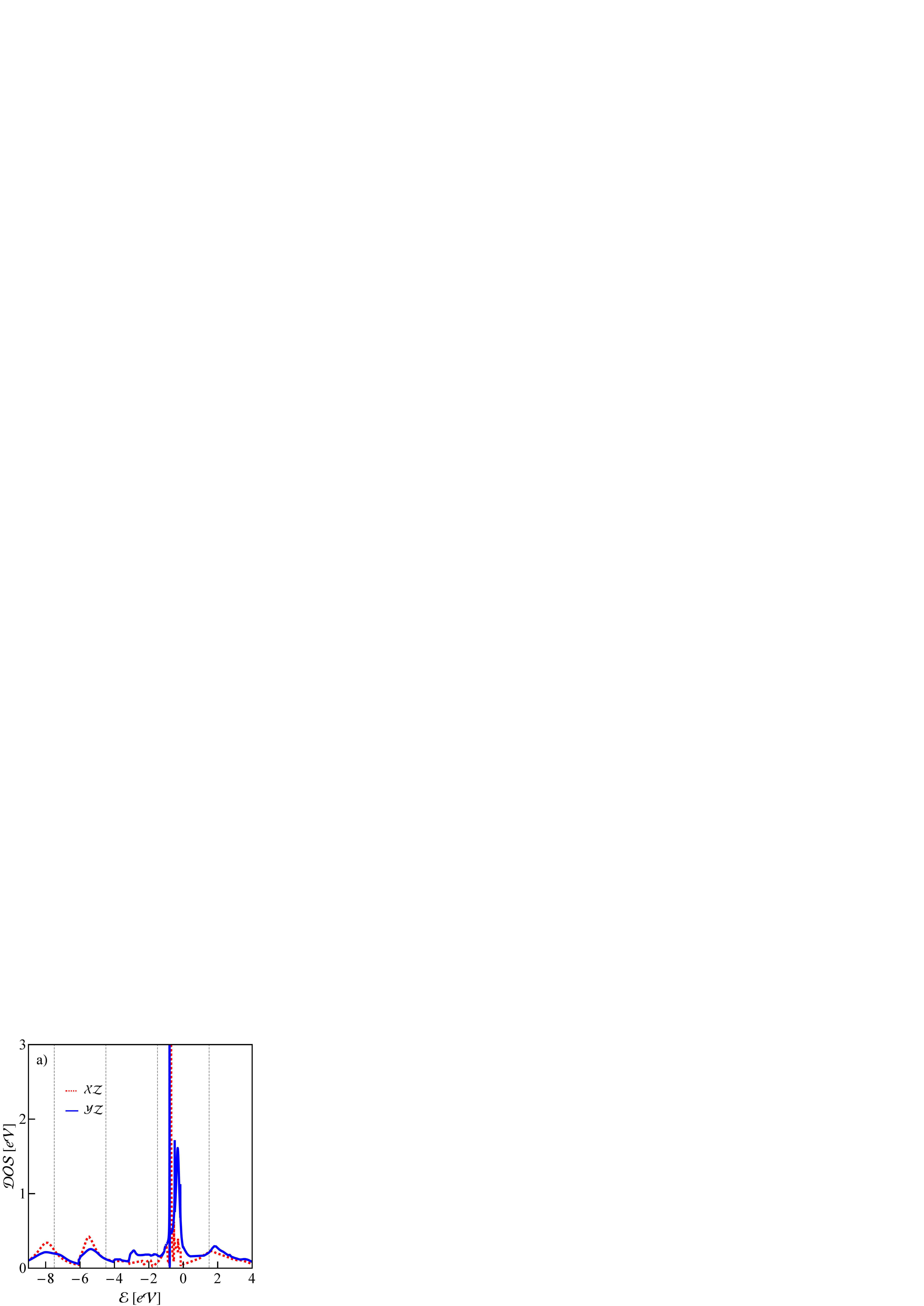}
\includegraphics[width=6.5 cm,bb=0 0 660 702,clip]{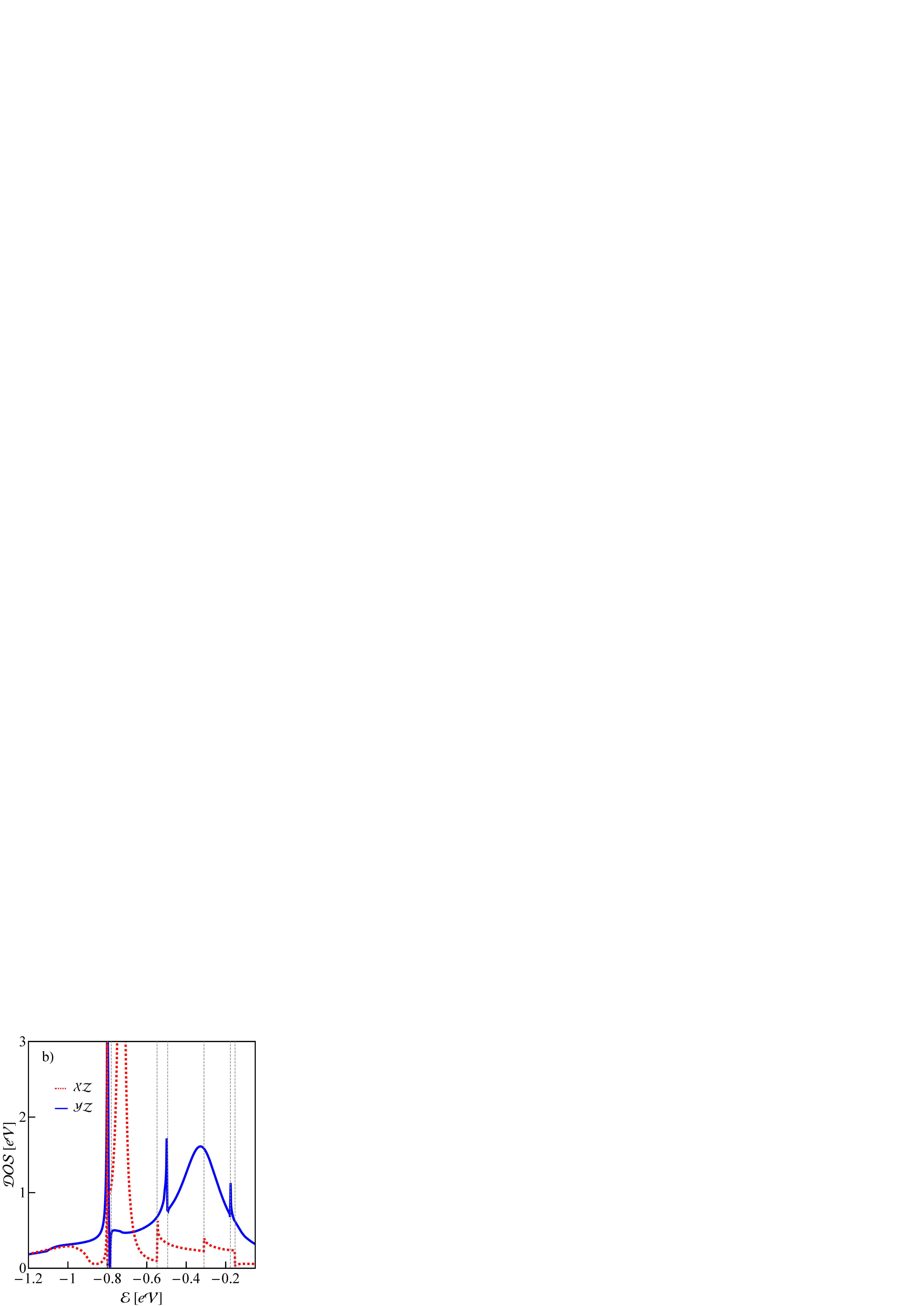}
\caption{\label{fig:epsart20}(Color online) Comparison of $xz$ (dotted line) and $yz$ (solid line) contributions to the DOS of Co adatom in $h_{1}$ position of 4ZGNR (${\cal{E}}_{{\cal{F}}}=-0.8$) (NCA). The vertical broken lines on Fig. a determine the positions of bare Coulomb peaks and on Fig. b the positions of singularities in the interacting self energies.}
\end{figure}
Despite the low temperature deficiencies the use of these  impurity solvers allow us to get a crude insight into the full spectrum of the one particle Green's functions and not just the quasi-particle contribution. In the present work we apply the NCA method for finite ${\cal{U}}$   \cite{Pruschke, Grewe, Haule, Gerace}. In NCA one takes into account only  diagrams  without noncrossing of substrate electron lines, what corresponds to simple hopping processes where electron or hole hops into the adatom at some time and then out at a later time. This  leads to a set of NCA integral equations for the fixed occupation self energies:
\begin{eqnarray}
&&\Sigma^{(0)}(z)=\sum_{m\sigma}\int\frac{d\varepsilon}{\pi}
\Delta_{m\sigma}(\varepsilon)f(\varepsilon){\cal{G}}^{(1)}_{m\sigma}(z+\varepsilon)\nonumber\\
&&\Sigma^{(1)}_{m\sigma}(z)=\int\frac{d\varepsilon}{\pi}
[\Delta_{m\sigma}(\varepsilon)f(-\varepsilon){\cal{G}}^{(0)}(z-\varepsilon)+
\Delta_{m\overline{\sigma}}(\varepsilon)f(\varepsilon){\cal{G}}^{(2)}_{m\sigma m\overline{\sigma}}(z+\varepsilon)\nonumber\\
&&+\sum_{\sigma'}\Delta_{\overline{m}\sigma'}(\varepsilon)f(\varepsilon){\cal{G}}^{(2)}_{m\sigma \overline{m}\sigma'}(z+\varepsilon)]\nonumber\\
&&\Sigma^{(2)}_{m\sigma m'\sigma'}(z)=\int\frac{d\varepsilon}{\pi}
[\Delta_{m\sigma}(\varepsilon)f(-\varepsilon){\cal{G}}^{(1)}_{m'\sigma'}(z-\varepsilon)+
\Delta_{m'\sigma'}(\varepsilon)f(-\varepsilon){\cal{G}}^{(1)}_{m\sigma}(z-\varepsilon)\nonumber\\
&&+
\left\{
  \begin{array}{ll}
    \Delta_{m'\overline{\sigma'}}(\varepsilon)f(\varepsilon){\cal{G}}^{(3)}_{m\sigma}(z+\varepsilon)+
\Delta_{m\overline{\sigma}}(\varepsilon)f(\varepsilon){\cal{G}}^{(3)}_{m'\sigma'}(z+\varepsilon), & m\neq m' \\
    \sum_{\sigma}\Delta_{\overline{m}\sigma}(\varepsilon)f(\varepsilon){\cal{G}}^{(3)}_{\overline{m}\sigma}(z+\varepsilon), & m = m'
  \end{array}
\right.]\nonumber\\
&&\Sigma^{(3)}_{m\sigma}(z)=\int\frac{d\varepsilon}{\pi}
[\Delta_{m\sigma}(\varepsilon)f(-\varepsilon){\cal{G}}^{(2)}_{\overline{m}\sigma\overline{m}\overline{\sigma}}(z-\varepsilon)+
\sum_{\sigma'}\Delta_{\overline{m}\overline{\sigma'}}(\varepsilon)f(-\varepsilon){\cal{G}}^{(2)}_{m\sigma\overline{m}\sigma'}(z-\varepsilon)\nonumber\\
&&+\Delta_{m\overline{\sigma}}(\varepsilon)f(\varepsilon){\cal{G}}^{(4)}(z+\varepsilon)]\nonumber\\
&&\Sigma^{(4)}=\sum_{m\sigma}\int\frac{d\varepsilon}{\pi}
\Delta_{m\overline{\sigma}}(\varepsilon)f(-\varepsilon){\cal{G}}^{(3)}_{m\sigma}(z-\varepsilon),
\end{eqnarray}
where ${\cal{G}}^{(p)}_{m\sigma}(z)=[z-{\cal{E}}_{(p)}-\Sigma^{(p)}_{m\sigma}(z)]^{-1}$ ($p = 1, 3$ with energies ${\cal{E}}_{(1)}=\varepsilon_{0}$ and ${\cal{E}}_{(3)}=3\varepsilon_{0}+3{\cal{U}}$) and ${\cal{G}}^{(2)}_{m\sigma m\overline{\sigma}}(z)=[z-{\cal{E}}_{2}-\Sigma^{(2)}_{m\sigma m\overline{\sigma}}(z)]^{-1}$ (${\cal{E}}_{2}=2\varepsilon_{0}+{\cal{U}}$), ${\cal{G}}^{(4)}(z)=[z-{\cal{E}}_{4}-\Sigma^{(4)}(z)]^{-1}$ (where ${\cal{E}}_{4}=4\varepsilon_{0}+6{\cal{U}}$) are pseudoparticle  fermion and boson propagators. Fermion resolvents correspond to odd occupancies of adatom and boson to even. $f(\varepsilon)$ is the Fermi distribution function and $\Delta_{m\sigma}(\varepsilon)=-Im[\Sigma_{m\sigma}(\varepsilon)]$. The retarded local Green's functions may be evaluated by analytic continuation from the corresponding imaginary time propagator and can be expressed as convolution of pseudoparticle Green's functions:
\begin{eqnarray}
&&{\cal{G}}_{m\sigma}(i\omega)=(1/{\cal{Z}})\oint_{{\cal{C}}}\frac{dz}{2\pi i}e^{-z/(k_{B}T)}
[{\cal{G}}^{(0)}(z){\cal{G}}^{(1)}_{m\sigma}(z+i\omega)+{\cal{G}}^{(1)}_{m\overline{\sigma}}(z){\cal{G}}^{(2)}_{m\sigma m\overline{\sigma}}(z+i\omega)\nonumber\\
&&+\sum_{\sigma'}{\cal{G}}^{(1)}_{\overline{m}\sigma'}(z){\cal{G}}^{(2)}_{m\sigma \overline{m}\sigma'}(z+i\omega)
+\sum_{\sigma'}{\cal{G}}^{(2)}_{m\overline{\sigma}\overline{m}\sigma'}(z){\cal{G}}^{(3)}_{\overline{m}\sigma'}(z+i\omega)\nonumber\\
&&+{\cal{G}}^{(2)}_{\overline{m}\sigma \overline{m}\overline{\sigma}}(z){\cal{G}}^{(3)}_{m\sigma}(z+i\omega)+
{\cal{G}}^{(3)}_{m\overline{\sigma}}(z){\cal{G}}^{(4)}(z+i\omega)],
\end{eqnarray}
where ${\cal{Z}}$ is the impurity partition function, i.e.,
\begin{eqnarray}
&&{\cal{Z}}=\oint_{{\cal{C}}}\frac{dz}{2\pi i}e^{-z/(k_{B}T)}
[{\cal{G}}^{(0)}(z)+\sum_{m\sigma}{\cal{G}}^{(1)}_{m\sigma}(z)+\sum_{m\sigma m'\sigma'}{\cal{G}}^{(2)}_{m\sigma m'\sigma'}(z)\nonumber
\\&&+\sum_{m\sigma}{\cal{G}}^{(3)}_{m\sigma}(z)+{\cal{G}}^{4}(z)]
\end{eqnarray}
It is known that noncrossing approximations  encounter difficulties in the case of broken symmetry, it can produce at low temperatures  spurious peaks in DOS \cite{Wingreen}, but we have not observed such artifacts for the examined case. The complementary method we use  EOM,  consists in differentiating the Green's functions with respect to time which generates the hierarchy of equations with higher order GFs (11).  For the discussed  ${\cal{N}} = 3$ case apart from single and two electron also three and four particle Green's functions play the role. In order to truncate  the series of  EOM equations, we use the generalized  procedure proposed by Lacroix \cite{Lacroix} which approximates the GFs involving two conduction electron operators by single particle correlations and the corresponding  Green's function of lower order:
\begin{eqnarray}
&&\langle\langle c^{+}_{kn\overline{\sigma}}c_{qn\sigma}d_{m'\overline{\sigma}}{\cal{O}};d^{+}_{m\sigma}\rangle\rangle \simeq -\langle c^{+}_{kn\overline{\sigma}} d_{m'\overline{\sigma}}\rangle \langle\langle c_{qn\sigma}{\cal{O}};d^{+}_{m\sigma}\rangle\rangle\nonumber\\
&&\langle\langle c^{+}_{kn\overline{\sigma}}d_{m\sigma}c_{qn\overline{\sigma}}{\cal{O}};d^{+}_{m\sigma}\rangle\rangle \simeq -\langle c^{+}_{kn\overline{\sigma}} c_{qn\overline{\sigma}}\rangle \langle\langle d_{m\sigma}{\cal{O}};d^{+}_{m\sigma}\rangle\rangle\nonumber\\
&&\langle\langle d^{+}_{m'\overline{\sigma}}c_{qn\sigma}c_{kn\overline{\sigma}}{\cal{O}};d^{+}_{m\sigma}\rangle\rangle \simeq -\langle d^{+}_{m'\overline{\sigma}} c_{kn\overline{\sigma}}\rangle \langle\langle c_{qn\sigma}{\cal{O}};d^{+}_{m\sigma}\rangle\rangle\nonumber\\
&&\langle\langle c^{+}_{qn\overline{\sigma}}c_{kn\overline{\sigma}}d_{m\sigma}{\cal{O}};d^{+}_{m\sigma}\rangle\rangle \simeq +\langle c^{+}_{qn\overline{\sigma}} c_{kn\overline{\sigma}}\rangle \langle\langle d_{m\sigma}{\cal{O}};d^{+}_{m\sigma}\rangle\rangle,
\end{eqnarray}
where $m'=xz(yz)$ and ${\cal{O}}=1,n_{m_{1}\sigma_{1}},n_{m_{1}\sigma_{1}}n_{m_{2}\sigma_{2}}$.
The correlations $\langle c^{+}_{kn\sigma} d_{m'\sigma}\rangle$ and $\langle c^{+}_{kn\sigma} c_{qn\sigma}\rangle$ occurring in Eq (11) play the leading role in Kondo effect. Upon calculating these averages  self-consistently using the spectral theorem and corresponding Green's functions the  EOM set  is closed and can be therefore solved. For detailed analysis of EOM hierarchy and decoupling schemes see e.g. Refs.\cite{Kashcheheyevs, Entin}.
\begin{figure}
\includegraphics[width=6.8 cm,bb=0 0 660 660,clip]{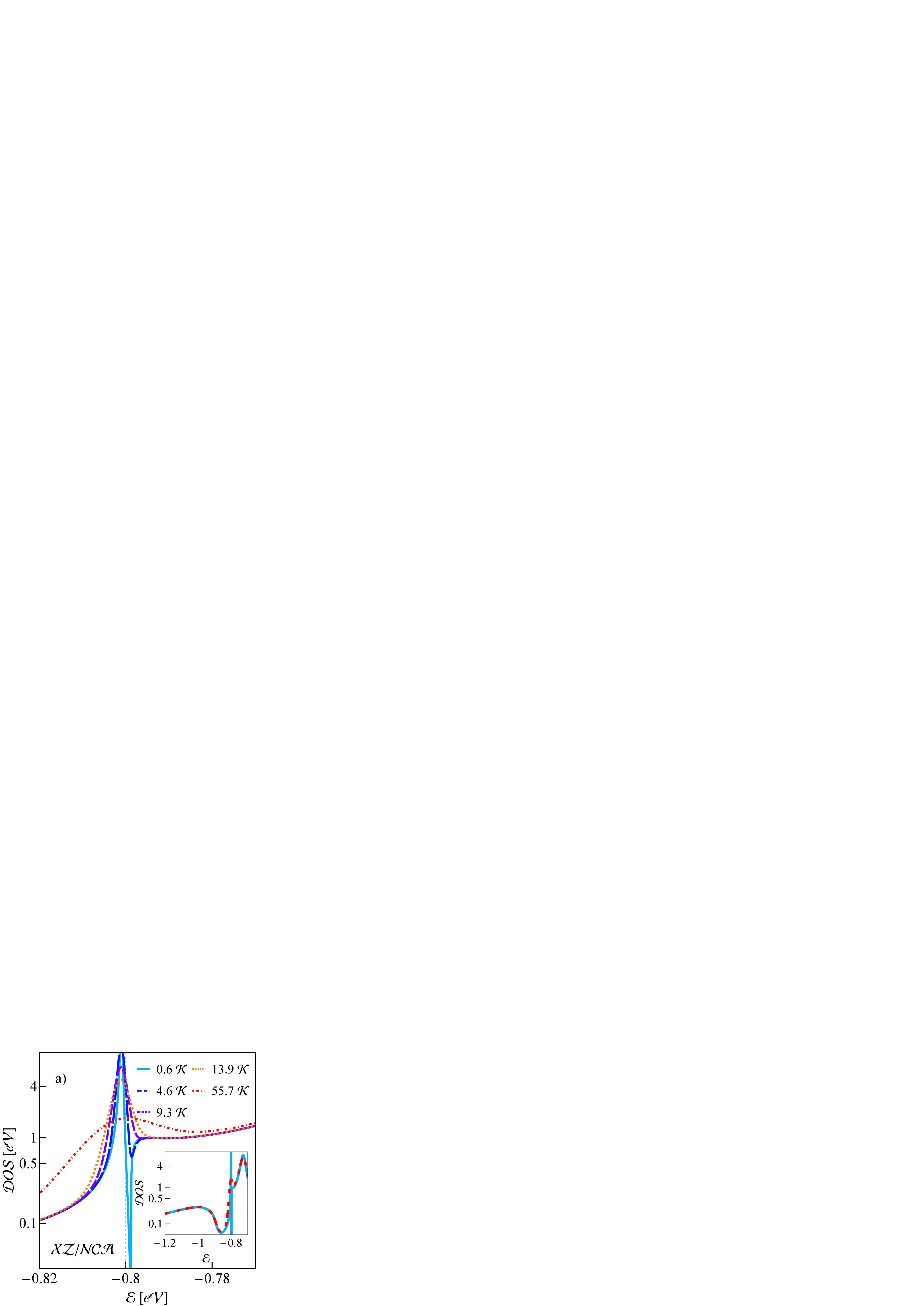}
\includegraphics[width=7 cm,bb=0 0 660 630,clip]{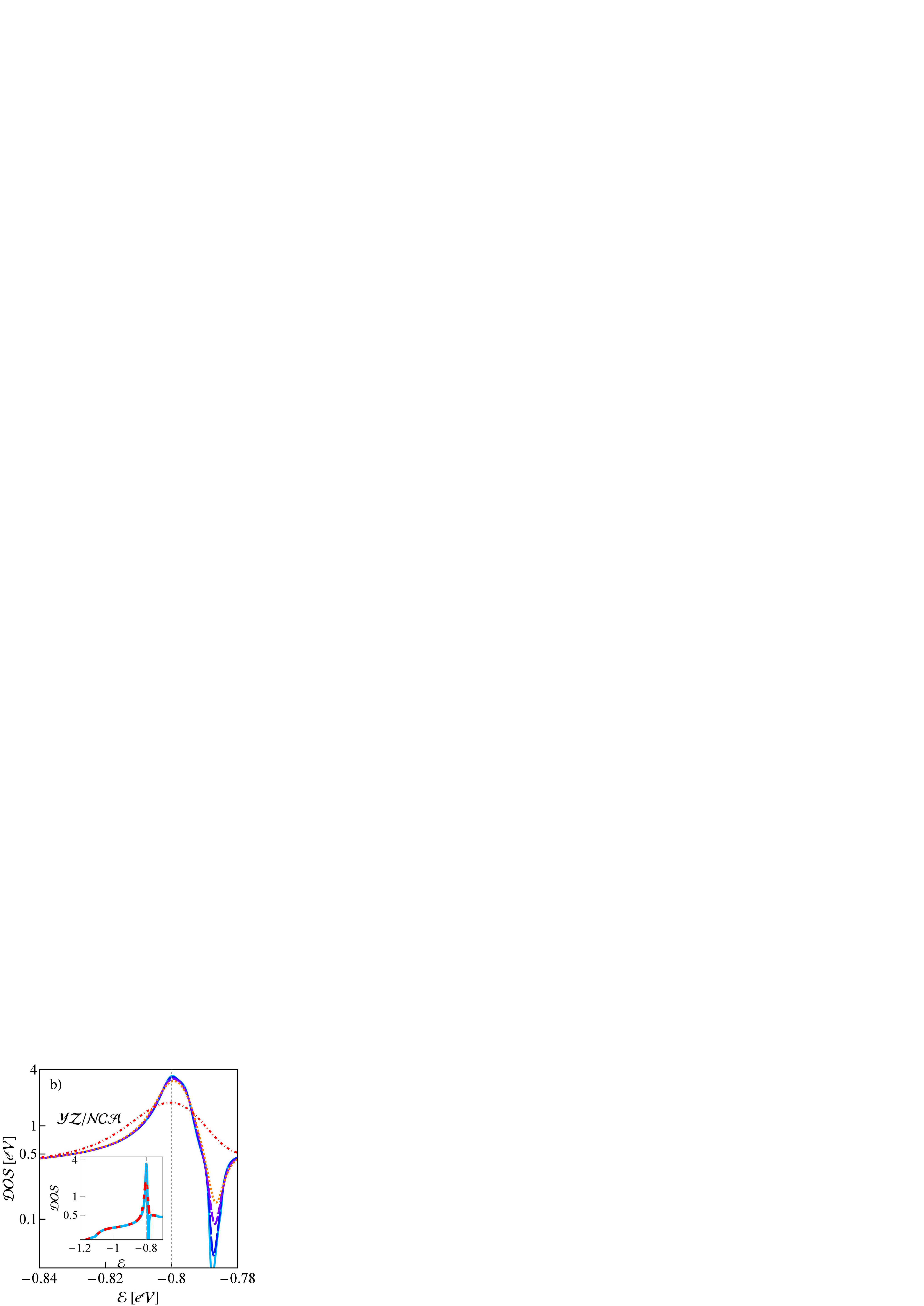}\\
\includegraphics[width=7 cm,bb=0 0 660 642,clip]{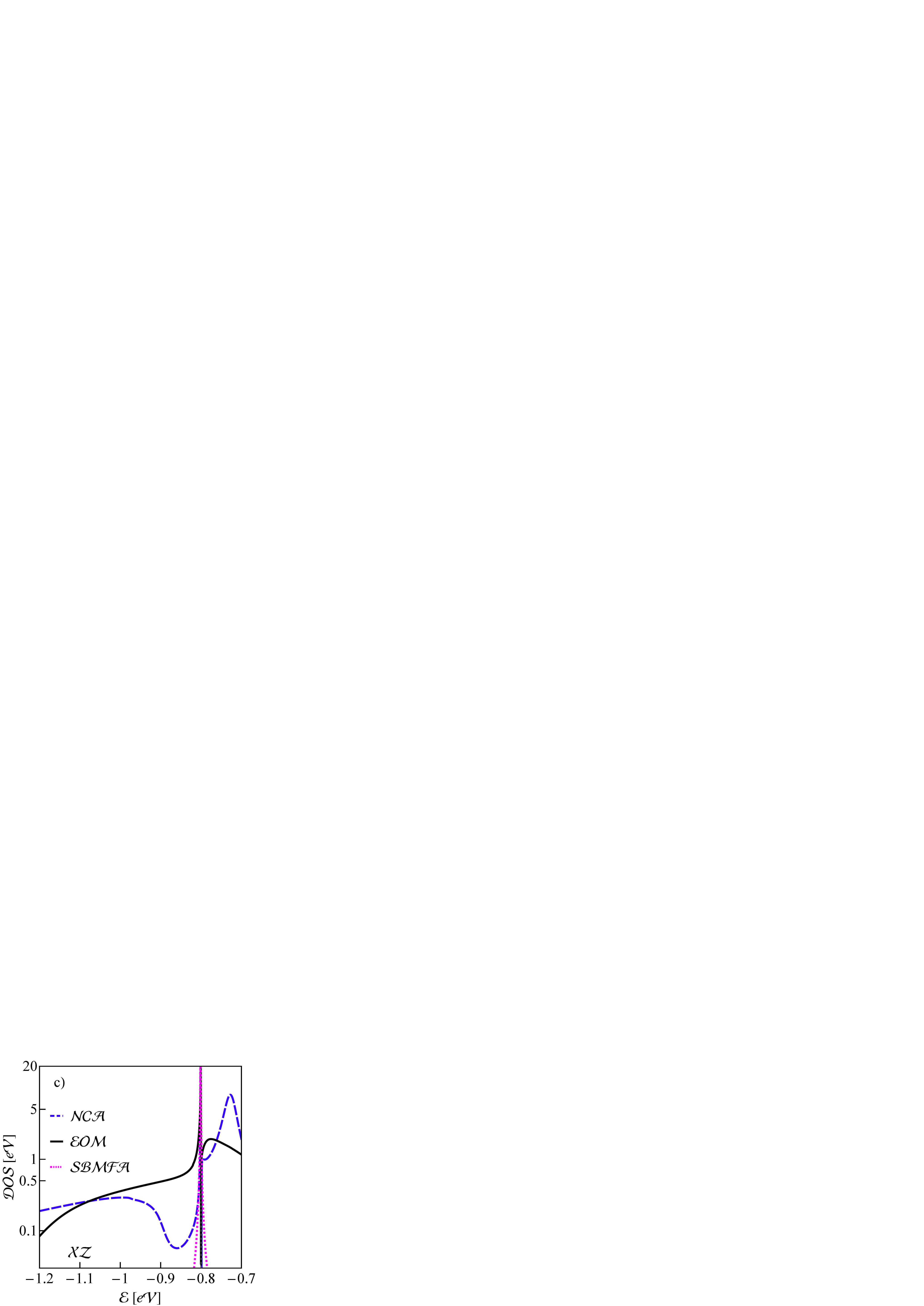}
\includegraphics[width=7 cm,bb=0 0 660 620,clip]{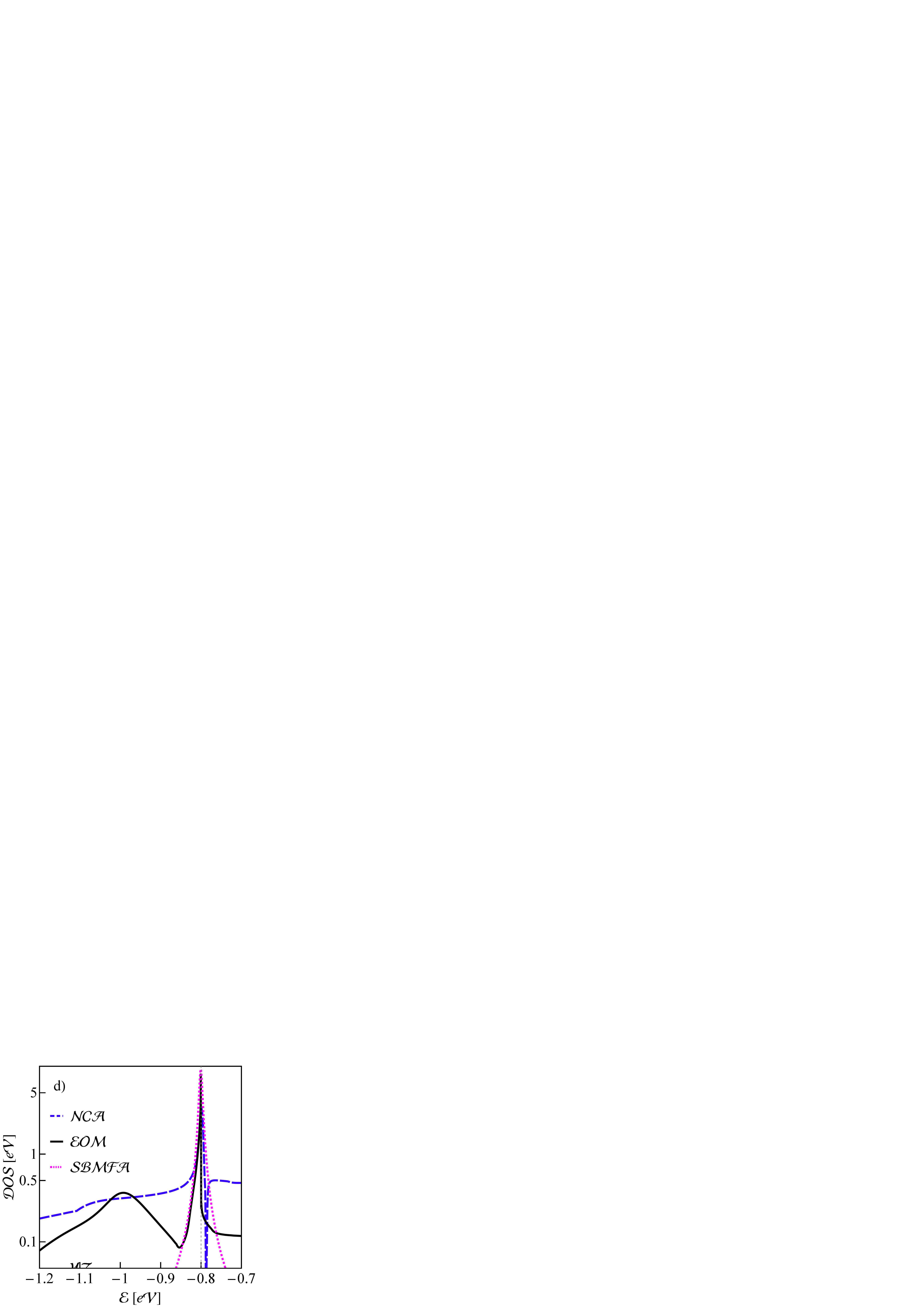}
\caption{\label{fig:epsart21}(Color online) Density of states of Co impurity at $h_{1}$ position (${\cal{E}}_{{\cal{F}}}=-0.8$) a, b) temperature evolution of orbital partial densities of states calculated with the use of NCA approximation (Insets are the extended views for $T=0.6$ K and $55.7$ K.) c, d) comparison of NCA DOS with EOM spectra and SBMFA densities of states.}
\end{figure}
The one particle NCA spectrum is assembled in Fig. 20. For the assumed position of the Fermi level (${\cal{E}}_{{\cal{F}}}=-0.8$) the main contribution comes from convolution of  ${\cal{G}}^{(2)}$ and ${\cal{G}}^{(3)}$   functions.   In addition to the many-body resonances located around ${\cal{E}}_{\cal{F}}$, also charge fluctuations peaks are visible reflecting fluctuations into fully, doubly, single occupied  and empty {$\{d_{xz}$, $d_{yz}\}$} shell.  Their positions are renormalized  and the peaks are broadened as a result of combined effect of hybridization and many body correlations.  The Coulomb peaks are  only weakly temperature dependent, whereas  significant  temperature evolution of many-body resonances is observed (Fig. 21a, b). The energy scales of spin-orbital fluctuations and charge fluctuations are not well separated and especially  (${\cal{N}} = 3$, ${\cal{N}} = 4$)  Coulomb resonances strongly perturb  quasiparticle resonances. The singularities of the nanoribbon spectrum can influence the physics around Fermi level despite the fact that they are not located close to ${\cal{E}}_{{\cal{F}}}$.  The observed dips are not direct  traces of singularities of hybridization function, they reflect singularities  of interacting self energies, which describe repeated conversion  of doubly (triple) occupied impurity into single and triple  (double and fully) occupied adatom by emitting or absorbing nanoribbon electron. The  singularities of interacting self energies however have as  source corresponding Van Hove singularities of density of states. Tracing   formal  generation of singularities  via equations (8) one can point out these connections. For example  one can identify that  singularities  at ${\cal{E}}=-0.79$ and ${\cal{E}}=-0.78$ originate from ${\cal{VH}}3\nu$ of hybridization function, these at ${\cal{E}}=-0.55$ and ${\cal{E}}=-0.5$ come from ${\cal{VH}}3c$ singularity  and at ${\cal{E}}=-0.31$ and ${\cal{E}}=-0.18$ in turn from ${\cal{VH}}4\nu$  etc. (see Figs. 11b, 20b). The main features of EOM spectrum are similar to NCA results.
\begin{figure}
\includegraphics[width=6.6 cm,bb=0 0 660 686,clip]{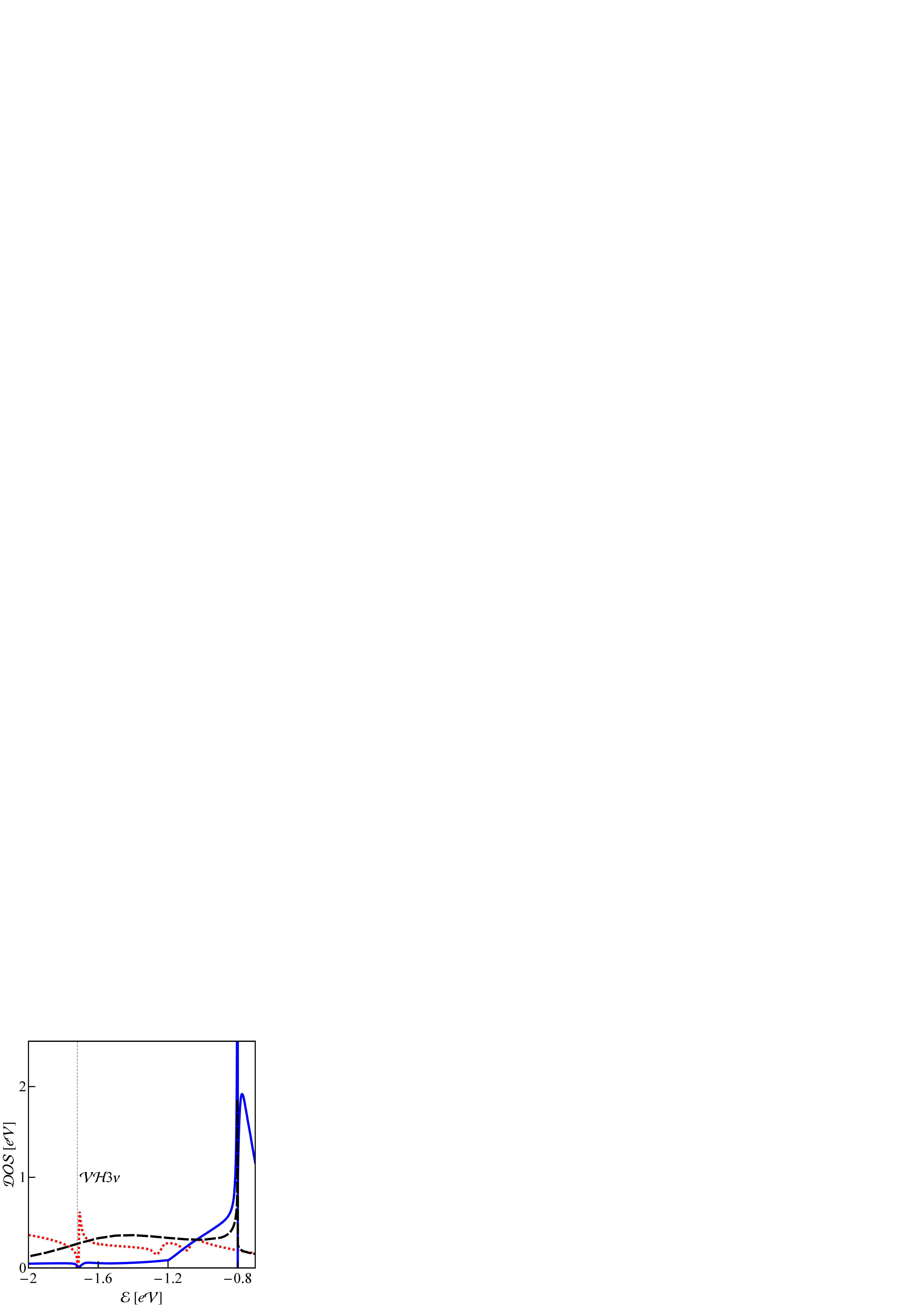}
\caption{\label{fig:epsart22}(Color online) Partial $xz$ DOS of Co adatom calculated by EOM with realistic (DFT) hybridization function of ZGNR with Lacroix's decoupling (solid blue line), without dynamical correlations (red dotted line). Broken black curve presents the corresponding DOS calculated with Lacroix's decoupling, but using energy independent, constant hybridization function.}
\end{figure}
Charge fluctuation peaks show up more clearly in EOM and the observed impact of charge fluctuation peaks next to the Fermi level on the many-body resonances is stronger than in NCA.
Similarly to NCA calculations also in EOM density of states a dip introduced by interacting self-energies, being a reminiscence of singularity of nanoribbon electronic structure is visible. The sharp dips appearing in the presented spectra would certainly be partially smoothed out if finite lifetime effects were take into account similarly to the presented temperature effects (Figs. 20b, c). This remark concerns mainly the impact of singularities on interacting self-energies, because they probe also electrons away from the Fermi level.
In some cases, in  addition to the dips, peculiarities of electronic structure of the host reflect also in spectral function of impurity  as additional peaks (see the peak slightly above the Fermi level in EOM and NCA  Co densities of states (Fig. 21).  This structure is due to a new pole of the Green's function - intersection of  $\omega - \varepsilon_{0}$  line with the real part of self-energy. Real part of interacting self energy dramatically changes between singularities taking values from a wide range of energy and thus the mentioned  intersection is likely in this interval. The occurrence of additional many-body structure is a combined effect of correlations and singularities of nanoribbon DOS.  In order to elucidate this point we present in Fig. 22  a comparison of DOS calculated in EOM considering the case of inclusion of dynamical correlations (Lacroix's decoupling) or neglect of correlations ($\langle c^{\dagger}_{kn\sigma}d_{m\sigma}\rangle = 0$) as well as comparing densities of states calculated with hybridization function from DFT with the results, where energy independent hybridization has been assumed. Additional peak above ${\cal{E}}_{\cal{F}}$ is only found when both correlations and full structure of hybridization function is taken into account. The interesting problem of the enriched structure of many-body resonances resulting from peculiarities of electronic  structure of the host has been only announced here and we leave a more detailed analysis of this problem as an open question for future work.

\section{CONCLUSIONS}
For  the  graphene or its nanostructures, a precise deposition of an atom in a selected position has not yet been implemented, but controlled adatom manipulation for the  open surface with the use of atomic force microscopy is  within reach of present-day technique \cite{Seo, Gross}. It is also possible to probe Kondo effect of magnetic adatoms on surfaces by scanning tunneling spectroscopy \cite{Madhavan, Manoharan, Mattos}. We have addressed in the present study potentially important problem for spintronic applications, the issue of geometrical and electric control of magnetic properties of Kondo impurity on ultranarrow zigzag graphene nanoribbon via peculiarities of its electronic structure. Experiments on Kondo physics in graphene nanoribbons are still missing, but we believe the results presented in this paper will stimulate the experimental effort in this direction. The presented  scheme of calculations, which is similar to some other slightly different earlier  approaches \cite{Korytar, Karolak, Wehling, Jacob} combines the first principles calculations with the addition of missing correlations by Hubbard type term and next solving the many-body problem by the well known impurity solvers. The basic input quantity for many-body analysis - hybridization function is determined by impurity - matrix hopping amplitudes and local nanoribbon DFT Green's functions, both quantities achievable from most output files of DFT programs (e.g. in VASP almost directly code from PROCAR file).

We have shown that Kondo effect of Co impurity in graphene nanoribbon is controlled not only by spin but also by the orbital degrees of freedom. Our DFT analysis showed that only two from five $d$ orbitals are responsible for magnetic properties of impurity. For the  preferred hollow positions of Co impurity and  chemical potential lying in the vicinity of the gap this role is played by  $d_{xz}$, $d_{yz}$  orbitals.   In nanoribbon the  $C_{6v}$ symmetry of pure graphene is  broken and  $d_{xz}$,   and $d_{yz}$ couple differently to nanoribbon matrix.    The presence of the edge states in ZGNR introduces   local magnetic polarization close to the edge and consequently breaks also impurity spin degeneracy  in this region.  The electronic structure of ZGNR is rich in Van Hove singularities and  this property can be exploited for electric control of magnetic properties.  If Fermi level crosses the singularities the drastic changes of hybridization functions result which  in turn reflect  in strong alternation of many-body resonances,  leading in some cases to transition from Kondo like behavior into mixed valence or even resulting in complete destroying of resonances.   For symmetry reasons the specific  singularity exhibits differently in  different spin and orbital channels and therefore not all channels are equally influenced by its presence. Crossing   the singularity by Fermi level results in some cases in an interchange of the roles of orbitals or spins leading to reversal of spin or orbital pseudospin.  Since the chemical potential can be  shifted by gate voltage, this opens a path of electric field control of these  properties. The described effects can be probed by STM. Similarly as the above reported drastic changes of Kondo correlations, also strong impact of singularities  on  coupling between magnetic impurities is expected \cite{Lipinski1}. This problem will be discussed in a forthcoming publication. Our present study shows, that the unconventional electronic and magnetic features of zigzag grapehene nanoribbons not only raise new fundamental issues in many-body physics of adatoms, but also that ZGNRs with impurities can be promising objects for potential applications in spintronics.

\begin{acknowledgments}
This work was supported by the Polish Ministry of Science and Higher Education as a research Project No. N N202 199239 for years 2010-2013.
Two of us (DK and JK) would like also to thank for support by the Institute of Molecular Physics, Polish Academy of Sciences under an internal grant for Young Scientists.
\end{acknowledgments}

\end{document}